\documentclass[12pt,preprint]{aastex}

\begin{document}
\newcommand{\kms}{\,km\,s$^{-1}$}

\title{Highly-Ionized Gas in the Galactic Halo: A FUSE Survey of \ion{O}{6} Absorption 
toward 22 Halo Stars}
\author{J. Zsarg\'{o}, K.R. Sembach, J.C. Howk\footnotemark[1]}
\affil{Department of Physics and Astronomy, The Johns Hopkins University, 
Baltimore, MD 21218}	
\author{B.D. Savage}
\affil{Department of Astronomy, University of Wisconsin, Madison, 
WI 53706}

\begin{abstract}
Far Ultraviolet Spectroscopic Explorer ($FUSE$) spectra of 22 Galactic halo stars are studied
to determine the amount of \ion{O}{6} in the Galactic halo between $\sim$0.3 and $\sim$10 kpc from
the Galactic mid-plane. Strong \ion{O}{6} $\lambda$1031.93 absorption was detected toward 21
stars, and a reliable 3 $\sigma$ upper limit was obtained toward HD~97991.  The weaker 
member of the \ion{O}{6} doublet at 1037.62~\AA\ could be studied toward only six stars because of stellar
and interstellar blending problems. The measured logarithmic total column densities
vary from 13.65 to 14.57 with $<$log$N$$>$= 14.17$\pm$0.28 (1$\sigma$). The observed columns are reasonably 
consistent with a patchy exponential \ion{O}{6} distribution with a mid-plane density of 
1.7$\times$10$^{-8}$ cm$^{-3}$ and scale height between 2.3 and 4 kpc. We do not see 
clear signs of strong high-velocity components in \ion{O}{6} absorption along the Galactic 
sight lines, which indicates the general absence of high velocity \ion{O}{6} within 2-5 kpc
of the Galactic mid-plane. This result is in marked contrast to the findings of Sembach {\it et al}.
who reported high velocity \ion{O}{6} absorption toward $\sim$60\% of the complete halo sight lines observed
by $FUSE$. 
The line centroid velocities of the \ion{O}{6} absorption does not reflect Galactic 
rotation well. The \ion{O}{6} velocity dispersions range from 33 to 78 \kms\ with an average of 
$<$$b$$>$= 45$\pm$11 \kms\ (1$\sigma$). These values are much higher than the value of $\sim$18 \kms\  
expected from thermal broadening for gas at $T \sim$ 3$\times$10$^5$ K, the temperature at which 
\ion{O}{6} is expected to reach its peak abundance in collisional ionization equilibrium. Turbulence, 
inflow, and outflow must have an effect on the shape of the \ion{O}{6} profiles.
Kinematical comparisons of \ion{O}{6} with \ion{Ar}{1} reveal that 8 of 21 sight lines are
closely aligned in LSR velocity ($|\Delta V_{LSR}| \leq$5 \kms\ ), while 9 of 21 exhibit significant 
velocity differences ($|\Delta V_{LSR}| \geq$ 15 \kms\ ). 
This dual behavior may indicate the presence of two different types of \ion{O}{6}-bearing environments
toward the Galactic sight lines. The correlation between the \ion{H}{1} and \ion{O}{6} intermediate 
velocity absorption is poor. We could identify the known \ion{H}{1} intermediate velocity components 
in the \ion{Ar}{1} absorption but not in the \ion{O}{6} absorption in most cases.
Comparison of \ion{O}{6} with other highly-ionized species suggests that the high ions are produced primarily by 
cooling hot gas in the Galactic fountain flow, and that turbulent mixing also has a significant contribution. 
The role of turbulent mixing varies from negligible to dominant. It is most important toward sight lines that sample
supernova remnants like Loop I and IV. The average N(\ion{C}{4})/N(\ion{O}{6}) ratios for the nearby halo 
(this work) and complete halo (Savage {\it et al}.) are similar ($\sim$0.6), but the dispersion is 
larger in the sample of nearby halo sight lines.  
We are able to show that the \ion{O}{6} enhancement toward the Galactic center region that was observed
in the $FUSE$ survey of complete halo sight lines (Savage {\it et al}.) is likely associated with 
processes occurring near the Galactic center by comparing the observations toward the nearby HD~177566 
sight line to those toward extragalactic targets.

\end{abstract}


\footnotetext[1]{Current Address: Center for Astrophysics and Space Sciences,
University of California at San Diego, C-0424, La Jolla, CA~92093}

\section{Introduction} \label{section:intro}

The existence of a hot, low density Galactic halo was theoretically 
postulated to provide pressure support for cool clouds
located at large distances from the Galactic plane (Spitzer 1956). 
Since that prediction, observational evidence has accumulated in support of 
this hot phase of the interstellar medium (ISM).  Detections of 
absorption lines of highly-ionized atoms, such as \ion{O}{6},
\ion{N}{5}, \ion{C}{4}, and \ion{Si}{4} (York 1974, 1977; Jenkins 1978a;
Savage \& Massa 1987; Sembach \& Savage 1992; Savage, Sembach, \& Lu 1997;
Savage et al. 2000; Savage {\it et al}. 2002; Sembach {\it et al}. 2002; 
Wakker {\it et al}. 2002) suggest the widespread, but patchy, distribution
of high temperature gas in the Galactic disk and halo. The detections of 
\ion{O}{6} are particularly important because the energy needed to ionize 
\ion{O}{5} is 113.9~eV, well above the 54.4~eV ionization potential of \ion{He}{2}. 
The amount of stellar flux capable of ionizing \ion{O}{5} 
in the ISM is limited by strong photospheric \ion{He}{2} absorption; therefore, 
it is likely that collisional ionization and not photoionization is the dominant 
source of \ion{O}{6}.  The presence of \ion{O}{6} in the Galactic
halo and the detection of soft X-ray emission at high Galactic latitudes
(Levine et al. 1976, 1977; Snowden et al. 1998) provide support to 
the existence of a hot Galactic halo.

Our best access to the interstellar \ion{O}{6}, the \ion{O}{6} 1031.93, 1037.62 
\AA\ resonance doublet, lies in a wavelength
range that is only now being explored in detail.   Apart from sporadic and 
short-lived programs to observe \ion{O}{6} with the Hopkins Ultraviolet 
Telescope (Davidsen 1993) and the {\it Orbiting and Retrievable Far and 
Extreme Ultraviolet Spectrometers} (Hurwitz \& Bowyer 1995; Hurwitz et al.
1998; Windmann et al. 1998; Sembach, Savage,
\& Hurwitz 1999), our understanding of the Galactic distribution 
of this ion was limited until recently to the interstellar 
\ion{O}{6} survey by the $Copernicus$ satellite (Jenkins 1978a,b).  
The situation has changed rapidly with the launch of the $FUSE$ satellite, which 
was specifically designed for high resolution and high sensitivity studies of 
the 905 to 1187~\AA\ spectral range.  A primary 
objective of the $FUSE$ program has been to quantify the distribution and
properties of hot gas traced by \ion{O}{6} in the Galaxy and nearby universe. 
Initial studies of the hot gas in the Magellanic Clouds (Howk {\it et al}. 2002b;
Hoopes {\it et al}. 2002) and the low-redshift universe (Sembach {\it et al}.
2001) have already been completed. 

A large number of Galactic ($\sim$200) and extragalactic ($\sim$100) targets have been observed
by the $FUSE$ Science Team for the purpose of mapping the \ion{O}{6} 
distribution in the Galactic halo and disk. 
The halo study is based primarily on a survey of extragalactic objects
(Savage {\it et al}. 2002; Sembach {\it et al}. 2002), while a sample of early-type stars is being 
used to study the properties of \ion{O}{6} in the Galactic disk at distances greater 
than those accessible with Copernicus
(Jenkins 1978a,b). Extragalactic objects -AGNs and QSOs- offer better sky coverage 
above $|b| \sim$~20$^{\circ}$ in the Galactic halo, because only a handful of suitable halo stars 
are available. Also, observations toward extragalactic objects sample the full extent of halo gas,
providing information on the outer regions of the halo where suitable Galactic targets are rare.
However, the kinematical structure of the \ion{O}{6} absorption can be confusing, and it is often
difficult to assess where the absorption occurs in the halo. 
The purpose of including early-type halo stars in the original sample was to provide 
information about the \ion{O}{6} distribution at intermediate distances away from the
Galactic plane ($|z| \sim$ 0.5-3~kpc). In the present work, we attempt to provide this information and 
fill in the distance gap between the disk and halo surveys. 

The $FUSE$ sample of Galactic halo stars is also well suited for comparisons of the highly-ionized atoms 
in regions of the Galaxy where many different physical processes may be occurring (see Shull \& Slavin 1994). 
Most of the stars in our sample have been observed previously with the International Ultraviolet Explorer 
($IUE$), the Goddard High Resolution Spectrograph (GHRS), or with the Space Telescope Imaging Spectrograph (STIS). 
Column density measurements for \ion{Si}{4}, \ion{C}{4}, and \ion{N}{5} are readily available in the literature 
(Sembach \& Savage 1992, 1994; Savage \& Sembach 1994; Savage, Meade, \& Sembach 2001; Savage, Sembach, 
\& Howk 2001) for comparison with \ion{O}{6}. The comparisons of the \ion{Si}{4}, \ion{C}{4}, \ion{N}{5}, 
and \ion{O}{6} columns are important because the models of the hot gas in the Galactic halo
predict different density ratios for the highly-ionized species.
Several physical processes have been proposed to explain the production of these species in or near the interfaces
between the hot and warm gas. These are the cooling Galactic fountain models (Shapiro \& Field 1976; 
Shapiro \&  Benjamin 1991), cooling supernova remnant (SNR) models (Slavin \& Cox 
1992, 1993; Shelton 1998), and models that involve interfaces between hot and warm gas with either turbulent 
mixing (Begelman \& Fabian 1990; Slavin, Shull, \& Begelman 1993) or conductive heat transfer occurring 
in the presence of a magnetic field (Borkowski, Balbus, \& Fristrom 1990). 
   
This paper is organized as follows.
We briefly describe the sample in \S\ref{section:Sample} and the observations 
in \S\ref{section:Obs}. We mention the major difficulties encountered in
the interpretation of the measurements and discuss the method used to derive column densities 
in \S\ref{section:AnMeth}. We present our findings on the distribution and kinematics of \ion{O}{6} 
in \S\ref{section:OVIresult}, and the \ion{O}{6} columns are compared with earlier and contemporary results on other
highly-ionized atoms in \S\ref{section:ionratios}. We comment on individual sight lines in 
\S\ref{section:peculiar}, discuss our results in \S\ref{section:Discussion}, and summarize our conclusions 
in \S\ref{section:Conclusion}.

\section{Selection of the Sample} \label{section:Sample}

The $FUSE$ Science Team obtained observations of 57 O- and B-type halo stars 
during the first two observing cycles of the mission (December 1999 - December 2001).
We examined all of these observations and selected a sub-sample to use in the 
present analysis based on the quality of the data and the complexity of the 
stellar photospheric and wind features in the vicinity of the \ion{O}{6} 1031.93, 1037.62
\AA\  doublet. In addition to these objects, the sample was expanded with four 
post asymptotic giant branch (PAGB) stars
that have strong interstellar \ion{O}{6} absorption and very straightforward continuum 
placement in the spectral region of the \ion{O}{6} doublet. These objects were
observed for Guest Investigator (GI) programs A026 (PI Heber) and A108 (PI Dixon) that
are not related to the present investigation. We are grateful to these investigators 
for their permission to use their data.

The primary set of halo stars observed by the $FUSE$ Team was designed 
to provide broad sky coverage, sampling as many directions through the Galactic
halo as possible at distances (z) greater than $\sim500$ pc from the 
Galactic plane.  Since the number of known early-type stars in the halo is 
relatively small, the sample included stars ranging in spectral type from
O6 to B3. The luminosity classes were dominated by giants and supergiants, 
although a few main sequence stars with large projected rotational 
velocities ($v$~sin~$i \gtrsim 100$ \kms) were also included in the sample.

\begin{figure}[htbp]
\begin{center}
\rotatebox{-90}{
\epsscale{1.0}
\plotone{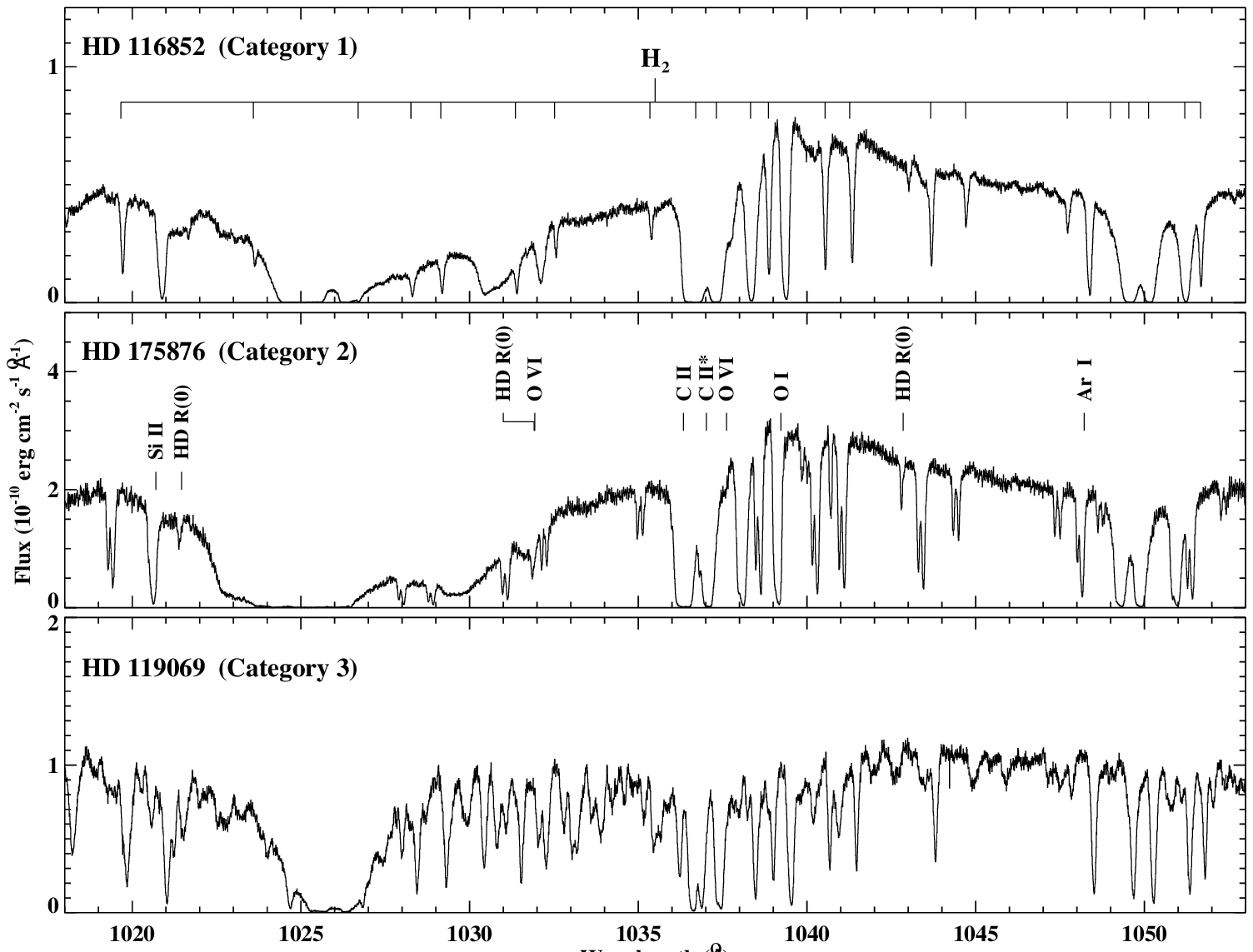}}
\figcaption{
Flux versus wavelength over the spectral range 1018-1053 \AA . The prominent H$_2$ lines are
marked  in the upper panel while \ion{O}{6} $\lambda\lambda$1031.93, 1037.62 and various
other atomic transitions are labelled in the middle panel. \ion{O}{6} can be studied in the
spectra of Category 1 objects (e.g., HD~116852) but the analysis becomes more difficult and
impossible for objects in Category 2 (e.g., HD~175876) and Category 3 (e.g., HD~119069),
respectively. 
\label{GoodandBad}}
\end{center}
\end{figure}

The inspection of the full halo star sample categorized many of the 
sight lines as being unsuitable for detailed interstellar \ion{O}{6} 
studies.  In general, supergiants often had very complex 
stellar absorption in the vicinity of the \ion{O}{6} lines, and many of them
had the added complication that their values of $v$~sin~$i$ were found to be less than 
$\sim100$~km~s$^{-1}$. We qualitatively categorized the objects into three classes based 
on an assessment of the stellar continua and line blending near 1032\,\AA: 
(1) good, with straight-forward 
continuum placement and a clear picture of line identification and component structure, 
(2) satisfactory, with detailed modeling required to fully assess the continuum 
placement or identify absorption lines and blending (e.g., the overlap of the 
HD~$R(0)_{6-0}$~$\lambda1031.91$ and \ion{O}{6}~$\lambda1031.93$ lines), and (3) poor,
with no reasonable prospects for accurate continuum placement or blending decontamination.  Of these,
we selected the category 1 (good) objects and those from category 2 (satisfactory) for which
we had confidence in the continuum placement, line identification, and blending decontamination. 
The 22 sight lines that were selected for the present investigation (including the objects from 
the GI programs) are listed in Table~\ref{tab1}. 
To help future investigations of the Galactic halo, we list the rejected stars 
in Table~\ref{tab6} with the reasons for the rejection and the categories we assigned to them.
Figure~\ref{GoodandBad} illustrates examples of category 1 (good), 2 (satisfactory), and 3 (poor) spectra. 
The interstellar absorption lines are clearly identifiable in the spectrum of HD~116852 (category 1),
with moderate absorption from H$_2$ and a well-developed stellar wind continuum around 1032 \AA.
The stellar continuum around \ion{O}{6} $\lambda$1031.93 is straightforward toward HD~175876 (category 2), but
the blending with \ion{Cl}{1}, HD, and H$_2$ lines made the assessment of the \ion{O}{6} absorption
at 1032 \AA\  difficult. The blending decontamination was further complicated by the presence of
multiple components in H$_2$ absorption. We will discuss our decontamination procedure in 
\S\ref{subsection:Cont} in more detail. Figure~\ref{GoodandBad} also shows that the 
spectrum of HD~119069 (category 3) contains numerous blended and often unidentified 
photospheric lines, making the continuum placement and the 
assessment of the interstellar \ion{O}{6} absorption futile. 

\section{Data Reduction and Calibration \label{section:Obs}}

The details of the $FUSE$ instrument design and inflight performance are discussed 
by Moos {\it et al}. (2000) and Sahnow {\it et al}. (2000). In brief, $FUSE$ involves
four co-aligned telescopes and Rowland-spectrographs that feed two microchannel-plate 
detectors. Two of these telescope/spectrograph channels are coated with SiC 
to provide reflectivity below 1000 \AA, while the other two mirrors and gratings
have Al:LiF coatings for sensitivity in the 1000 - 1187 \AA~range. We will refer 
to these pairs as SiC and LiF channels in the following discussion, respectively.
The data can be taken either in time-tag mode (TTAG) in which each photon event is recorded 
by its position and arrival time, or in histogram mode (HIST) in which the data are binned
to form a spectral image.

Table~\ref{tab2} describes the most important characteristics of the $FUSE$ observations 
of the 22 stars used in the present investigation.
The spectra of 14 stars were obtained in histogram mode, with the rest obtained in 
time-tagged mode. The light of the stars was centered in the large 
($30\arcsec\times30\arcsec$; LWRS) aperture of the LiF1 channel
in 17 cases. The small ($1.25\arcsec\times20\arcsec$; HRS) aperture was used
for observations of HD~100340, and the $4\arcsec\times20\arcsec$ (MDRS) aperture 
was used for the remaining sight lines. Since the LiF1 channel 
provides the highest effective area in the wavelength region of the
\ion{O}{6} doublet, we have restricted our analysis to data obtained in
this channel. Other channels were used to verify the presence of 
some weak features (e.g., \ion{Cl}{1} $\lambda$1031.51) as well as to 
check for fixed-pattern noise introduced by the microchannel-plate detectors.

We followed the basic data handling procedures discussed by Sembach et al.
(2000) to reduce the individual exposures using the $FUSE$ pipeline calibration
software {\tt CALFUSE} (v1.8.7). For time-tagged data, the photon event lists for 
an observation were merged before processing with CALFUSE. If the data were taken 
in histogram mode, the extracted spectra for the exposures were cross-correlated and co-added. 
When multiple observations were available, the spectra for the observations
were also cross-correlated and co-added. The spectral resolution of the fully-reduced data is 
$\sim20$ \kms\ (FWHM) for all observations. The S/N ratios per resolution element ranged from
$\sim$10 to $\sim$90 and varied from sight line to sight line since the measured fluxes and 
exposure times were different.    

The nominal reference frame for the $FUSE$ wavelength calibration should be
heliocentric; however, the absolute $FUSE$ wavelength zero-point is typically quite
uncertain. Corrections of 40-50 \kms\ to the CALFUSE calibration are not uncommon.  
Howk {\it et al}. (2002b) and Danforth {\it et al}. (2002), for example, found that 
a consistent shift of -47$\pm$5~km~s$^{-1}$ from the CALFUSE v1.8.7 wavelength solution
was needed to bring the $FUSE$ spectra of LMC/SMC stars onto a proper velocity scale.
The magnitude of this error varies from sight line to sight line because the CALFUSE
wavelength calibration errors include a sign error in the application of heliocentric 
correction in all CALFUSE versions prior to v2.0.5. Fortunately, most of the stars in our halo
sample were observed with $IUE$, so we used the measurements of Savage {\it et al}. (2001a) 
to place the $FUSE$ spectra into the proper Local Standard of Rest (LSR) frame. 
The corrections were calculated by matching the velocity centroids of \ion{Si}{2} $\lambda$1808.01 
in the $IUE$ spectra and \ion{Si}{2} $\lambda$1020.70 in the $FUSE$ spectra. For consistency, 
we applied this procedure for 
every sight line when both $IUE$ and $FUSE$ measurements were available, even if no wavelength 
calibration error was apparent in the $FUSE$ spectra. Thompson, Turnrose, \& Bohlin (1982) estimated
a $\pm$4 \kms\ (2$\sigma$) uncertainty for the absolute wavelength zero-point in $IUE$ spectra when
the object is properly centered in the $IUE$ aperture. Therefore, the contribution of the
$IUE$ uncertainty to the wavelength zero-point uncertainty of the corrected $FUSE$ spectra 
should be similar.

There were six cases, HDE~225757, HDE~233622, NGC~6397-ROB~162, NGC~5904-ZNG~1, NGC~5139-ROA~5342, and
NGC~6723-III~60, for which no $IUE$ observations were available. We used STIS observations of the 
\ion{Fe}{2} 1608.45~\AA~line (measurement done by the authors) and $FUSE$ spectra of \ion{Fe}{2} 
$\lambda$1055.26 to place the $FUSE$ observations into the heliocentric reference frame for 
HDE~233622, while 
observations of \ion{H}{1} emission (Hartmann \& Burton 1997) and the \ion{Ar}{1} $\lambda$1048.22
line in the $FUSE$ range were used for NGC 5904-ZNG~1. 
For the remaining sight lines, we could only estimate the magnitude of the errors in the heliocentric
correction by comparing the $FUSE$ spectra of the \ion{Ar}{1} 1048.22 \AA~line processed by 
CALFUSE v1.8.7 to those processed by CALFUSE v2.0.5. 
A further correction for the motion of the Sun with respect to the LSR is necessary to place 
the heliocentric velocity scales into the conventional LSR frame. We assumed that the Sun is moving 
in the direction of $l$= 56$^\circ$, $b$= 23$^\circ$ at a speed of +19.5~km~s$^{-1}$ 
(Mihalas \& Binney 1981).

Some of the analysis in our investigation depends on the comparison of absorption
features at different wavelengths (e.g., the \ion{O}{6} doublet 
and several HD lines), so we examined the relative wavelength scale 
for each object carefully. The observations were binned by 4 pixels, resulting in
a velocity sampling of $\sim$7.8~km~s$^{-1}$ ($\sim$0.027~\AA).  
The standard pipeline processing generally provides
relative wavelengths accurate to $\sim6-8$ \kms\ ($1\sigma$) when comparing absorption 
features separated by intervals of more than a few \AA ngstroms. Given these limitations 
for the reliability of velocity differences measured in our spectra, we 
concluded that only velocity differences greater than 10-15~km~s$^{-1}$ could be 
considered significant for lines several \AA ngstroms apart.
A particularly important example of the $FUSE$ wavelength calibration uncertainties is
the consistent shift between the absorption profiles of the two \ion{O}{6} lines in
spectra reduced by CALFUSE versions prior to v2.0.5. 
In their survey toward extragalactic sight lines, Wakker et al. (2002)
found that a consistent $\sim$+10~km~s$^{-1}$ velocity correction to the \ion{O}{6} 1037.62~\AA~
line was necessary to match the velocity profiles of the \ion{O}{6} doublet.
In our work, we saw a similar consistent velocity difference between the line centroids of H$_2$ 
transitions from the same rotational level but with central wavelengths above and below 1040 \AA~in 
all LWRS observations; therefore, we applied a +10 km~s$^{-1}$ correction not only to the \ion{O}{6} 
1037.62 \AA\  line but also to \ion{Ar}{1} $\lambda$1048.22 when comparisons 
with the stronger \ion{O}{6} line or with \ion{Si}{2} $\lambda$1020.70 were made.

\section{Analysis and Methodology \label{section:AnMeth}}

\subsection{Continuum Determination and Contaminating Features \label{subsection:Cont}}

One of the major challenges in our investigation was to find reasonable stellar 
continua around the two \ion{O}{6} lines, especially in the spectra of 
B giants and supergiants. We were able to establish the 
stellar continuum around the $\lambda$1031.93 line for all stars listed in
Table~\ref{tab1}. The continuum fits and the resulting normalized spectra are 
displayed in Figure~\ref{OVI1032}. We were unable to fit the stellar continuum near the \ion{O}{6} 
$\lambda$1037.62 transition for 16 stars due to the presence of strong, overlapping 
\ion{C}{2}, \ion{C}{2}*, and H$_2$ lines in this spectral region. For these stars, we could only perform a
consistency check by removing the expected \ion{O}{6} contribution at 1037.62~\AA~and making sure that
the residuals were reasonable for the sight lines. We scaled the strength of the \ion{O}{6} absorption at 
1031.93~\AA\  by 0.5 to estimate the \ion{O}{6} $\lambda$1037.62 absorption. The weaker member 
of the \ion{O}{6} doublet was extracted toward HD~18100, HD~121968, HDE~225757, 
HDE~233622, NGC~6397-ROB~162, and NGC~5904-ZNG~1. Figure~\ref{OVI1037} shows the \ion{O}{6} 
$\lambda$1037.62 continuum fits and normalized spectra for these sight lines. Generally, these continuum
fits were unreliable due to the presence of the aforementioned atomic and molecular lines around
1037~\AA. We used these profiles only to estimate the uncertainties in the total \ion{O}{6} column densities 
and to assess the saturation effects in the \ion{O}{6} absorption. In the subsequent analysis, 
the description of the \ion{O}{6} absorption and the comparison with other highly-ionized atoms are 
primarily based on the quantities inferred from the \ion{O}{6} 1031.93 \AA~line.

The difficulty in finding suitable continuum around the \ion{O}{6} doublet seriously
limited our ability to assess the uncertainties introduced by the continuum placement. It is 
usually one of the major sources of uncertainties in studies of interstellar \ion{O}{6} toward stars 
and is often estimated by investigating alternative, but still acceptable, continua. Howk {\it et al}. 
(2002b) and Hoopes {\it et al}. (2002) give examples of such procedures and discuss the issues
that affect the continuum placement toward early-type stars. Unfortunately, exploring alternative continua
was not always possible for our sight lines and of limited use in many cases. As noted in 
\S\ref{section:Sample}, viable continuum placement was one criterion for sample selection. Many
sight lines in our sample were borderline cases, for which finding any acceptable continuum is 
difficult. We found it necessary to base our error estimates only on the S/N of the spectra and the 
quality of the HD and \ion{Cl}{1} decontamination for these sight lines. In Figure~\ref{OVI1032}, we display
several possible continuum fits for the sight lines where we explored alternative continua.
In these cases, we include the uncertainties introduced by the continuum placement in our error estimates.

The other major difficulty of our study was posed by the presence of molecular and 
atomic lines around the stronger member of the \ion{O}{6} doublet. The HD~R(0) 
$\lambda$1031.91 line was the most important ``contaminant'' for our sight lines. Occasionally,
blending with the \ion{Cl}{1} $\lambda$1031.51 and H$_2$ $R(4)$ $\lambda$1032.35 lines also 
caused problems (like in the case of HD~175876, see below), but the \ion{O}{6} absorption
profiles rarely extended to $|V_{LSR}| \geq$ 120 km~s$^{-1}$ to interfere with these lines. 
Fortunately, we could use HD R(0) $\lambda$1042.85 or $\lambda$1021.46, H$_2$ $R(4)$ $\lambda$1057.38,
and \ion{Cl}{1} $\lambda$1004.67 to estimate and remove the HD, H$_2$, and \ion{Cl}{1} contaminations 
from the \ion{O}{6} absorption at 1031.93~\AA . We scaled the absorption profiles of these
transitions by the appropriate $f \lambda$ ratios to calculate the HD, H$_2$, and \ion{Cl}{1} contributions. 
Unresolved components that might be present in the HD, H$_2$, or \ion{Cl}{1} absorption did not 
seriously affect the \ion{O}{6} decontamination, because the line strengths of the
transitions used in the calculation are very similar to those of the offending lines. 
Since we estimate the contamination of \ion{O}{6} absorption around 1032 \AA\ by comparing transitions 
several \AA ngstroms apart, we compare the profiles of \ion{Cl}{1} $\lambda$1004.67, H$_2$ $R(4)$ 
$\lambda$1057.38, HD~R(0) $\lambda$1042.85, and $\lambda$1021.46 to those of H$_2$ $P(3)$ $\lambda$1031.19 
or $R(4)$ $\lambda$1032.35, and look for relative wavelength calibration errors. 
The comparison never reveals significant LSR velocity differences between the last two 
H$_2$ lines and \ion{Cl}{1} $\lambda$1004.67 or HD~R(0) $\lambda$1021.46.
In the cases of HD~R(0) $\lambda$1042.85 and H$_2$ $R(4)$ $\lambda$1057.38, we detect and 
correct for velocity shifts of $\sim$10-15 km~s$^{-1}$. 

The complex nature of extracting the \ion{O}{6} column density can be illustrated by 
the case of HD~175876. Figure~\ref{profiles} shows that there is considerable 
HD and \ion{Cl}{1} contamination in the \ion{O}{6} $\lambda1031.93$ absorption. 
The broad feature (FWHM $\sim$ 100-150 \kms\ ) near $V_{LSR} \sim$ 0 \kms\ 
in Figure~\ref{profiles} was
initially thought to be of stellar origin; however, when we calculated the
nearby \ion{Cl}{1}, HD J=0, and H$_2$ J=4 profiles (see Figure~\ref{profiles}), 
this assessment became ambiguous. 
Unfortunately, the weaker member of the \ion{O}{6} doublet is hopelessly 
blended with H$_2$ lines and is thus unavailable for comparison. After the 
removal of the \ion{Cl}{1}, HD, and H$_2$ contributions, we conclude that there are 
two components in interstellar \ion{O}{6} absorption,  more or less corresponding 
to the ones visible in the H$_2$ absorption.

\subsection{Column Density Determination}

The \ion{O}{6} halo cloud absorption is generally broad enough to be fully 
resolved by $FUSE$.  In such cases, a very simple and efficient method for
extracting information about the number of absorbing ions along the sight line 
is the apparent column density method (Savage \& Sembach 1991). The observed flux 
from an astronomical source can be represented by

\begin{equation}
F( \lambda )= \int_0^{\infty} F_* ( \lambda - \lambda' ) \, exp \left( -\tau 
\left( 
\lambda - \lambda' \right) \right) \Phi ( \lambda' ) \, \delta \lambda' \; . 
\label{eqn:intensity}
\end{equation} 
For $FUSE$ data, the stellar continuum ($F_* ( \lambda )$) varies slowly over 
the width of the line spread function ($\Phi ( \lambda )$).  Therefore, it is 
possible to estimate the stellar continuum at the wavelengths of the 
interstellar lines without having to explicitly deconvolve the stellar 
continuum from the observed spectrum. Thus, we can transform the observed flux 
to apparent optical depth by

\begin{equation}
\tau_a ( \lambda ) = \, ln \left( \frac{F_* ( \lambda )}{F( \lambda )} \right) 
\; .
\label{eqn:optdepth}
\end{equation}
The apparent column density per unit velocity is defined as

\begin{eqnarray}
N_a ( v ) = \, 3.768 \times 10^{14} \, \frac{\tau_a( v )}{\lambda_0 f};  & & v= c \, 
\frac{\lambda - \lambda_0}{\lambda_0} \; , 
\label{eqn:apColDen}
\end{eqnarray}
where $\lambda_0$, $f$, and $c$ are the wavelength at the line center (in \AA), the 
oscillator strength of the given transition, and the speed of light, 
respectively. When the line is fully resolved, the apparent column density
per unit velocity  is a valid representation of the true column density
per unit velocity (see Savage \& Sembach 1991; Jenkins 1996). 
In cases where the line is not fully resolved, the values of $N_a(v)$ may
underestimate the true $N(v)$ if saturated structure exists within the 
profiles. This can be checked by comparing the  $N_a(v)$ profiles for 
several lines of the same species having different values of $f \lambda_0$. Figure~\ref{OVIcomp} 
shows the apparent
column density comparison of the \ion{O}{6} doublet in the six cases when we could extract $N_a(v)$ 
profiles for both members of the doublet. Five of the 6 profiles in Figure~\ref{OVIcomp} reveal 
little or no saturation near the line centers. Significant differences occur on the
edges of the profiles ($|v_{LSR}| \geq$~50~km~s$^{-1}$), but this was expected since
blending with strong \ion{C}{2}, \ion{C}{2}*, and H$_2$ lines limited the range where the weaker 
\ion{O}{6} line could be adequately extracted. HD~121968 is the only sight line that shows
a modest level of saturation ($\leq$20 \%) near the line center. We explored whether alternative
continua around \ion{O}{6} $\lambda$1037.62 could improve the correspondence between the \ion{O}{6} profiles,
and found that we could nearly eliminate the saturation with a marginally acceptable continuum. Since
the level of possible saturation toward HD~121968 is small and it can be an artifact of the continuum 
placement, we did not make any correction for this saturation. The apparent column density profile
comparison of the \ion{O}{6} doublet reveals that unresolved saturated structures do not affect 
most of the sight lines in Figure~\ref{OVIcomp}.
We expect similar results to apply for the rest of our sight lines since
the \ion{O}{6} profile widths and depths are similar for all sight lines in our sample.

We also used the apparent column density method to extract information on the 
kinematical structure of \ion{Ar}{1} and \ion{Si}{2} lines. These lines may be
affected by unresolved saturated structures, but their kinematical characteristics are not
altered significantly. We used the \ion{Ar}{1} and \ion{Si}{2} 
lines for kinematical comparisons only and not for detailed column density determinations.

\section{\ion{O}{6} in the Low Galactic Halo \label{section:OVIresult}}

\subsection{Distribution of \ion{O}{6} in the Low Halo \label{section:OVIdist}}

The $FUSE$ halo survey of Savage {\it et al}. (2002) mapped the \ion{O}{6} distribution
in the Galactic halo (thick disk by their designation) using a sample of 100 extragalactic 
and 2 Galactic objects, covering much of the sky above $|b| \sim$ 20$^{\circ}$. Our sample of 22 stars is
too sparsely distributed over the sky to make a similar analysis of the halo at lower 
$|z|$ distances from the Galactic plane. Nevertheless, we can look for the phenomena and 
tendencies found by Savage {\it et al}. (2002) and provide important insights on 
remaining open questions.

Table~\ref{tab4} lists the total \ion{O}{6} columns ($N$) along our lines of 
sight, and other useful characteristics of the \ion{O}{6} absorption. The values of $N$ were calculated 
by integrating the apparent column density profiles over the velocity intervals given in column 
3 of Table~\ref{tab4}. The logarithm of the total columns range from 13.65 to 14.57, with a median of 14.25.
The logarithmic and  linear averages are $<$log$N$$>$= 14.17$\pm$0.28 (1~$\sigma$) and log$<N>$= 
14.24$^{+0.19}_{-0.36}$ (1~$\sigma$), respectively. We also list the projections of the 
total column densities perpendicular to the Galactic plane, log$N_{\perp}$= log(Nsin$|b|$), in column 
6 of Table~\ref{tab4}. These vary from 13.13 to 14.48 with $<$log$N_{\perp}$$>$= 13.77$\pm$0.37 (1~$\sigma$), 
log$<N_{\perp}>$= 13.92$^{+0.28}_{-0.95}$ (1~$\sigma$), and a median of 13.74.
The averages and medians are smaller while the standard deviations are greater than the respective 
values for the extragalactic sample ( $< $log$ N $$>$= 14.36~$\pm$~0.18 (1 $\sigma$) and a median of 14.38; 
$ < $log$ N_{\perp}$$>$= 14.21~$\pm$~0.23 (1 $\sigma$) and a median of 14.23; see Savage {\it et al}. 2002).
The scatter of our measured \ion{O}{6} columns reflects not only the general patchiness of the 
\ion{O}{6}-bearing gas, but also a general increase of the \ion{O}{6} column density 
with distance. One can account for this effect in a simplified way by calculating the average densities 
($\overline{n}= N_{tot}$/d)
along the sight lines, which we list in column 7 of Table~\ref{tab4}. The median, average, and standard
deviation of the average densities are 1.35$\times$10$^{-8}$~cm$^{-3}$, 1.56$\times$10$^{-8}$~cm$^{-3}$ and
0.75$\times$10$^{-8}$~cm$^{-3}$, respectively.

The large-scale distribution of \ion{O}{6} throughout the Galactic halo is often 
represented by an exponential function of height (z) above the Galactic disk:

\begin{equation}
n(|z|)= n_0 \; exp \left( - \frac{|z|}{h} \right) \; , \label{eqn:expon}
\end{equation}
where $n_0$ is the mid-plane density of \ion{O}{6}, $h$ is the scale height, and $z$ is the height above or 
below the Galactic plane. 
This is a very rudimentary approximation, but nevertheless useful in describing the \ion{O}{6} distribution
within a few kiloparsecs from the Sun. From Equation~\ref{eqn:expon} the vertical projection of
the column density is

\begin{equation}
N_{\perp} \equiv \, N sin |b| = \, n_0 h \, \left( 1 - exp \left( - \frac{|z|}{h} 
\right) \right) \; . 
\label{eqn:Nvert}
\end{equation}
Jenkins (1978b) estimated the mid-plane density and scale height to be 
$n_0$=~2.8~$\times$~10$^{-8}$~cm$^{-3}$ and $h$= 0.3 kpc, using $Copernicus$ 
observations of hot stars (Jenkins 1978a; Jenkins \& Meloy 1974) mainly in the Galactic
disk. A reanalysis of these data by Shelton \& Cox (1994) resulted in a new estimate 
of $n_0$=~1.3-1.5~$\times$~10$^{-8}$~cm$^{-3}$, after taking the Local Bubble into 
account, and hinted at a larger scale height ($h$= 3 kpc). The extragalactic survey by Savage 
{\it et al}. (2002) 
confirms that \ion{O}{6} is not confined to the Galactic plane and suggests a scale 
height of $h \sim$ 2.5 kpc. The combined analysis of the Copernicus and $FUSE$ disk star
sample by Jenkins {\it et al}. (2002) yields a mid-plane density of $n_0$=~1.7~$\times$~10$^{-8}$~cm$^{-3}$. 

In Figure~\ref{OVIhalo} we plot our values of $N_{\perp}$ (triangles in the upper panel) as a 
function of $|z|$ on a logarithmic scale, together with the values from the main $FUSE$ \ion{O}{6} 
survey (circles in the lower panel; Savage {\it et al}. 2002) and the Copernicus measurements toward 
disk stars (squares in the lower panel; Jenkins 1978a). The $|z|$ values of the extragalactic targets 
that were used to produce Figure~\ref{OVIhalo} are arbitrary; their actual numerical values 
are irrelevant as long as they are much larger than the 
\ion{O}{6} scale height. In Figure~\ref{OVIhalo} we also plot three 
exponential distributions with $n_0$= 1.7$\times$~10$^{-8}$~cm$^{-3}$ and three scale heights. 
The curves correspond to $h$= 1~kpc (dashed line), $h$= 2.5~kpc (solid line), and $h$= 4~kpc (dotted line). 
All data are reasonably consistent with a patchy exponential distribution with 
$n_0$= 1.7$\times$~10$^{-8}$~cm$^{-3}$ and a scale height between 2.3 and 4~kpc. 
The average $N_{\perp}$ of the extragalactic sample should be a very good measure of the \ion{O}{6} scale 
height since $<N_{\perp}> \sim$~$n_0 h$ from Equation~\ref{eqn:Nvert}; however, Savage {\it et al}. (2002) 
found an intriguing asymmetry in the average \ion{O}{6} columns of the two Galactic hemispheres. In particular, 
$<$log$N_{\perp}$$>$ is $\sim$0.25 dex higher in the northern Galactic sky than it is in the southern sky. 
To visualize this asymmetry in $<$log$ N_{\perp}$$>$, we separated the northern and southern extragalactic 
targets in Figure~\ref{OVIhalo} by increasing $|z|$ tenfold toward each sight line with $b \geq$ 0$^{\circ}$. 
Figure~\ref{OVIhalo} shows that the subsample of northern extragalactic sight lines supports a larger \ion{O}{6} 
scale height ($h \sim$ 4 kpc) than the one suggested by the southern extragalactic targets. It is also
apparent in the upper panel of the figure that the halo star measurements are also favoring an 
exponential distributions with higher values of $h$ ($\sim$4~kpc). 
Savage {\it et al}. (2002) chose to describe the \ion{O}{6} distribution in the Galactic halo by a 
combination of an exponential distribution and an excess of \ion{O}{6} in the northern 
Galactic sky. To examine whether the excess \ion{O}{6} in the Northern sky is reflected by a separation of 
the measurements made toward northern and southern Galactic stars we display all the data toward Galactic
targets with solid symbols if their Galactic latitude is greater than 0$^{\circ}$ and $|z|$ is greater 
than 250~pc. Figure~\ref{OVIhalo} shows that the high $|z|$ region of the low halo sample is dominated
by stars in the northern Galactic hemisphere while there are more southern sight lines at lower $|z|$.
Unfortunately, the low halo sight lines are not well distributed to assess where the \ion{O}{6} excess 
in the northern sky occurs. Measurements toward southern stars with higher $z$ are necessary resolve
this question.

In Figure~\ref{OVIgalaxy} we display a Hammer-Aitoff projection of the measured \ion{O}{6} column 
densities on the sky. The Galactic Center is in the middle of the figure and Galactic longitude
increases to the left. The diameter of each circle is scaled by the corresponding \ion{O}{6} column
density. Open and closed symbols represent measurements for stars with $d >$ 3 kpc and $d \leq$ 3 kpc,
respectively. We see in Figure~\ref{OVIgalaxy} that there is a slight excess of \ion{O}{6} in the 
northern sky; however, this is an artifact of Galactic sight lines sampling material to different 
distances.

\subsection{Kinematics of \ion{O}{6} in the Low Halo \label{section:Result}}

Figure~\ref{lines} displays the observed normalized absorption profiles of the \ion{O}{6} doublet for 
comparison with those of the \ion{Ar}{1}~$\lambda$1048.22 and \ion{Si}{2} $\lambda$1020.70 lines. 
We see multiple \ion{O}{6} components clearly along only four sight lines: HD~175876, HD~177989, 
NGC~6397-ROB~162, and NGC~6723-III~60; there are also strong indications for multiple components in the 
\ion{O}{6} absorption toward HD~121800 and HDE~233622. We do not see strong high-velocity 
($|V_{LSR}| \geq$ 100 \kms\ ) \ion{O}{6} absorption toward any of the halo stars. 
The general absence of high-velocity features greatly simplified the analysis since the absorption from 
\ion{O}{6} was rarely blended with the H$_2$ $P(3)$ $\lambda$1031.19 or H$_2$ $R(4)$ $\lambda$1032.35
lines. It is still possible, however, that weak 
high-velocity absorption was overlooked due to continuum placement or simply because it was masked by 
H$_2$ absorption at $\sim$~-214~km~s$^{-1}$ and +124~km~s$^{-1}$ relative to the \ion{O}{6} $\lambda$1031.93 line. 
There were three sight lines, HD~88115, HD~100340 and NGC~5139-ROA~5342, with extended wings in the 
\ion{O}{6} absorption at 1032 \AA. 
They are the best candidates for showing weak high-velocity absorption in our sample. However, the continuum 
placement around \ion{O}{6} $\lambda$1031.93 is ambiguous toward HD~100340 and NGC~5139-ROA~5342. Therefore,
the strongest evidence for high velocity absorption is the negative velocity absorption wing toward HD~88115
extending to $\sim$ -150 \kms\ .  

In Table~\ref{tab3} we compare the relevant kinematical characteristics of the \ion{O}{6} $\lambda$1031.93 and 
\ion{Ar}{1}~$\lambda$1048.22 lines to those of \ion{C}{4}~$\lambda$1548.20, and \ion{Si}{4}~$\lambda$1393.76 
from Savage {\it et al}. (2001a). We use \ion{Ar}{1} for comparison to highlight the differences
(or similarities) between the truly neutral and the highly-ionized gas. The \ion{Si}{2} absorption shares many 
but not all the features displayed by the \ion{Ar}{1} absorption in Table~\ref{tab3} since \ion{Si}{2} 
traces both the neutral and weakly ionized gas. 
The average line centroids and velocity dispersions of the species were calculated by the expressions

\begin{equation}
<v>= \frac{\int v N_a (v) dv}{\int N_a (v) dv}
\label{eq:<v>}
\end{equation}

and

\begin{equation}
b= \sqrt{2} \sqrt{ \frac{\int \left( v - <v> \right) ^2 N_a (v) dv}{\int N_a (v) dv} }
\label{eq:bvalue} \; ,
\end{equation}  
where $N_a(v)$ is the measured apparent column density per unit velocity. The quantity
$b$ is only formally related to the well-known Doppler parameter, and it becomes the $b$-value only if 
the distribution is Gaussian. It includes all the instrumental
effects, thermal and turbulent broadening, and can be affected by the presence of multiple components. 

Table~\ref{tab3} reveals close alignment ($|\Delta <$$v$$>| \leq$ 5 \kms\ ) between the line centroid of 
the \ion{O}{6} absorption and that of the \ion{Ar}{1} in 8 of 21 cases (HD~97991 is not included since 
\ion{O}{6} is not detected). Velocity differences greater than 15~\kms\  are observed in 9 cases. 
Only four sight lines exhibit velocity differences that are between 5 and 15 \kms\ . The dual behavior 
of \ion{O}{6} with respect to \ion{Ar}{1} 
may indicate the presence of two different types of \ion{O}{6}-bearing environments in the low halo.
We see significant shifts between the centroids of the \ion{O}{6} and \ion{Ar}{1} absorption toward the 
sight lines with multiple \ion{O}{6} components and toward BD+38~2182, HD~100340, HD~121800, and JL~212. 
These velocity differences are greater than 15 \kms\ .
The shift of the \ion{O}{6} line or the position of the extra component is usually toward positive velocities
with respect to the \ion{Ar}{1} lines. We see blueshifted components or significant blueshifts 
of the \ion{O}{6} centroids in three cases: the HD~175876, HDE~233622, and vZ~1128 sight lines. 
 
Table~\ref{tab3} also reveals significant differences between the average velocities of \ion{C}{4} and 
\ion{Si}{4} and those of \ion{O}{6}. 
These differences are occasionally very large. Discrepancies in the average velocities 
can occur if the distributions of the highly-ionized species are different along the line of sight. 
If there are multiple absorbing components along a sight line with different \ion{O}{6}, \ion{C}{4}, 
and \ion{Si}{4} abundances, then the average velocities calculated over the entire line profiles could 
be significantly different. Such observations are not uncommon; for example, Savage {\it et al}. (2001b) 
measured different high ion ratios in the absorbing 
components toward HD~177989. Unfortunately, the uncertainties of our wavelength calibration 
could contribute to the velocity differences, but are not likely to be 
the explanation for all of them since the differences are larger than 10-15~km~s$^{-1}$ in many cases.

To examine whether Galactic rotation has an imprint on the observed LSR velocities, 
we created a simple model to predict the expected rotational velocities of the \ion{O}{6} 
along these sight lines. We used the Galactic rotation curve of Clemens (1985) for the disk and 
assumed that the halo and disk corotate. 
The \ion{O}{6} was assumed to be exponentially distributed perpendicular to the disk with a
scale height of 2.5~kpc. A turbulent velocity dispersion of $b$= 60~\kms\ (Savage {\it et al}. 2002)
was assumed for the non-rotational motions in the lower halo.
Figure~\ref{clemens} displays the observed and predicted \ion{O}{6} line centroid velocities for our sight lines. 
There is no apparent correlation between the predicted Galactic rotation and the observed velocities.
The observed average velocities range from $\sim$-30 to $\sim$30 km~s$^{-1}$ for sight lines with $|V_{exp}| 
\leq$ 20 km~s$^{-1}$, which implies that the Galactic rotation does not play a significant role 
in shaping of the average velocities along the sight lines. 
The imprint of Galactic rotation in the LSR velocity distribution is limited by the fact 
that most of our targets are relatively nearby and at higher latitudes, 
resulting in low expected rotation velocities. In this case, the non-rotational motions like inflow/outflow 
in the high-z region dominate the average motion of the absorbing material. The fact that we 
see both excess positive and negative velocities with respect to the predicted rotational 
velocities indicates that no large scale motion dominates the low halo 
within 3-5 kpc of the Sun, and that the \ion{O}{6}-bearing gas participates both in inflow and outflow.

Table~\ref{tab3} also gives measures of the line-widths ($b$-values) for the 
high-ionization species. The typical line width of \ion{O}{6} is much broader than that expected 
from a pure thermal broadening in gas at $T_{kin} \sim$ 3$\times$10$^5$~K ($b \sim$ 18 km~s$^{-1}$). 
We find that the line dispersion for \ion{O}{6} varies from $b \sim$~33~km~s$^{-1}$ to $\sim$~78~km~s$^{-1}$.
The average, median, and the standard deviation of the line breadths are 45~km~s$^{-1}$, 44~km~s$^{-1}$, and
11~km~s$^{-1}$, respectively. Savage {\it et al}. (2002) found a $<$$b$$>$= 61$\pm$15 km~s$^{-1}$ (1$\sigma$) and 
a median of 59 km~s$^{-1}$ for the full extragalactic sample. The values derived for the Galactic disk by 
Jenkins (1978a) vary from 10.7 to 56 km~s$^{-1}$ with a median of 27~km~s$^{-1}$. The line dispersions in
all three samples are larger than the $b$-value expected from thermal broadening in a 
gas at $T \sim$ 3$\times$10$^5$ K, which suggests that \ion{O}{6} is produced in multiple environments 
dominated by inflow, outflow, and turbulence.
The progressive increase of the average line widths from the disk to the Galactic halo likely reflects 
an increase in the number of the \ion{O}{6}-bearing environments.  
Table~\ref{tab3} also shows that the widths of \ion{Si}{4}, \ion{C}{4}, and \ion{O}{6} are similar, with a 
tendency for the \ion{O}{6} line widths to be slightly larger than those of \ion{C}{4} and \ion{Si}{4}.
The \ion{C}{4} and \ion{Si}{4} line widths track each other closely. The average values of 
$<$$b$(\ion{C}{4})$>$= 37$\pm$10 km~s$^{-1}$ (1$\sigma$) and $<$$b$(\ion{Si}{4})$>$= 34$\pm$9 km~s$^{-1}$ 
(1$\sigma$) are nearly identical. The linear 
Pearson correlation coefficient of the \ion{Si}{4} and \ion{C}{4} $b$-values in Table~\ref{tab3} is 0.89, 
while the same statistical test between the \ion{O}{6} and \ion{Si}{4}/\ion{C}{4} breadths produce 
correlation coefficients less than 0.5. 
It is possible, therefore, that some of the \ion{O}{6} along the halo star sight lines sample environments 
different
from those where the majority of \ion{Si}{4} and \ion{C}{4} reside. Even though the effects of instrumental 
broadening of $IUE$ and $FUSE$ on the calculated $b$-values are similar, caution is required when assessing the 
significance of this difference between the \ion{C}{4}/\ion{Si}{4} ($IUE$) and \ion{O}{6} ($FUSE$) $b$-values.

\section{Comparison of \ion{O}{6}, \ion{C}{4}, \ion{Si}{4}, and \ion{N}{5} Column
Densities \label{section:ionratios}}

\ion{O}{6} is a good tracer of gas in the transitional state between
the hot (T$\sim$~10$^6$~K) and warm (T$\sim$~10$^4$~K) phases of the ISM. 
The observed quantities of \ion{O}{6} are most readily produced by collisional 
ionization at temperatures $T\sim(3-5)\times10^5$~K (Sutherland \& Dopita 
1993).  In this temperature regime, the gas experiences
rapid radiative cooling, and the observed amount of \ion{O}{6} should be 
eliminated in a relatively short time. Therefore, it is likely that hot 
($T \gtrsim 10^6$ K) gas is 
continuously injected into the Galactic halo, replenishing the gas at ``transition 
temperatures'' ($T \sim 10^5-10^6$~K), as it cools or comes into contact with colder 
material. The most likely source of the hot, low 
density gas is the temporally and spatially correlated supernova bursts in 
OB associations of the Galactic disk. The superbubbles created by these 
supernovae are thought to break through the denser material of the disk and eject the 
hot gas in their interiors into the Galactic halo (Norman \& Ikeuchi 1989). 
\ion{O}{6} could then be produced either within the cooling hot gas of the supernova 
remnants or at the interfaces between the hot gas of the remnant and 
the ambient interstellar medium. 

More can be learned if \ion{O}{6} is studied in conjunction with  other highly-ionized species, 
such as \ion{N}{5}, \ion{C}{4} and \ion{Si}{4}.
The theories that are proposed to explain the observed quantities of 
these highly-ionized species, predict different abundance ratios of these ions.
Sembach, Savage, \& Tripp (1997) summarize these differences in their Table~10 for the most widely 
accepted models. Of these, the 
turbulent mixing layer models produce ratios that are the most model specific. 
Predictions of the other models are fairly similar, usually 
too close to favor any one model when only a single ion ratio is available and the uncertainties of the 
respective densities are large. 
 
The behavior of interstellar \ion{N}{5} and \ion{O}{6} are similar in some respects. The source 
of \ion{N}{4} and \ion{O}{5} ionization is likely to be primarily collisional. 
Due to photospheric shielding by \ion{He}{2} in normal hot stars, the available stellar flux above 
$h \nu \geq$~54.4~eV is 
too low to produce the observed \ion{O}{6} and \ion{N}{5} columns in the quiescent warm phase of the ISM.
Unfortunately, information about interstellar \ion{N}{5} is very limited. The \ion{N}{5} 
doublet around 1240 \AA~is often very weak and not readily seen in low resolution or low 
signal-to-noise data. 
The \ion{Si}{4} and \ion{C}{4} doublets around 1400~\AA~and 1550 \AA, on the other hand, are almost 
always measurable and can offer a valuable opportunity for comparison with \ion{O}{6}. Attention
is required in this comparison, however, because \ion{Si}{4} and, to a lesser extent, \ion{C}{4} can be
produced by photoionization in an environment different from that in which \ion{O}{6} resides. 
Unlike \ion{O}{5} and \ion{N}{4}, the ionization potentials of both \ion{Si}{3} and 
\ion{C}{3} are lower than that of \ion{He}{2}. 

Savage {\it et al}. (2001a) analyzed $IUE$ spectra of the \ion{C}{4} and \ion{Si}{4} doublets and presented
total column densities and $N_a(v)$ profiles for 164 early-type stars, including 15 of our sight lines. 
In addition, high resolution GHRS measurements of the \ion{Si}{4}, \ion{C}{4}, and \ion{N}{5}
doublets are available for HD~18100 and HD~100340 (Savage \& Sembach 1994), and for HD~116852 
(Sembach \& Savage 1994).
Furthermore, Savage {\it et al}. (2001b) measured column densities of these ions toward HD~177989 using
both GHRS and STIS spectra. With the available data, a detailed comparison of \ion{C}{4}, \ion{Si}{4}, 
\ion{N}{5}, and \ion{O}{6} is possible.

\subsection{Column Density Ratios of the Highly-ionized Species \label{subsection:denrat}}

\subsubsection{Sight-line-averaged Values \label{section:totCD}}

Traditionally, the number density ratios of highly-ionized species are approximated by the ratios of the measured
total column densities. Table~\ref{tab5} shows the observed $N$(\ion{C}{4})/ $N$(\ion{Si}{4}), 
$N$(\ion{C}{4})/ $N$(\ion{O}{6}), $N$(\ion{Si}{4})/ $N$(\ion{O}{6}), and $N$(\ion{N}{5})/ $N$(\ion{O}{6}) ratios 
with their 1~$\sigma$ errors for those sight lines in our sample that have \ion{C}{4}, \ion{Si}{4}, or 
\ion{N}{5} measurements available in the literature. The sample averages and the 1~$\sigma$ dispersions of 
the respective ion ratios, together with the various model predictions, are also shown in the table. 
The sample of $N$(\ion{C}{4})/ $N$(\ion{Si}{4}) ratios in column 2 of Table~\ref{tab5} is not well enough 
constrained to study the processes that are responsible for the production of highly-ionized species in 
the Galactic halo.
We have only lower limits for this ratio in five cases that are consistent with all proposed models. 
The allowed values and limits are generally less than 5 with a sample average of 
3.5$\pm$1.1. This is somewhat smaller than the Galactic average of 4.3$\pm$1.9 (Sembach {\it et al}. 1997), 
but the two values are still in reasonable agreement. 
The values of the $N$(\ion{C}{4})/ $N$(\ion{Si}{4}) ratio in Table~\ref{tab5} 
are within the predicted ranges of the cooling Galactic fountain or the turbulent mixing layer models, 
while the cooling SNR or conductive interface models predict ratios that are much greater
than the observed values. 

The $N$(\ion{C}{4})/ $N$(\ion{O}{6}) ratios in column 3 of Table~\ref{tab5} range from 0.2 to 3.08 with most of the
values below unity and often less than 0.5. Without the lower and upper limits the sample 
average is 0.60$\pm$0.47, which is essentially identical to the 
average ratio  of 0.62$\pm$0.10 that was found for complete paths through the thick disk/halo 
(Savage {\it et al}. 2002). The large dispersion on our values suggests that \ion{C}{4} and \ion{O}{6} 
are produced 
by a diverse set of environments along the sight lines and that the ratios cannot be explained by a single model. 
We can classify the $N$(\ion{C}{4})/ $N$(\ion{O}{6}) ratios in Table~\ref{tab5} into three groups. The first group 
involves eight sight lines with ratios that are between 0.1 and 0.5. These values are 
the most consistent with the predictions from the cooling Galactic fountain models or with those from
the conductive interface models. The cooling SNR model predicts ratios that are somewhat lower than these observations. 
The second group consists of three sight lines with ratios greater than unity. Such values are predicted only by 
the turbulent mixing layer models. The third group of the remaining sight lines has values between 
0.5 and 1 that cannot be reconciled with any one of the model predictions. 
This apparent contradiction can be resolved easily if one considers that the proposed models are not 
exclusive. Observations that sample multiple \ion{C}{4} and \ion{O}{6}-bearing environments can exhibit 
total column density ratios that are intermediate between the values predicted by the models. For example, 
a sight line that samples cooling hot gas in a superbubble and mixing hot and warm gas within the bubble 
walls will most likely exhibit a $N$(\ion{C}{4})/ $N$(\ion{O}{6}) ratio that is between 0.5 and 1. 

The $N$(\ion{Si}{4})/ $N$(\ion{O}{6}) ratios in column 4 of Table~\ref{tab5} can be most readily produced by the
cooling Galactic fountain or turbulent mixing layer models. The values are generally between 0.03 and 0.5 with the 
average of $N$(\ion{Si}{4})/ $N$(\ion{O}{6})= 0.20$\pm$0.13. These ratios are too high to be produced by the cooling 
SNR or conductive interface models. Column 5 of Table~\ref{tab5} shows that the $N$(\ion{N}{5})/ $N$(\ion{O}{6}) ratios 
are generally between 0.06 and 0.41 with an average of 0.12$\pm$0.07. This is marginally lower than the average
ratio of 0.18$\pm$0.05 measured for complete paths through the Galactic halo (Savage {\it et al}. 2002).
Since \ion{O}{6} and \ion{N}{5} behave similarly in the transition temperature gas, the models predict similar values 
for their ratio. The only exception is the turbulent mixing 
layer model that predicts the highest $N$(\ion{N}{5})/ $N$(\ion{O}{6}) ratios, somewhat distinct from the predictions
of the other models. Our sample of $N$(\ion{N}{5})/ $N$(\ion{O}{6}) ratios shows that the cooling Galactic fountain and 
the turbulent mixing layer models are very successful in reproducing the observed $N$(\ion{N}{5})/ $N$(\ion{O}{6}) 
ratios, but the cooling SNR and conductive interface models are also capable of explaining many of the 
observed values.  

The synthesis of the results on the $N$(\ion{C}{4})/ $N$(\ion{Si}{4}), $N$(\ion{C}{4})/ $N$(\ion{O}{6}), 
$N$(\ion{Si}{4})/ $N$(\ion{O}{6}), and $N$(\ion{N}{5})/ $N$(\ion{O}{6}) ratios suggests that \ion{O}{6} primarily 
traces cooling hot gas along most of the sight lines, and that there are contributions from 
other processes, most likely
from turbulent mixing in the interfaces between the hot and warm ISM phases. This contribution varies from sight line 
to sight line. The high ion ratios toward approximately half of the targets in Table~\ref{tab5} are consistent with 
the predictions of the cooling Galactic fountain model of R. Benjamin (2002, private communication) without any 
contributions from turbulent mixing. 
On the other hand, turbulent mixing seems to be the primary high ion producer toward HD~121968 
and HD~177989. The ratios toward the remaining sight lines require comparable contributions from cooling hot 
gas and from turbulent mixing. 
These conclusions are well illustrated by Figures~\ref{CIVvsNV}-\ref{SiIVvsCIV} where we plot the various 
high ion ratios against each other. Figure~\ref{CIVvsNV} shows the $N$(\ion{C}{4})/ $N$(\ion{O}{6}) ratios
as a function of the $N$(\ion{N}{5})/ $N$(\ion{O}{6}) ratios. In the figure, we display the values allowed by 
the various theoretical models by rectangles. Strictly speaking, this representation is incorrect because
the ratios of the highly-ionized species are not independent from each other. 
However, the available information 
on the various model predictions did not permit us to explore such correlations; therefore, we decided to use
this simple method to visualize the model predictions. Three groups of the sight lines are immediately 
apparent in the figure. 
The $N$(\ion{N}{5})/ $N$(\ion{O}{6}) and $N$(\ion{C}{4})/ $N$(\ion{O}{6}) ratios toward HD~100340, JL~212, and probably 
toward HDE~233622 are well described by the cooling Galactic fountain model, while turbulent mixing is the dominant 
high ion producing mechanism toward HD~121968 and HD~177989. Ratios toward the HD~18100, HD~116852, and  
HD~148422 sight lines indicate mixed environments with varying contributions from the two aforementioned processes. 
Figures~\ref{SiIVvsNV} and \ref{SiIVvsCIV} support these conclusions. Sight lines 
like HD~121968 and JL~212 systematically have $N$(\ion{C}{4})/ $N$(\ion{O}{6}), $N$(\ion{Si}{4})/ $N$(\ion{O}{6}), 
and $N$(\ion{N}{5})/ $N$(\ion{O}{6}) ratios that are in agreement with the same model in all figures. 
Figures~\ref{CIVvsNV}-\ref{SiIVvsCIV} also show no difference in the behavior of the high ion ratios measured 
toward northern and southern sight lines and suggest that the high ion-bearing environments are similar in the 
two Galactic hemispheres.

There is an important difference between the behavior of the ratios in Figure~\ref{CIVvsNV} and those in
Figures~\ref{SiIVvsNV} and \ref{SiIVvsCIV}. The conductive interface and cooling SNR models are viable models 
in the first figure while their predictions could not even be displayed in the last two. Figure~\ref{CIVvsNV} 
displays the $N$(\ion{C}{4})/ $N$(\ion{O}{6}) and the $N$(\ion{N}{5})/ $N$(\ion{O}{6}) ratios that are the least 
affected by photoionization. The $N$(\ion{Si}{4})/ $N$(\ion{O}{6}) ratio that is very sensitive to the 
photoionization effects are shown in the other two figures. Among the models we used in our study, only the 
cooling Galactic fountain model of R. Benjamin (2002, private communication) and the turbulent mixing layer 
model of Slavin {\it et al}. (1993) take photoionization into account. Contributions from photoionization 
could be very important for the cooling SNR and the conductive 
interface models because a hybrid model of photoionization and any of the two may also reproduce those 
$N$(\ion{C}{4})/ $N$(\ion{O}{6}), and $N$(\ion{Si}{4})/ $N$(\ion{O}{6}) ratios that are well described by the 
cooling Galactic fountain model. This phenomenon is also apparent in the $N$(\ion{C}{4})/ $N$(\ion{Si}{4}) ratios 
in Table~\ref{tab5}. Since \ion{Si}{4} is more readily produced by photoionization than \ion{C}{4}, the contribution
from photoionization could lower the $N$(\ion{C}{4})/ $N$(\ion{Si}{4}) ratio and explain why the conductive interface
and cooling SNR model consistently overestimate this ratio. It is possible, therefore, that the success of 
the cooling Galactic fountain model in reproducing many of our high ion ratios may be the result of the more 
sophisticated state of the modeling (included photoionization) rather than the actual role of cooling Galactic 
fountains in the high ion production. Incorporating photoionization into the cooling SNR model could be especially
promising since this model underestimates \ion{C}{4} but correctly estimates \ion{N}{5} for those sight lines that
are consistent with the cooling Galactic fountain model (see Figure~\ref{CIVvsNV}). The same is not true for the
conductive interface model. Clearly, further theoretical work is necessary to fully assess the role of
cooling SNRs and conductive interfaces in the production of highly-ionized atoms.

There are two sight lines in our sample, HD~177989 and HD~121968, that show high ion ratios that suggest
a dominant role for turbulent mixing in the production of \ion{Si}{4}, \ion{C}{4}, \ion{N}{5}, and \ion{O}{6}.
HD~121968, for example, displays very large $N$(\ion{C}{4})/ $N$(\ion{O}{6}), $N$(\ion{Si}{4})/ $N$(\ion{O}{6}),
and $N$(\ion{N}{5})/ $N$(\ion{O}{6}) ratios. Both HD~121968 and HD~177989 sample special regions on the sky. 
HD~121968 is behind the Radio Loop I and IV supernova remnants, while HD~177989 is behind the Scutum supershell. 
Sembach {\it et al}. (1997) observed anomalous $N$(\ion{C}{4})/ $N$(\ion{Si}{4}) and 
$N$(\ion{C}{4})/ $N$(\ion{N}{5}) ratios toward Loops I and IV using GHRS observations of HD~119608 and 3C 273. 
The observed values were very different 
from the respective Galactic averages. Savage {\it et al}. (2001b) saw a similar deviant
behavior in the gas associated with the Scutum supershell. They used high resolution STIS and GHRS spectra of 
HD~177989 to identify the component associated with the supershell and measure $N$(\ion{C}{4})/ $N$(\ion{Si}{4}) 
and $N$(\ion{C}{4})/ $N$(\ion{N}{5}) ratios. 
A hybrid model, with equal contribution from turbulent mixing, cooling hot gas, and conductive heating was 
necessary to explain the deviant ratios toward both directions. The high values for $N$(\ion{C}{4})/ $N$(\ion{O}{6})
and $N$(\ion{N}{5})/ $N$(\ion{O}{6}) ratios toward HD~121968 and HD~177989 show that the atypical behavior extends to
\ion{O}{6} and confirms the increased importance of turbulent mixing toward these directions.  

\subsubsection{Column Density Ratios as a Function of LSR Velocity}

To test the conclusions that were drawn from the comparison of the total columns, we also compared the column 
densities in the velocity bins of the respective absorption profiles. If the observed amounts of high ions 
are produced by mixed environments then one would expect significant variation in the ratios as a 
function of LSR velocity. Figure~\ref{CDratio} shows such a comparison of \ion{C}{4} and \ion{O}{6} for our 
sight lines. The \ion{C}{4} $\lambda$1548.20 profiles are $IUE$ observations from Savage {\it et al}. 
(2001a), except for HDE~233622 and JL~212, which were extracted from STIS archival spectra.
Detailed profile comparisons of high resolution STIS spectra of \ion{C}{4} and $FUSE$ \ion{O}{6} 
observations also exist for HD~116852 (Fox {\it et al}. 2002) and for HD~177989 (Sterling {\it et al}. 2002). 
The upper panels in Figure~\ref{CDratio} display the apparent column density profiles of the 
\ion{O}{6} $\lambda$1031.93 and \ion{C}{4} $\lambda$1548.20 lines rebinned to 10 \kms\  velocity bins, while the 
lower panels show the $N$(\ion{C}{4})/ $N$(\ion{O}{6}) ratios in each velocity bin. The ratios are
displayed only if their uncertainties were less than the ratios themselves, which
excluded  HD~3827, HD~88115, HD~97991, and HD~121800 from the detailed profile comparison.  

The detailed comparison of $N_a(v)$ profiles could introduce additional 
difficulties not present in the comparison of total column densities. 
The resolution differences between the $IUE$, STIS and $FUSE$ spectra, for example, could cause problems. 
Fortunately, we saw little or no saturation in the profiles of the \ion{O}{6} doublet 
whenever we could extract both members. Similarly, \ion{C}{4} appeared well-resolved for the 
sight lines presented here (Savage {\it et al}. 2001a), except in the case of HD~177989. 
Also, the STIS profile of \ion{C}{4} $\lambda$1548.20 toward HDE~233622 showed no significant change after
convolving with a line spread function that is appropriate for $FUSE$ (FWHM$\sim$20-25~\kms\ ). 
Therefore, we conclude that the different resolutions of $IUE$, STIS, and $FUSE$ should not significantly affect 
the detailed comparison of \ion{C}{4} and \ion{O}{6} $N_a(v)$ profiles in most cases. The other difficulty
arises from the relative wavelength calibration errors between the $FUSE$ and the $IUE$ spectra. Since
we expect that these uncertainties are generally less than $\pm$10 km~s$^{-1}$, we estimated the column density
errors due to the wavelength calibration uncertainties by the magnitude of $N_a(v)$ changes that occurred 
in each velocity bin after moving the profiles by $\pm$10 \kms\ . These uncertainties, together
with the statistical errors, are included in the error estimates of the $N$(\ion{C}{4})/ $N$(\ion{O}{6}) 
ratios displayed in the lower panels of Figure~\ref{CDratio}. Generally, the ratios at the line edges and at 
LSR velocities where abrupt changes in $N_a(v)$ occur are the most affected by velocity errors.

Most sight lines in Figure~\ref{CDratio} show at least a modest level of variation in the 
$N$(\ion{C}{4})/ $N$(\ion{O}{6}) ratios, which may indicate the presence of multiple 
components with different ratios toward these sight lines. HD~121968 and HD~177989, which have very 
high integrated $N$(\ion{C}{4})/ $N$(\ion{O}{6}) ratios, clearly show the presence of several 
components, including one unusually high ($\geq$2) ratio. The very high $N$(\ion{C}{4})/ $N$(\ion{O}{6}) 
ratios around $V_{LSR} \sim$ -40 \kms\  toward HD~121968 are probably the reflections 
of the turbulent edge of Loop IV that this sight line passes through. Likewise, the component toward HD~177989 
at $\sim$ 45 \kms\ is associated with the turbulent environment of the Scutum supershell and has high 
$N$(\ion{C}{4})/ $N$(\ion{O}{6}) ratios. Sight lines like HD~18100 and HD~219188 that show 
moderate level of variations in the ratio across the line profiles also have total 
$N$(\ion{C}{4})/ $N$(\ion{O}{6}) ratios between 0.5 and 1. 
A combined model of turbulent mixing and cooling hot gas is necessary to explain these ratios. 
The sight lines whose total $N$(\ion{C}{4})/ $N$(\ion{O}{6}) ratios are most consistent with a single 
model (cooling Galactic fountain -- HD~100340, HDE~233622, and JL~212) display relatively 
featureless $N$(\ion{C}{4})/ $N$(\ion{O}{6}) ratios as a function of LSR velocity.
The detailed comparison of \ion{C}{4} and \ion{O}{6} columns, therefore, supports the conclusions drawn 
by the comparison of the total column densities in \S\ref{section:totCD}.

\subsection{High Ion Ratios as a Function of Distance from the Galactic Midplane \label{subsection:ionvsz}}

The vertical distribution of \ion{Si}{4}, \ion{C}{4}, and \ion{N}{5} has been extensively studied in the past.
Savage {\it et al}. (1997) examined a mixed sample of Galactic and extragalactic sight lines and found 
that the distributions of these ions are reasonably well-characterized by patchy exponential 
stratifications in the Galactic halo. 
The mid-plane densities and scale heights of \ion{Si}{4}, \ion{C}{4}, and \ion{N}{5} are  
$n_0$(\ion{Si}{4})=2.3~$\times$~10$^{-9}$~cm$^{-3}$, $h$(\ion{Si}{4})=5.1~kpc, 
$n_0$(\ion{C}{4})= 9.2~$\times$~10$^{-9}$~cm$^{-3}$, $h$(\ion{C}{4})= 4.4~kpc, and $n_0$(\ion{N}{5})= 
2.6~$\times$~10$^{-9}$~cm$^{-3}$, $h$(\ion{N}{5})= 3.3~kpc, respectively. Our sample of high ion ratios 
toward Galactic halo stars also offers a valuable opportunity to study
the vertical distribution of the highly-ionized species and completes the results of Savage {\it et al}. (1997) 
by including the \ion{O}{6} distribution in the investigation. Figures~\ref{CIVpSiIVvsZ}-\ref{NVvsZ} display 
the different high ion ratios as functions of $|z|$ height of the background stars. Besides our measurements, 
we also show the ratios measured toward 6 disk stars (Spitzer 1996) and the available 
ratios toward extragalactic targets (Savage {\it et al}. 2002; Hoopes {\it et al}. 2002). The vertical
distributions that are expected from the exponential stratifications of high ions are also plotted.
We used the mid-plane densities and scale heights of Savage {\it et al}. (1997) for \ion{Si}{4}, \ion{C}{4}, 
and \ion{N}{5}, as well as $n_0$(\ion{O}{6})= 1.7~$\times$~10$^{-8}$~cm$^{-3}$ and $h$(\ion{O}{6})= 2.5~kpc for 
\ion{O}{6} to calculate these predicted curves.

The exponential distributions of Savage {\it et al}. (1997) suggest that the $N$(\ion{C}{4})/ $N$(\ion{Si}{4}) 
ratios are relatively constant at most $|z|$. Figure~\ref{CIVpSiIVvsZ} shows that the observed ratios are in 
agreement with this prediction up to the distant regions of the Galactic halo. The largest discrepancies 
occur in the disk sample, which involves nearby stars ($d \leq$ 0.3 kpc) with low and unreliable \ion{C}{4} 
and \ion{Si}{4} column densities. Despite the considerable dispersion in the $N$(\ion{C}{4})/ $N$(\ion{Si}{4}) 
ratio, Figure~\ref{CIVpSiIVvsZ} shows that \ion{Si}{4} and \ion{C}{4} trace each other well and may be 
produced together in the same environment. The $N$(\ion{C}{4})/ $N$(\ion{O}{6}) ratios in Figure~\ref{CIVvsZ}, 
on the other hand, do not follow the distribution predicted by the exponential stratifications. 
The observed ratios are much lower in the disk than the predicted value of $\sim$0.54.  
At intermediate $z$ (-0.5$\leq$log$|z| \leq$0.5) the low halo sight lines display a ``chaotic'' behavior in
the $N$(\ion{C}{4})/ $N$(\ion{O}{6}) ratio. The average value of 0.6$\pm$0.47 is close to the ratios 
predicted by the exponential distributions of \ion{C}{4} and \ion{O}{6}, but the dispersion is very large. 
The values range from 0.2 to 3.08 with no obvious correspondence to the $|z|$ heights of the background stars. 
The $N$(\ion{C}{4})/ $N$(\ion{O}{6}) ratios toward the complete sight lines of Savage {\it et al}. (2002) 
do not differ significantly from those in the Galactic sample, but their scatter is smaller. Also, no 
extragalactic sight line shows a ratio much lower than 0.5. 
The behavior of the $N$(\ion{Si}{4})/ $N$(\ion{O}{6}) and $N$(\ion{N}{5})/ $N$(\ion{O}{6}) 
ratios in Figure~\ref{SiIVvsZ} and \ref{NVvsZ} are similar to that of $N$(\ion{C}{4})/ $N$(\ion{O}{6}) 
in Figure~\ref{CIVvsZ}. The ratios in the disk are generally very low while those in the halo are closer 
to the predictions, but with large dispersions. The scatter decreases with increasing $|z|$, and the 
extremely low ratios are rare toward the extragalactic targets. 

It is difficult to understand the vertical distribution of the highly-ionized species without accurate 
information on the disk gas. Spitzer (1996) reported an average $N$(\ion{C}{4})/ $N$(\ion{O}{6}) ratio 
of 0.15$^{+0.11}_{-0.07}$ and 0.93$^{+0.97}_{-0.48}$ in the disk and halo, respectively. 
This result is in marked contrast with the preliminary findings of E. Jenkins 
(2002, private communication) who found an average $N$(\ion{C}{4})/ $N$(\ion{O}{6}) ratio of 0.62$^{+0.68}_{-0.32}$ 
(shown by a cross-hatched region in Figure~\ref{CIVvsZ}) in the disk based on a sample of 54 stars with 
$|z| \leq$ 0.4 kpc. Since the later work is a more global assessment of the disk gas and involves a large 
sample of disk stars, it is likely that the sample of Spitzer (1996) is biased by selection effects. 
The high fraction of nearby stars in this sample indicates that the high ion ratios displayed for the 
disk in Figures~\ref{CIVpSiIVvsZ}-\ref{NVvsZ} reflect the conditions in the local ISM and are not 
representative for the average disk gas. Unfortunately, the results of the Jenkins 
{\it et al}. (2002) survey are not 
yet available for detailed comparison. If their preliminary result for the $N$(\ion{C}{4})/ $N$(\ion{O}{6}) 
ratios hold, then Figure~\ref{CIVvsZ} would show similar average $N$(\ion{C}{4})/ $N$(\ion{O}{6})
ratios in the disk, low halo, and distant halo. Then, the only apparent difference between the halo and 
disk ratios is the gradual decrease of scatter and the disappearance of the extremely low ratios. 

The $N$(\ion{Si}{4})/ $N$(\ion{O}{6}) ratios in Figure~\ref{SiIVvsZ} also indicate a possible peak around 
log$|z| \sim$ -0.15. Such a feature would be a strong departure from the distribution predicted by the simple 
exponential stratifications of \ion{Si}{4} and \ion{O}{6}. A closer inspection of Figure~\ref{CIVvsZ} reveals 
the possibility of a corresponding peak in the $N$(\ion{C}{4})/ $N$(\ion{O}{6}) ratios. A similar enhancement
in the $N$(\ion{Si}{4})/ $N$(\ion{N}{5}) and $N$(\ion{C}{4})/ $N$(\ion{N}{5}) ratios was observed by Savage 
{\it et al}. (1997), albeit at larger z heights (log$|z| \sim$0.5). Unfortunately, it is 
difficult to assess the exact nature of these features in the $N$(\ion{Si}{4})/ $N$(\ion{O}{6}) and 
$N$(\ion{C}{4})/ $N$(\ion{O}{6}) ratios since only limits on both ratios are available at the most 
crucial heights. If the peaks are real, they may indicate that ultraviolet radiation from OB associations 
is leaking out of superbubbles and producing \ion{C}{4} and \ion{Si}{4} without corresponding \ion{O}{6} 
and \ion{N}{5}, as was predicted by Ito \& Ikeuchi (1988). 

Finally, to look for any large-scale structure, or longitude or latitude dependence of the high ion ratios, 
we displayed the Hammer-Aitoff projection of the Galactic distribution of the $N$(\ion{C}{4})/ $N$(\ion{O}{6}) 
ratios in Figure~\ref{CIV_per_OVIgalaxy}. No large scale longitudinal or latitudal distribution is apparent 
in the figure. 
The absence of latitude dependence is further supported by the fact that Figures~\ref{CIVpSiIVvsZ}-\ref{NVvsZ} 
display no obvious difference in the behavior of the ratios measured toward the northern (solid symbols) and 
the southern (open symbols) sight lines.

\section{Comments on Individual Sight Lines \label{section:peculiar}}

In this section we comment on interesting sight lines, and compare our measurements to those of Savage 
{\it et al}. (2002) toward specific directions. Table~\ref{tab7} shows the characteristics of the thick 
disk and high velocity \ion{O}{6} absorption for those Galactic and extragalactic sight lines that are
separated in direction by 8$^{\circ}$ or less. In general, the measured thick disk absorptions toward
halo stars are less than those measured toward the extragalactic objects. This behavior is expected
if the large-scale halo/thick disk \ion{O}{6} distribution follows an exponential stratification (see
\S\ref{section:OVIdist}).
We note, however, that the comparisons of these sight lines can be very problematic. 
The recent study of the Galactic \ion{O}{6} absorption toward the Magellanic Clouds (Howk {\it et al}. 2002c) revealed 
large column density variations on sub-degree scales, comparable to those observed over large angular scales 
(Savage {\it et al}. 2002). It is not yet clear whether this is a local phenomenon or applies to the entire 
\ion{O}{6} halo. One should therefore be very careful when comparing observations even with small angular separations. 
Despite these caveats, we believe that such comparisons can still provide useful insights on the distribution of 
\ion{O}{6} along a path, especially if more than two sight lines in the same general direction are involved. We 
limited the comparisons to those opportunities when three or more closely aligned targets were available in a given 
direction, and when these observations did not reveal significant column density variations on small angular scales. 
For example, the HD~219188 sight line is within $\sim$6$^{\circ}$ of the directions toward NGC~7469 and NGC~7714 
from the extragalactic sample. The measured thick disk \ion{O}{6} columns and the kinematical structures are very
similar for these sight lines which make them suitable for comparison (see \S\ref{subsection:weak}). 
This is in contrast with
the situation toward HDE~233622, Mrk~106, and Mrk~116, another set of closely aligned Galactic and extragalactic
sight lines in the general direction of l$\sim$160$^{\circ}$ and b$\sim$45$^{\circ}$. Here, a difference of 
$\sim$0.24 dex between the measured \ion{O}{6} columns for the two extragalactic targets indicates the presence 
of significant small scale variations.

\subsection{BD+38~2182, HD~121800, HDE~233622, and the Intermediate Velocity Clouds}

BD+38~2182 is a sight line in the direction of high-velocity cloud Complex M in the northern
Galactic hemisphere. Danly, Albert, \& Kuntz (1993) constrained the distance to Complex M 
by using BD+38~2182 and the much closer star HD~93521, which is $\sim$1$^{\circ}$ away from the direction of 
BD+38~2182. Unfortunately, $FUSE$ has not observed HD~93521 as of August 2002. 
We used BD+38~2182 alone to compare the low- and 
high-ionization states toward Complex M. Ryans {\it et al}. (1997) detected several high and intermediate
velocity components in \ion{Ca}{2} and \ion{Na}{1} absorption. 
Of these, we see the intermediate velocity components near $\sim$-50~km~s$^{-1}$ 
in \ion{Ar}{1} and \ion{Si}{2} absorption (see Figure~\ref{lines}). They are probably present in \ion{C}{4}
since the \ion{C}{4} lines have extended blue wings in Figure~\ref{CDratio}. There is no indication, however, 
that intermediate velocity components at these or any other velocities are present in \ion{O}{6} absorption. 
The average velocity of \ion{O}{6} is $\sim$3~km~s$^{-1}$ smaller than the predicted velocity from 
Galactic rotation. It is likely that the \ion{O}{6} absorption toward BD+38~2182 is not related to 
the Intermediate Velocity Arch.

A similar phenomenon is apparent in the spectra of HD~121800, located behind the Intermediate Velocity Arch 
near Complex C.
Absorption from intermediate velocity clouds (IVC9 and IVC19 in Wakker 2001) is present in 
the low-ionization states, but not in those of \ion{O}{6} or \ion{C}{4}. The average velocity of the
\ion{O}{6} absorption is redshifted by $\sim$18~km~s$^{-1}$ with respect to the predicted value from the Galactic 
rotation, and there is a strong and broad \ion{O}{6} component at $\sim$12 km~s$^{-1}$ that has no counterpart in
the absorption of the low-ionization states. The nature of this \ion{O}{6} component is not clear. 
However, we can conclude that the \ion{O}{6} absorption toward HD~121800 is probably not related to the 
Intermediate Velocity Arch. This is in contrast with the observation toward PG~1351+640, an 
extragalactic sight line near the direction of HD~121800 (see Table~\ref{tab7}), where \ion{O}{6} absorption is 
detected at the velocities of both IVC9 and IVC19 (see Savage {\it et al}. 2002). 

HDE~233622 lies toward the extension of the Low Latitude Intermediate Velocity Arch. Wakker (2001) reported
an \ion{H}{1} intermediate velocity cloud at LSR velocity of $\sim$-40~\kms\ toward this sight line. This intermediate
velocity cloud is very prominent in \ion{Si}{2} absorption but weak in \ion{Ar}{1}. In this case, there is a corresponding
component in \ion{O}{6} absorption, visible in the spectrum of both member of the \ion{O}{6} doublet at $\sim$-40~\kms\ . 
The correspondence between \ion{O}{6} and \ion{H}{1} intermediate velocity absorption seems
plausible toward this sight line. This conclusion is further reinforced by the detection of \ion{O}{6} absorption
at $\sim$-40~\kms\ toward two nearby extragalactic targets, Mrk~106 and Mrk~116 (see Table~\ref{tab7}). 

The absence of intermediate velocity \ion{O}{6} absorption toward BD+38~2182 and HD~121800 implies that 
the neutral and the weakly ionized intermediate velocity gas in these directions may not interface with hot
($T >$ 10$^6$~K) gas. If such an interface existed we would expect to see transition temperature gas
(10$^4$~K $< T <$ 10$^6$~K) traced by \ion{O}{6} in the interface, as the case is toward HDE~233622.

\subsection{HD~175876, HD~177989, and the Scutum Supershell}

The measured \ion{O}{6} column densities toward two stars in our sample, HD~175876
and HD~177989, provide an interesting opportunity to do some in-depth mapping of the 
Galactic \ion{O}{6} distribution. Both sight lines lie in the general direction of the 
Scutum Supershell (GS~018-04+44), which is centered at $l$= 17.5$^\circ$, $b$= -4$^\circ$ 
at a distance of d$\sim$ 3.5 kpc and which spans about 5$^\circ$ on the sky 
(Callaway {\it et al}. 2000). HD~177989 is situated behind the Scutum supershell, while HD~175876 is 
in front of it (see Table~\ref{tab7}). Sterling {\it et al}. (2002) concluded that the component at 
$V_{LSR}$= +42 km~s$^{-1}$, having $N$(\ion{O}{6})= 7.76~$\times$~10$^{13}$~cm$^{-2}$, is associated 
with the Scutum supershell by simultaneous component fits to the \ion{O}{6} and \ion{C}{4} profiles. 
A closer inspection of Figure~\ref{lines} and \ref{CDratio} shows that 
this component is essentially absent in the \ion{O}{6} profiles of HD~175876 and is very prominent 
in those of HD~177989, which is in accord with the picture that the supershell 
is located between the two stars.

We measured a total \ion{O}{6} column of 2.04~$\times$~10$^{14}$~cm$^{-2}$ toward HD~177989
using the apparent column density method. If ones subtracts  $N$(\ion{O}{6})= 
1.38~$\times$~10$^{14}$~cm$^{-2}$, the value for HD~175876, a column density of 
6.60~$\times$~10$^{13}$~cm$^{-2}$ can be associated with the Scutum supershell. 
This is in reasonable agreement with the value of N(\ion{O}{6}) estimated by Sterling {\it et al}. 
(2002). 

There is also an interesting component toward HD~175876 at $V_{LSR} \sim$~-50~km~s$^{-1}$ that is present in
\ion{O}{6} and \ion{Ar}{1}, and possibly in \ion{C}{4} absorption (see Figure~\ref{lines} and \ref{CDratio}).
Pottasch, Wesselius \& Arnal (1980) observed a high-velocity cloud at $V_{LSR} \sim$~-95~km~s$^{-1}$ in the
spectrum of HD~175754, just $\sim$1.5$^{\circ}$ away from HD~175876 and roughly at the same distance.
We could clearly see the high-velocity feature in \ion{C}{2}~$\lambda$1036.34 and \ion{Fe}{2}~$\lambda$1144.94
toward HD~175754,
but the complex stellar wind features and strong H$_2$ absorption prevented us from extracting the
\ion{O}{6} doublet (therefore, HD~175754 was not part of our survey). On the other hand, no high or intermediate 
velocity feature is apparent in the spectrum of HD~177989 at $V_{LSR}\sim$-50 to -95~km~s$^{-1}$. It would 
be useful to clarify whether these features in the spectrum of HD~175754 and HD~175876 are related.

\subsection{Sight Lines Toward the Galactic Center}

An important finding of Savage {\it et al}. (2002) is the enhancement of \ion{O}{6} toward the region
$l$= 330$^{\circ}$ to 36$^{\circ}$ just south of the Galactic disk. They measured large \ion{O}{6} 
columns toward PKS~2005-489, Tol~1924-416, Mrk~509, and ESO~141-G55 that could be either the result of
processes occurring near the Galactic center or they may be associated with the Loop I supernova remnant. 
It was not possible to resolve this question based on the extragalactic sight line measurements alone. 

In Table \ref{tab7}, we list three Galactic and extragalactic sigh lines that are closely aligned and
lie in this general direction.
HD~177566 and NGC~6723-III~60 are two sight lines in our sample that lie within $\sim$4$^{\circ}$ and 
$\sim$8$^{\circ}$ of the direction toward Tol~1924-416, respectively. 
The separation between HD~177566 and NGC~6723-III~60 is $\sim$5$^{\circ}$. The measured \ion{O}{6} column 
toward HD~177566 is log$N$= 13.65$^{+0.06}_{-0.08}$, the lowest in our sample, which is consistent with the very 
small distance to the background star (d= 1.1 kpc). We found a much higher column density of log$N$= 
14.37$^{+0.10}_{-0.12}$
toward NGC~6723-III~60, which is a PAGB star in a globular cluster roughly 8.8 kpc away. Savage {\it et al}. (2002) 
measured a log$N$(\ion{O}{6})= 14.62$\pm$0.05 for Tol~1924-416, the highest among the three sight lines. 
Surface brightness-diameter relations (Berkhuijsen 1973) resulted in an estimate of 130$\pm$75 pc for the distance
and 230$\pm$135 pc for the diameter of Loop~I; therefore, all of the three sight lines in question are behind the 
supernova remnant. Since we measure a very low \ion{O}{6} column for HD~177566 it is likely that the 
strong thick disk \ion{O}{6} absorption toward Tol~1924-416 and NGC~6723-III~60 is not related to the Loop I 
supernova remnant and probably arises near the central regions of the Galaxy. 

There are two possible caveats with the above analysis. One is, of course, the possibility of small scale
angular variation in the \ion{O}{6} column density. However, the difference between the observed \ion{O}{6}
columns toward HD~177566 and Tol~1924-416 is almost tenfold. The probability for such a large variation over
4$^{\circ}$ is small, less than 10\% (Howk {\it et al}. 2002c). The other problem is that the distance to HD~177566
is not well defined. Despite the uncertainties in its distance, we do not expect HD~177566 to be in front of 
the Loop I supernova remnant.

\subsection{Directions with Weak \ion{O}{6} Absorption \label{subsection:weak}}

The $FUSE$ survey of \ion{O}{6} toward extragalactic targets (Savage {\it et al}. 2002) also found that 
the general region around l= 80$^{\circ}$ to 175$^{\circ}$ and b= -30$^{\circ}$ to -60$^{\circ}$ is 
deficient in \ion{O}{6}. Two of the sight lines with the smallest measured \ion{O}{6} columns in the extragalactic 
sample, NGC~7469 and NGC~7714, are toward this region (see Table~\ref{tab7}). 
One of our sight lines, HD~219188 lies near the directions to NGC~7469 and NGC~7714, with an angular separation 
of $\sim$6$^{\circ}$ from each. The observed thick disk \ion{O}{6} columns are log$N$= 13.96$\pm$0.09, 13.85$\pm$0.15, 
and 13.97$\pm$0.06, for NGC~7469, NGC~7714, and HD~219188, respectively. 
The kinematical characteristics of the \ion{O}{6} 
absorptions, average LSR velocities, and line dispersions are also similar. This correspondence is remarkable 
since an exponential \ion{O}{6} stratification with $h$(\ion{O}{6})= 2.5 kpc and a $z$= -1.789 kpc for HD~219188 
predicts a $\sim$0.29 dex column density difference between the extragalactic and the HD~219188 sight lines. 
There may be little \ion{O}{6} absorption beyond $|z| \sim$ 2 kpc in this general direction, which would explain 
the \ion{O}{6} deficiency observed by Savage {\it et al}. (2002).

\section{Discussion \label{section:Discussion}}

\subsection{The Absence of High-Velocity Absorption}

The lack of any evidence for strong or even weak high-velocity component in the \ion{O}{6} absorption 
toward our sight lines is in striking contrast with the results toward extragalactic targets. 
Sembach {\it et al}. (2002) studied the high-velocity \ion{O}{6} absorption toward 100 extragalactic objects 
and 2 distant halo stars. 
High velocity absorption was found toward $\sim$60\% of these targets. The average logarithmic
column densities of these features was found to be 13.95$\pm$0.34, comparable to the
lowest \ion{O}{6} columns in our sample. If high-velocity \ion{O}{6} absorption was as common toward halo stars
as toward extragalactic targets one would expect 10-12 detections of high-velocity absorption in our sample. 
The general absence of high-velocity \ion{O}{6} absorption in the halo star sample is  
very important because it justifies the association of the thick disk \ion{O}{6} with the low velocity 
($|V_{LSR}| \leq$100 \kms\ ) gas.
The difference between the high-velocity cloud statistics of the nearby halo and that of the more 
distant halo can be illustrated by the sight lines listed in Table~\ref{tab7}. Primary examples are 
the closely aligned HD~121800, PG~1351+640, and Mrk~279 sight lines, or the direction toward HD~100340 and Mrk~734. 
The angular separation between the first three objects is $\sim$3$^{\circ}$, but we do not see the 
high velocity \ion{O}{6} absorption toward HD~121800 that is present in the spectra of PG~1351+640 and Mrk~279.
Similarly, there is a very strong high velocity \ion{O}{6} absorption toward Mrk~734 that is certainly not 
seen toward HD~100340. The general direction toward HD~219188, NGC~7469, 
and NGC~7714, or the direction toward JL~212 and Fairall~9 are also very interesting. Strong high-velocity 
\ion{O}{6} absorption from the Magellanic Stream has been detected toward the extragalactic
objects, but no absorption at similar LSR velocities has been seen in the spectra of the halo stars.
The non-detection toward the halo stars supports the association of these high velocity features with 
the Magellanic Stream. 
 
The possibility remains, however, for the presence of weak high-velocity absorption in our sample. Such features 
could be overlooked due to the complex nature of the stellar continua around the \ion{O}{6} 1031.93 \AA~line or 
because they were blended with the H$_2$ $P(3)$ $\lambda$1031.19 or H$_2$ $R(4)$ $\lambda$1032.35 lines. 
HD~88115 and HD~100340 are the most promising candidates to show weak high velocity \ion{O}{6} absorption. 
Toward HD~100340, for example, we observed a red wing in the \ion{O}{6} profile that extends to $V_{LSR} 
\sim$ 140 km~s$^{-1}$ and maybe related to the strong redshifted high velocity feature that is seen 
toward Mrk~734 (see Table~\ref{tab7}). 
If the positive velocity wing toward HD~100340 is real it supports the idea advanced by Sembach {\it et al} 
(2002) that the positive high velocity wings are due to the outflow of gas from the disk into the halo.

\subsection{The Origin of the Highly-Ionized Gas in the Galactic Halo}

Our results confirm that the origin and distribution of \ion{Si}{4}, \ion{C}{4}, \ion{N}{5}, and \ion{O}{6} 
in the Galactic halo cannot be described in the framework of a single physical process. Observations 
toward halo stars reveal the diversity of the high ion producing environments. Most of the observed
high ion ratios suggest the presence of cooling hot gas along the sampled path with a varying
contribution from turbulent mixing in the interfaces of hot and warm material. Sight lines toward
special regions in the sky (e.g., HD~121968 toward Loop I and IV, and HD~177989 toward the 
Scutum supershell) require a dominant role for turbulent mixing in order to reconcile the observations 
with the predictions. The observed similarities in the kinematical structures of \ion{Si}{4} and 
\ion{C}{4} absorption, as well as the kinematical differences between \ion{C}{4}/\ion{Si}{4} and
\ion{O}{6} profiles suggest that photoionization may also contribute significantly to the \ion{Si}{4} 
and \ion{C}{4} production. Photoionization could be especially important in the superbubble walls where 
substantial UV flux is available from the OB associations within the bubble. Our observations 
highlight the need for hybrid models that predict the combined results of multiple physical
processes. 
There have been several efforts to create such models. Ito \& Ikeuchi (1988), for example, employed
a combination of cooling fountain flow from superbubbles and photoionization by the EUV radiation 
from OB associations in the superbubbles to predict the \ion{Si}{4}, \ion{C}{4}, and
\ion{N}{5} columns. Another example is the hybrid model of Shull \& Slavin (1994) which combines
turbulent mixing and cooling SNR gas. The role of turbulent mixing was assumed to be increasing 
with $|z|$ to explain the different \ion{C}{4} and \ion{N}{5} scale heights.
Any of these models could be a promising candidate to describe our observations if their predictions
for \ion{O}{6} columns were available.
 
The dual behavior of the \ion{O}{6} line centroids in comparison with the line centroids of the \ion{Ar}{1} 
lines is a further evidence that several high ion producing mechanisms are at operation in the Galactic halo. 
We observe significantly different \ion{O}{6} and \ion{Ar}{1} line centroids in 9 of 21 cases (excluding 
HD~97991). A poor correspondence between the LSR velocities of \ion{O}{6} and those of the colder gas are
expected when the majority of the \ion{O}{6} is produced in cooling hot gas.  
However, we see good agreement between the line centroid velocities of the low- and high-ionization states 
in 8 of 21 cases, which is difficult to understand if \ion{O}{6} is only produced in cooling hot gas. Therefore,
significant amount of \ion{O}{6} must be produced by turbulent mixing or thermal conduction in the interfaces 
between the hot and warm gas since such an arrangement would naturally explain the kinematical correspondence 
between the low- and high-ionization states (Cowie {\it et al}. 1979). 

Because of its large cross-section for photoionization (Sofia \&
Jenkins 1998; Jenkins {\it et al}. 2000) , \ion{Ar}{1} is easily 
ionized in the ISM and it therefore is
not a good tracer of low column density neutral gas structures or of
warm ionized gas structures. Such structures could also give rise to 
\ion{O}{6} absorption if the structures are surrounded by hot 10$^6$ K gas. An
inspection of the line profiles in Figure~\ref{lines} reveals that \ion{O}{6} 
is better aligned with \ion{Si}{2} $\lambda$1020.70 than with \ion{Ar}{1} 
$\lambda$1048.22.  The \ion{Si}{2} ion may represent  a better comparison ion 
for \ion{O}{6} since it traces cool and warm  gas with wider ranges of conditions 
than \ion{Ar}{1}.  Detailed intercomparisions of the kinematical relationships 
among \ion{O}{6} and the many other tracers of the cold neutral, warm neutral, 
and warm ionized gas along the 22 sight lines studied here would  represent 
a valuable extension of this investigation. Such an investigation should be
pursued with spectra extracted with the best possible wavelength
calibrations. 

\section{Summary \label{section:Conclusion}}

We have measured the \ion{O}{6} $\lambda\lambda$1031.93, 1037.62 absorption along partial
path lengths through the Galactic halo toward 22 Galactic halo stars, and examined the \ion{O}{6} distribution 
within $\sim$3-5 kpc of the Galactic mid-plane. Our results are compared to the findings of Savage {\it et al}. 
(2002) and Sembach {\it et al}. (2002) toward 102 complete sight lines through the Galactic halo. 
We also study the role of different physical processes in the production of highly-ionized species by 
comparing the observed N(\ion{C}{4})/N(\ion{Si}{4}),  N(\ion{C}{4})/N(\ion{O}{6}),  N(\ion{Si}{4})/N(\ion{O}{6}), 
and N(\ion{N}{5})/N(\ion{O}{6}) ratios to the predictions of the current models. Our main results are as follows:

1. Strong \ion{O}{6} absorption is observed at 1031.93 \AA\  in all of the directions, except toward HD~97991. 
We were able to extract the \ion{O}{6} absorption at 1037.62 \AA~ for 6 directions. 
We find that the total \ion{O}{6} logarithmic column densities vary from 13.65 to 14.57 with an average of 
$<$log$N$$>$= 14.17$\pm$0.28 and a median of 14.25. The logarithm of the column density perpendicular
to the Galactic mid-plane for these partial paths varies between 13.13 and 14.48 with an average of 
$<$log$Nsin|b|$$>$= 13.77$\pm$0.37.

2. The \ion{O}{6} column densities toward the 22 halo stars are reasonably well described by a patchy 
exponential distribution. Our measurements are the most consistent with $n_0$= 1.7$\times$10$^{-8}$ cm$^{-3}$ 
and a scale height between 2.3 and 4 kpc. 

3. We do not see any strong or weak high-velocity component in the \ion{O}{6} absorption along our
sight lines. Sembach {\it et al}. (2002) reported high velocity \ion{O}{6} absorption toward $\sim$60\%
of the complete halo sight lines. The non-detection of high-velocity absorption in our sample suggests 
that these features originate in the distant halo and confirms the association of the thick disk gas 
with the low velocity ($|V_{LSR}| \leq$ 100 \kms\ ) absorption in the \ion{O}{6} survey of 
Savage {\it et al}. (2002). 

4. Comparison of the \ion{O}{6} and \ion{Ar}{1} line centroid velocities reveals a mixed picture.
Significant velocity differences ($\geq$ 15 \kms\ ) are observed in 9 cases. 
\ion{Ar}{1} and \ion{O}{6} are closely aligned ($|\Delta V_{LSR}| \leq$ 5 km~s$^{-1}$)
in 8 cases and there are 5-15 \kms\  differences in the remaining 4 cases. The dual behavior
of \ion{O}{6} with respect to \ion{Ar}{1} may indicate the presence of two types of \ion{O}{6}-bearing
environments along the sight lines studied. 
  
5. There is no obvious correlation between Galactic rotation and the measured LSR velocities
of the \ion{O}{6} absorption. We see deviations from the predicted rotational velocities in positive
and negative directions with equal frequency. The \ion{O}{6}-bearing gas in the low halo participates 
both in inflow and outflow toward the Galactic plane.

6. The velocity dispersions ($b$-values) of the \ion{O}{6} profiles vary from 33 to 78 \kms\ with
an average of 45$\pm$11 \kms\ . These values are much larger than the profile widths $b \sim$ 18 
\kms\ expected at T$\sim$ 3$\times$10$^5$ K, the temperature at which \ion{O}{6} peaks in abundance
in collisional ionization equilibrium. A considerable amount 
of turbulent broadening or the presence of multiple components is necessary to explain the observed 
line breadths. The average \ion{O}{6} line width is smaller toward the halo stars than it is along
complete sight lines through the halo (Savage {\it et al}. 2002), reflecting the larger number of 
\ion{O}{6}-bearing structures along the complete sight lines.

7. Kinematical comparisons of the \ion{Si}{4}, \ion{C}{4}, and \ion{O}{6} line profiles reveal that
the \ion{Si}{4} and \ion{C}{4} line centroids and line widths correlate with each other well but not with
those of the \ion{O}{6} absorption. There may be a substantial amount of photoionized \ion{C}{4} and \ion{Si}{4} 
without corresponding \ion{O}{6} along the sight lines.

8. The origin of the highly ionized species in the halo cannot be described by a 
single physical process. The observed N(\ion{C}{4})/N(\ion{Si}{4}), N(\ion{C}{4})/N(\ion{O}{6}), 
N(\ion{Si}{4})/N(\ion{O}{6}), and N(\ion{N}{5})/N(\ion{O}{6}) ratios along the sight lines suggest that 
the highly-ionized species are most readily produced by cooling gas in a Galactic fountain flow with 
contributions from the turbulent interfaces between hot and warm gas. The role of turbulent mixing varies 
from sight line to sight line. 

9. The high N(\ion{C}{4})/N(\ion{O}{6}) ratios toward HD~121968 behind the Loop I and IV
supernova remnants, and toward HD~177989 behind the Scutum supershell suggest that turbulent mixing
has a dominant role in producing \ion{C}{4} and \ion{O}{6} toward the disturbed environments of 
supernova remnants and supershells.

10. Detailed comparisons of the \ion{C}{4} and \ion{O}{6} profiles reveal modest to large variations in
the N(\ion{C}{4})/N(\ion{O}{6}) ratios across the line profiles. We see the largest fluctuations when turbulent
mixing has an important role in the production of \ion{C}{4} and \ion{O}{6}.

11. The vertical distribution of the high ion ratios toward the halo stars (present analysis) 
and the complete (Savage {\it et al}. 2002) sight lines suggest that \ion{C}{4} and \ion{Si}{4} trace 
each other well in the Galactic halo, while \ion{C}{4} and \ion{O}{6} exhibit large variations. 
The average N(\ion{C}{4})/N(\ion{O}{6}) ratio is $\sim$0.6 throughout 
the entire halo with a decreasing dispersion toward larger $z$ heights. 

12. The \ion{O}{6} absorption does not correlate well with the known \ion{H}{1} intermediate velocity clouds.
We can identify known intermediate velocity features in the \ion{O}{6} absorption toward HDE~233622, but 
we see no correspondence toward BD+38~2182 and HD~121800. In the case of HD~121800, the component structure
of the \ion{O}{6} absorption is drastically different from that of the low-ionization states.
 
13.  Observations toward Galactic stars located in the general direction of $l$= 355$^{\circ}$ to 360$^{\circ}$ and 
$b$= -16$^{\circ}$ to -21$^{\circ}$ suggest that the large \ion{O}{6} columns observed by Savage {\it et al}. (2002) 
toward the region $l$= 330$^{\circ}$ to 36$^{\circ}$ and $b$= -24$^{\circ}$ to -33$^{\circ}$ are related to 
processes occurring near the Galactic center, and are not associated with the Loop I supernova remnant.

14. Savage {\it et al}. (2002) found that the general region in the sky toward $l$= 80$^{\circ}$ to 175$^{\circ}$ and 
$b$= -30$^{\circ}$ to -60$^{\circ}$ is deficient in \ion{O}{6}. Our observation toward HD~219188 in this
direction reveals an \ion{O}{6} column similar to those of the nearby extragalactic sight lines. This suggests that 
the \ion{O}{6} may be confined to within 2 kpc of the Galactic mid-plane in this direction.

\acknowledgements

This work is based on data obtained for the Guaranteed Time Team by
the NASA-CNES-CSA $FUSE$ mission operated by the Johns Hopkins
University. Financial support to U. S. participants has been provided by
NASA contract NAS5-32985. KRS acknowledges support through NASA contract NAS5-32985 and 
Long Term Space Astrophysics grant NAG5-3485. We thank Ulrich Heber and Van Dixon for 
their permission to use their $FUSE$ guest investigator data, and Alex Fullerton
for providing the $FUSE$ spectra processed by CALFUSE v1.8.7.

\clearpage

\clearpage
\begin{figure}[htbp] 
\begin{center}
\rotatebox{0}{
\epsscale{1.0}
\plotone{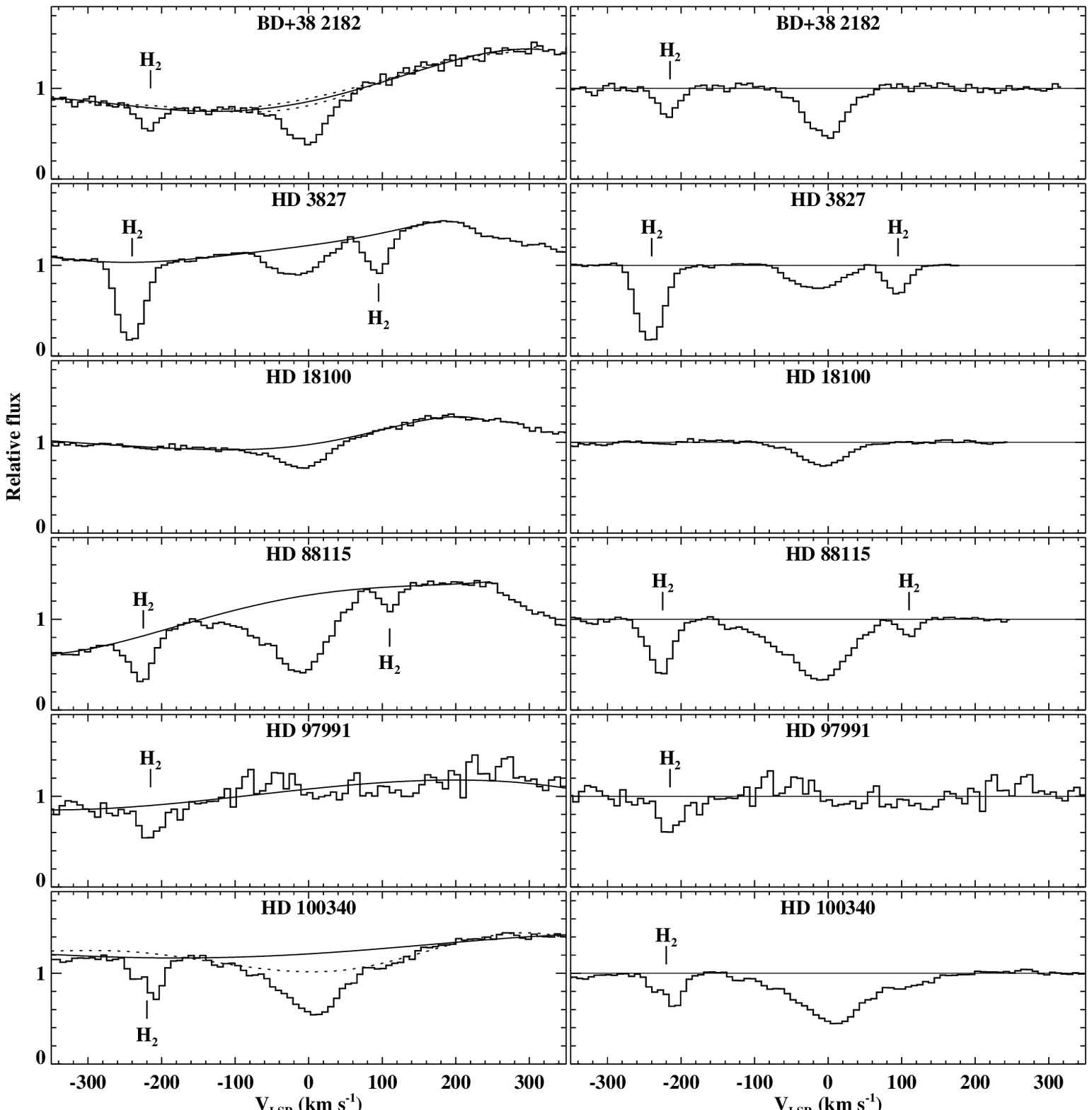}}
\figcaption{}
\end{center}
\end{figure}
\addtocounter{figure}{-1}

\clearpage
\begin{figure}[htbp] 
\begin{center}
\rotatebox{0}{
\epsscale{1.0}
\plotone{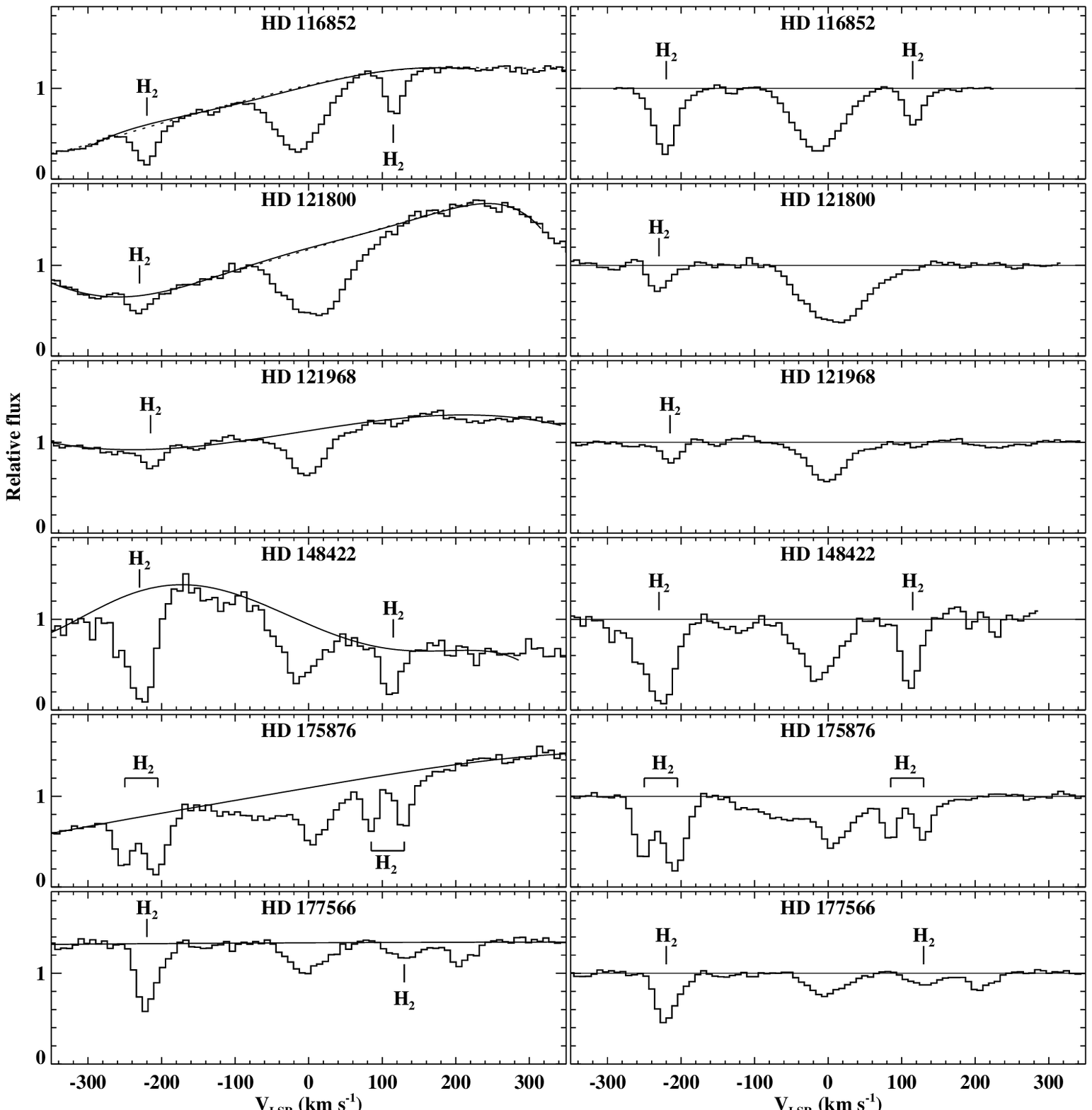}}
\figcaption{}
\end{center}
\end{figure}
\addtocounter{figure}{-1}

\clearpage
\begin{figure}[htbp] 
\begin{center}
\rotatebox{0}{
\epsscale{1.0}
\plotone{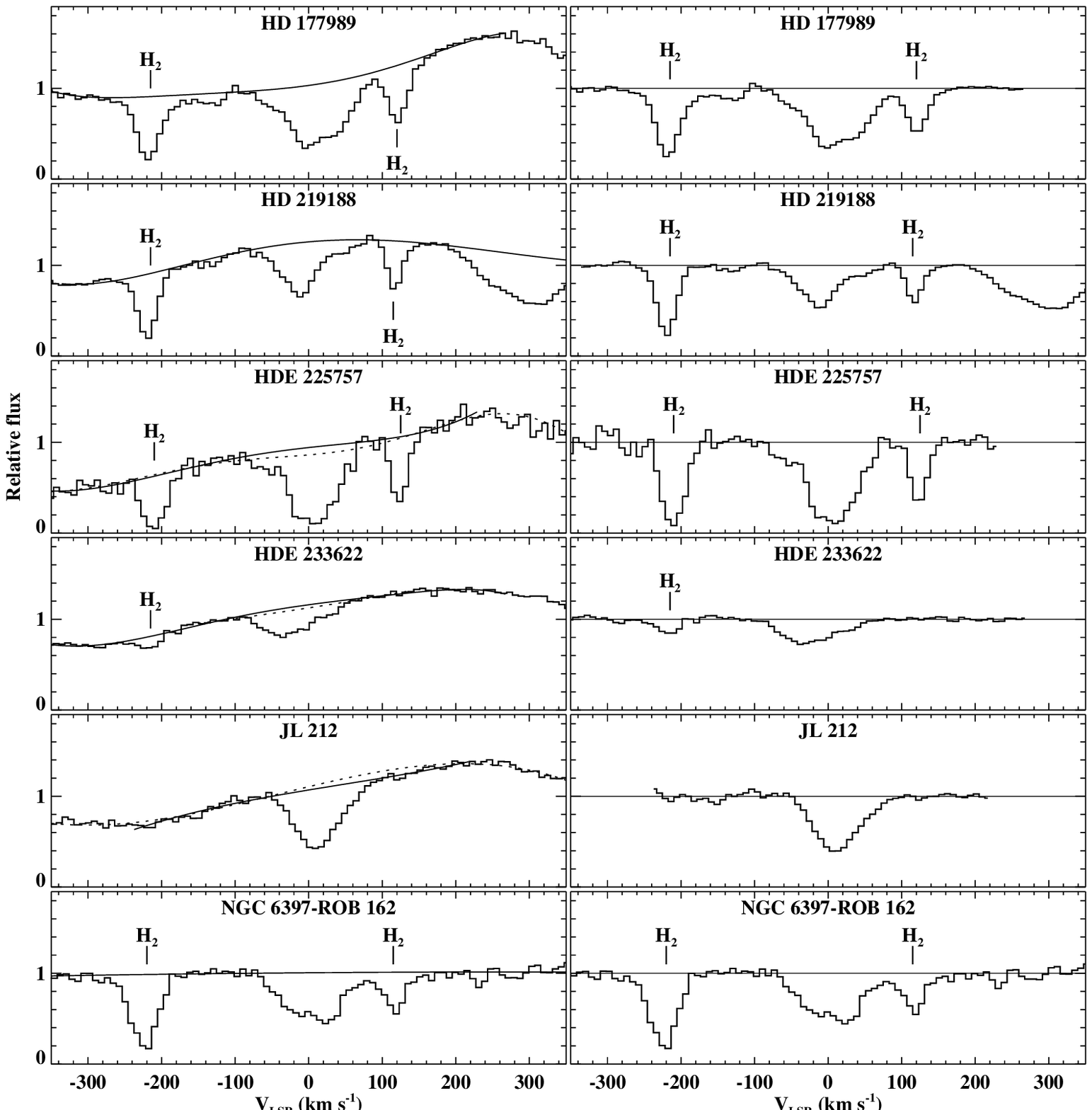}}
\figcaption{  }
\end{center}
\end{figure}
\addtocounter{figure}{-1}

\clearpage
\begin{figure}[htbp] 
\begin{center}
\rotatebox{0}{
\epsscale{1.0}
\plotone{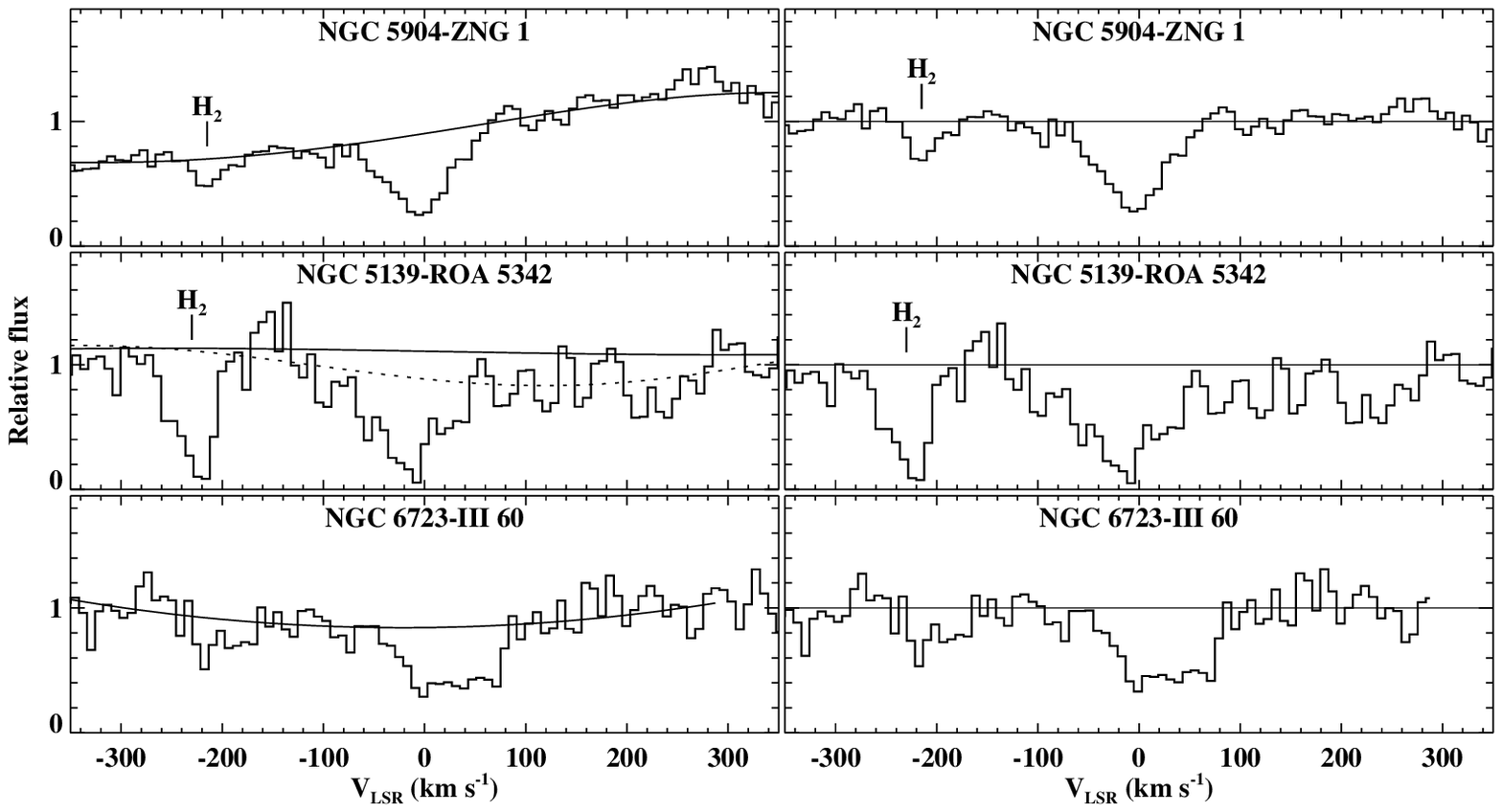}}
\figcaption{The relative flux around the \ion{O}{6} 1031.93 \AA\  line toward the sight lines
listed in Table~\ref{tab1}. Left: The observed relative fluxes plotted against the LSR velocity. 
The continua are overplotted by thin lines. 
If alternative continuum placements were considered, then they are shown by dotted lines.
Right: The resulting normalized spectra as a function of LSR velocity. 
We marked the position of the H$_2$ $P(3)$ 1031.19 \AA\  and $R(4)$ 1032.35 \AA\  lines 
if they were present.
\label{OVI1032}}
\end{center}
\end{figure}

\clearpage
\begin{figure}[htbp] 
\begin{center}
\rotatebox{0}{
\epsscale{1.0}
\plotone{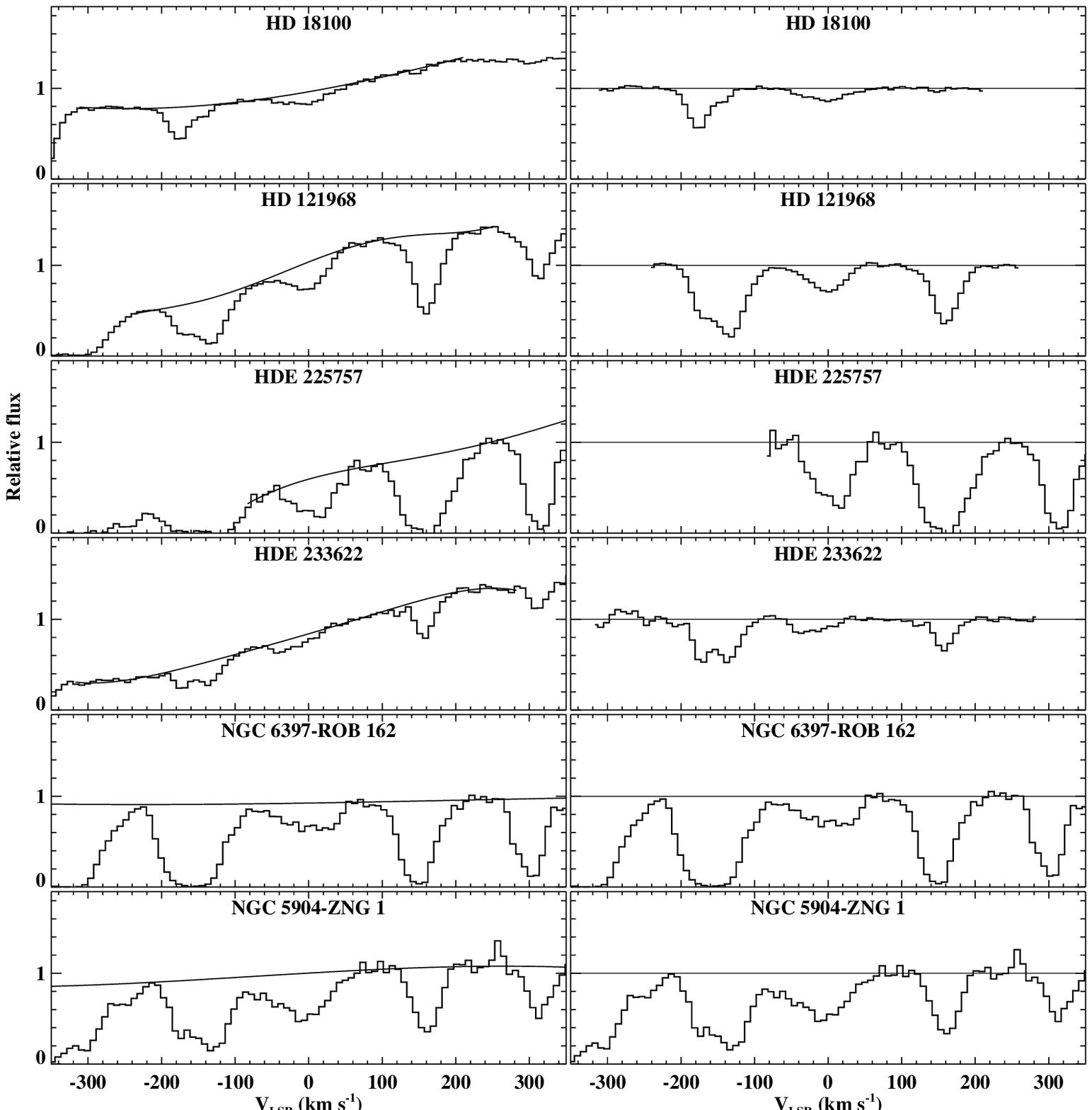}}
\figcaption{Same as Figure~\ref{OVI1032}, but for \ion{O}{6} $\lambda$1037.62 toward our sight lines.
Only 6 of the 22 profiles are shown. Severe interstellar and stellar blending makes measurements of
the \ion{O}{6} $\lambda$1037.62 absorption uncertain for the remaining 16 cases.
\label{OVI1037}}
\end{center}
\end{figure}

\clearpage
\begin{figure}[htbp] 
\begin{center}
\rotatebox{0}{
\epsscale{0.9}
\plotone{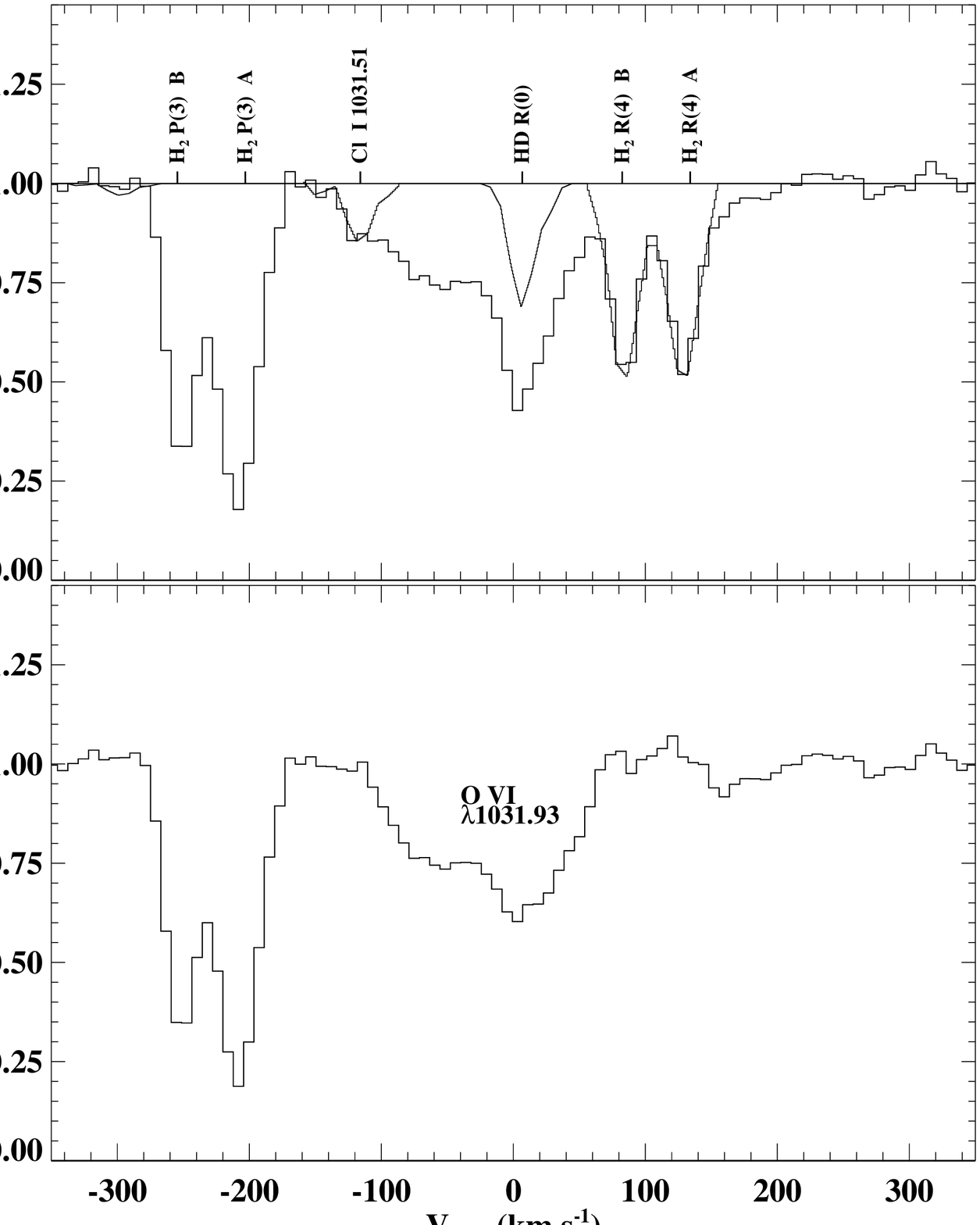}}
\figcaption{The complex system of blended lines around \ion{O}{6} $\lambda$1031.93 toward 
HD~175876. The spectra are plotted against LSR velocity. The normalized spectrum is shown with 
a thick solid line (histogram) in the upper panel, and the calculated \ion{Cl}{1}, HD, and 
H$_2$ profiles are overplotted with thin lines. 
The final spectrum cleared of HD, \ion{Cl}{1}, and  H$_2$ contributions is plotted in the lower panel.
There are two components in the H$_2$ absorption: component A, which is also visible in \ion{Cl}{1} and 
HD absorption, and component B, which is shifted by $\sim$ -50 \kms\  with respect
to component A. 
\label{profiles}}
\end{center}
\end{figure}

\clearpage
\begin{figure}[htbp] 
\begin{center}
\rotatebox{0}{
\epsscale{1.0}
\plotone{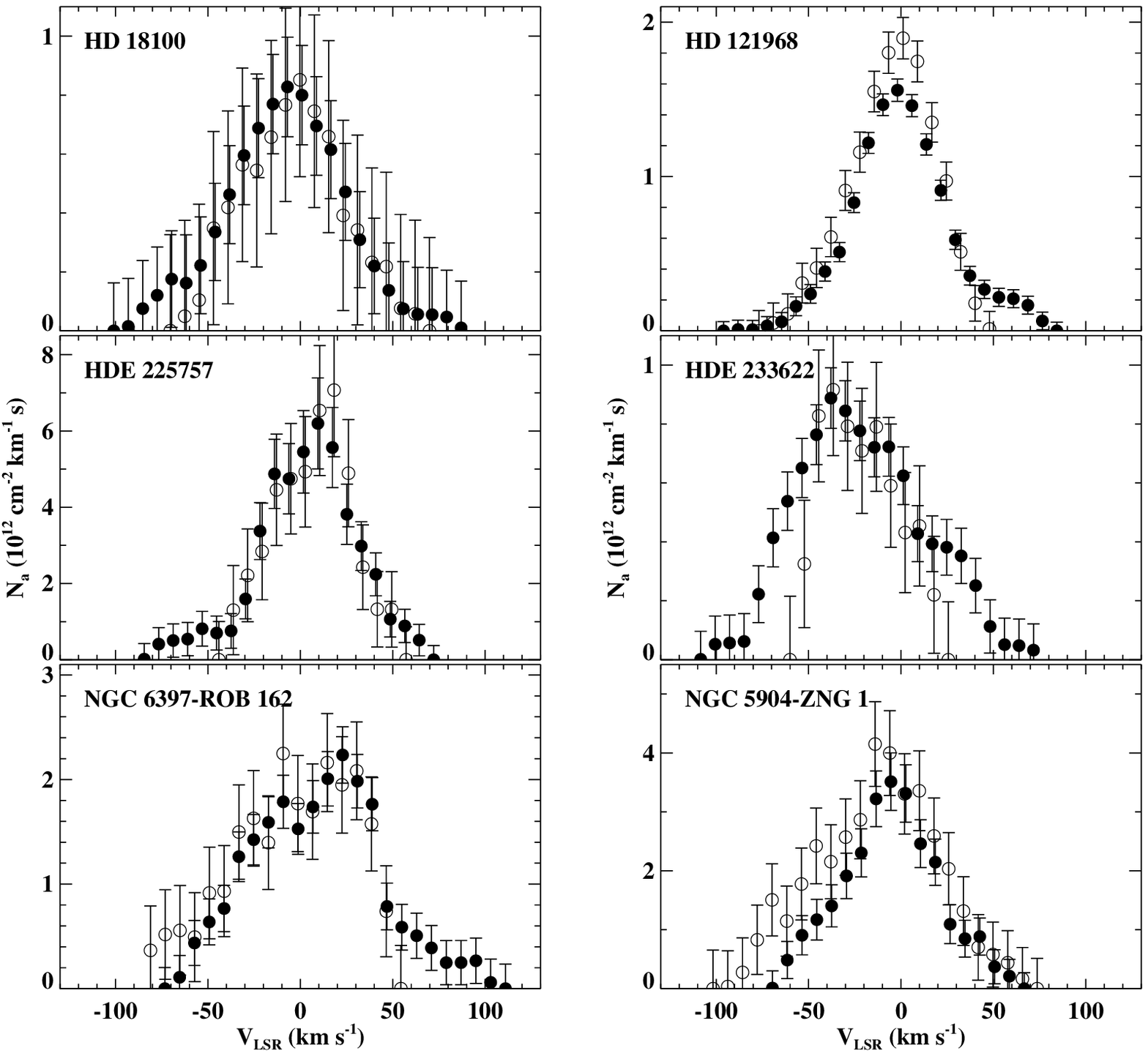}}
\figcaption{The apparent column density profiles of \ion{O}{6} $\lambda$1031.93 (closed circles)
and $\lambda$1037.62 (open circles) toward the 6 sight lines for which both were extracted. The error
bars reflect only the statistical uncertainties. 
The profiles do not reveal significant saturation effects at $|V_{LSR}| \leq$~50~km~s$^{-1}$, except
in the case of HD 121968 where a modest $\sim$20\% saturation may be present. There are
differences at $|V_{LSR}|$ greater than $\sim$50 \kms\  where strong \ion{C}{2}, \ion{C}{2}*, and
H$_2$ absorption limited our ability to find accurate stellar continua around the \ion{O}{6} 
1037.62 \AA\ lines.
\label{OVIcomp}}
\end{center}
\end{figure}

\clearpage
\begin{figure}[htbp] 
\begin{center}
\rotatebox{0}{
\epsscale{0.7}
\plotone{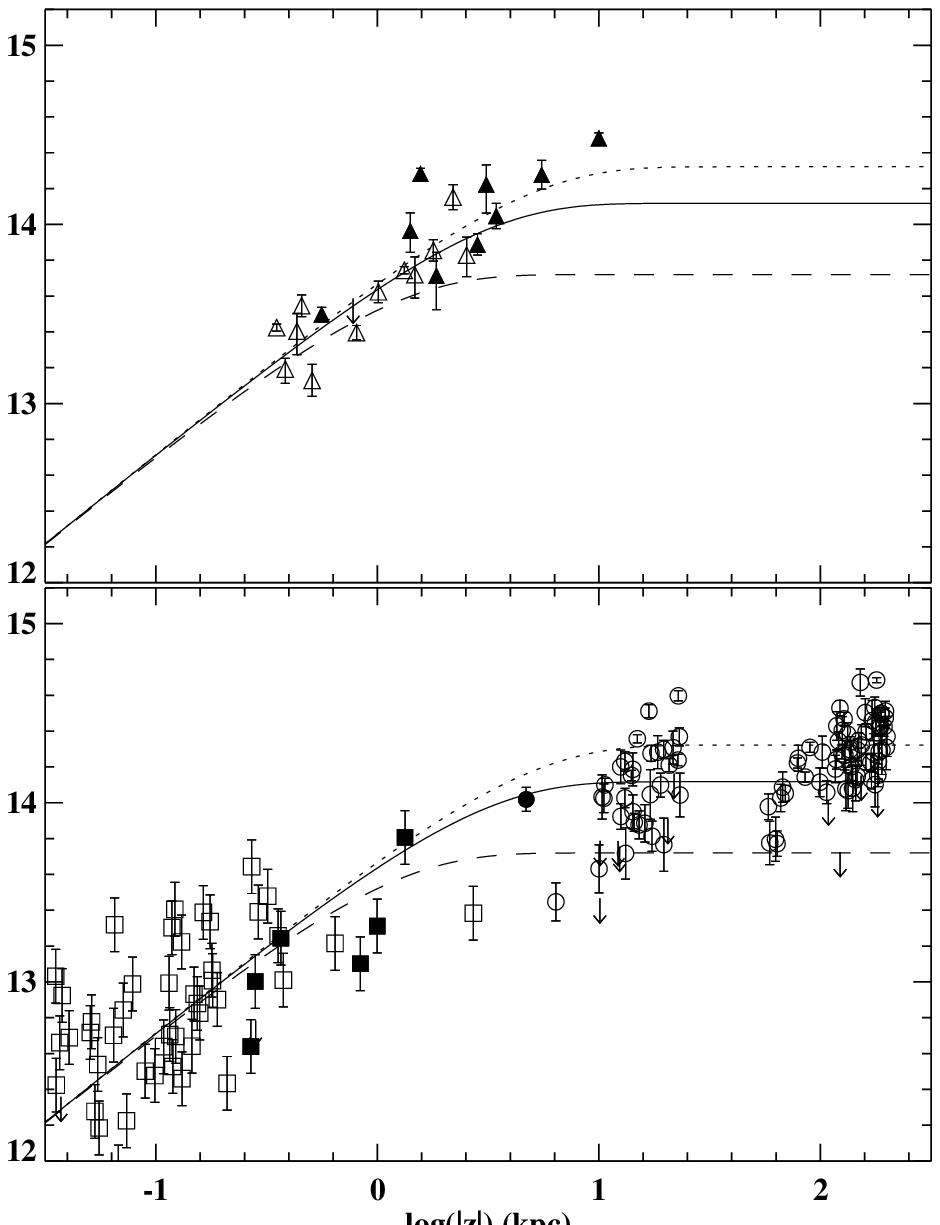}}
\figcaption{The total \ion{O}{6} columns perpendicular to the Galactic plane
as a function of vertical distance from the plane ($z$), plotted on a log-log 
scale. The panels are identical apart from the datasets plotted. 
Triangles, circles, and squares are measurements from the present analysis, Savage {\it et al}. (2002), 
and Jenkins (1978a), respectively. 
Downward-pointing arrows indicate upper limits. We display Galactic measurements (disk and halo) with 
solid symbols if their Galactic latitude and $|z|$ were greater than zero and 0.25~kpc, respectively. 
The measurements toward extragalactic targets with $b\geq$ 0$^{\circ}$ are artificially shifted by +1~dex 
in log$|z|$ with respect to those with $b <$ 0$^{\circ}$. 
The north-south asymmetry, observed by Savage {\it et al}. (2002), is well illustrated by the different 
values of $<N$sin$|b|>$ for the southern (log$|z|\sim$1.2) and northern (log$|z|\sim$2.2)
extragalactic sight lines.
The distributions calculated by Equation~5 using a mid-plane 
density of $n_0$= 1.7~$\times$~10$^{-8}$ cm$^{-3}$ and scale heights $h$(\ion{O}{6})= 2.5~kpc (solid line),
$h$(\ion{O}{6})= 1~kpc (dashed line), and 4~kpc (dotted line) are also plotted. 
\label{OVIhalo}}
\end{center}
\end{figure}

\clearpage
\begin{figure}[htbp] 
\begin{center}
\vspace{2cm} 
\rotatebox{0}{
\epsscale{1.0}
\plotone{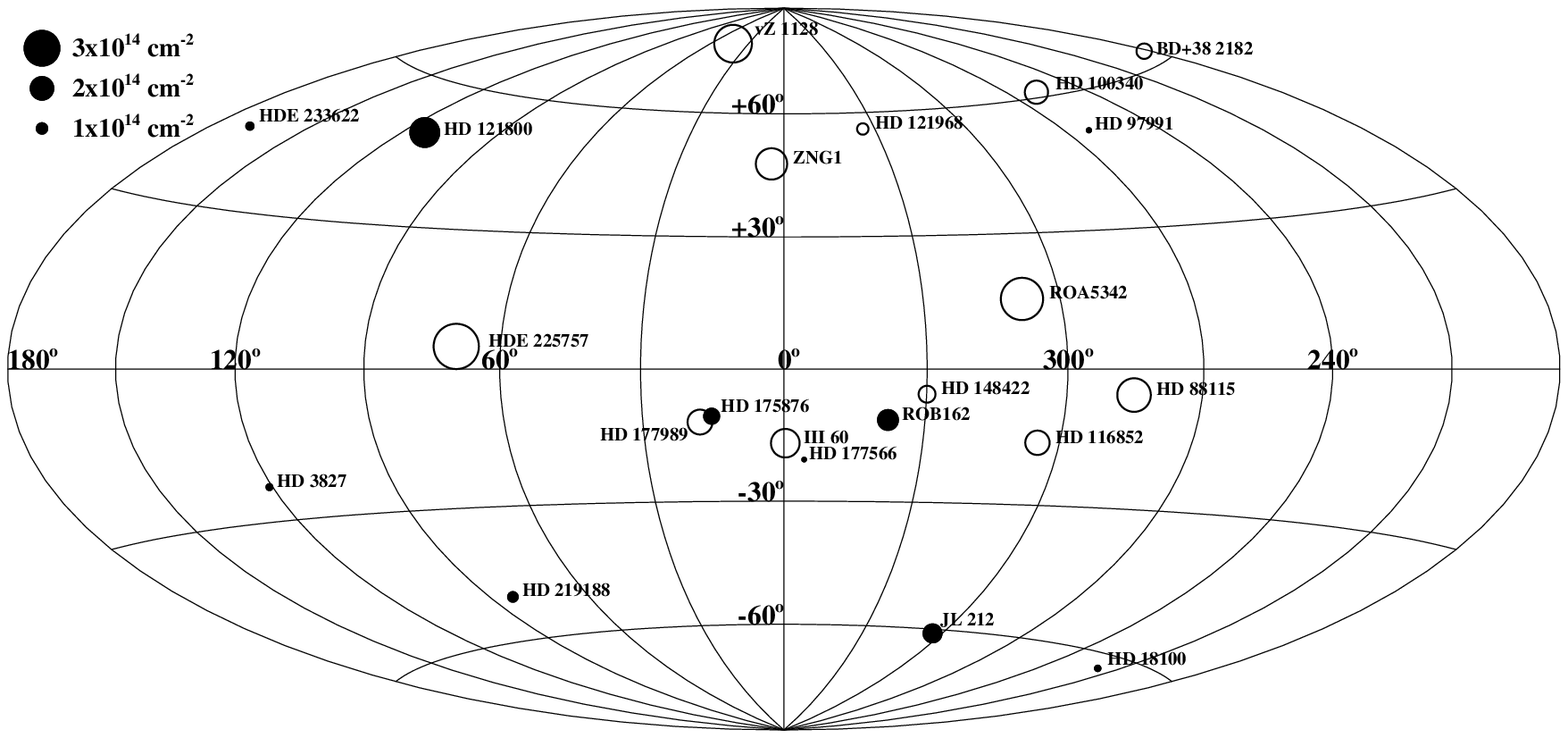}}
\vspace{2cm} 
\figcaption{Total \ion{O}{6} column densities toward our sight lines, displayed in a Hammer-Aitoff 
projection. The Galactic Center is in the middle of the Figure and Galactic longitude increases to the left. 
The diameter of each circle is linearly proportional to the value of the column density, and the 
scaling is displayed in the upper left corner. The upper limit toward HD~97991 was treated as an observation. 
Open and closed symbols represent measurements for stars with $d >$ 3~kpc and $d \leq$ 3~kpc, respectively.
The name of each object is printed beside its location. NGC~6397-ROB~162, NGC~5904-ZNG~1, NGC~5139-ROA~5342, 
and NGC~6723-III 60 were shortened to ROB~162, ZNG~1, ROA~5342, and III 60, respectively.
\label{OVIgalaxy}}
\end{center}
\end{figure}

\clearpage
\begin{figure}[htbp] 
\begin{center}
\rotatebox{0}{
\epsscale{0.8}
\plotone{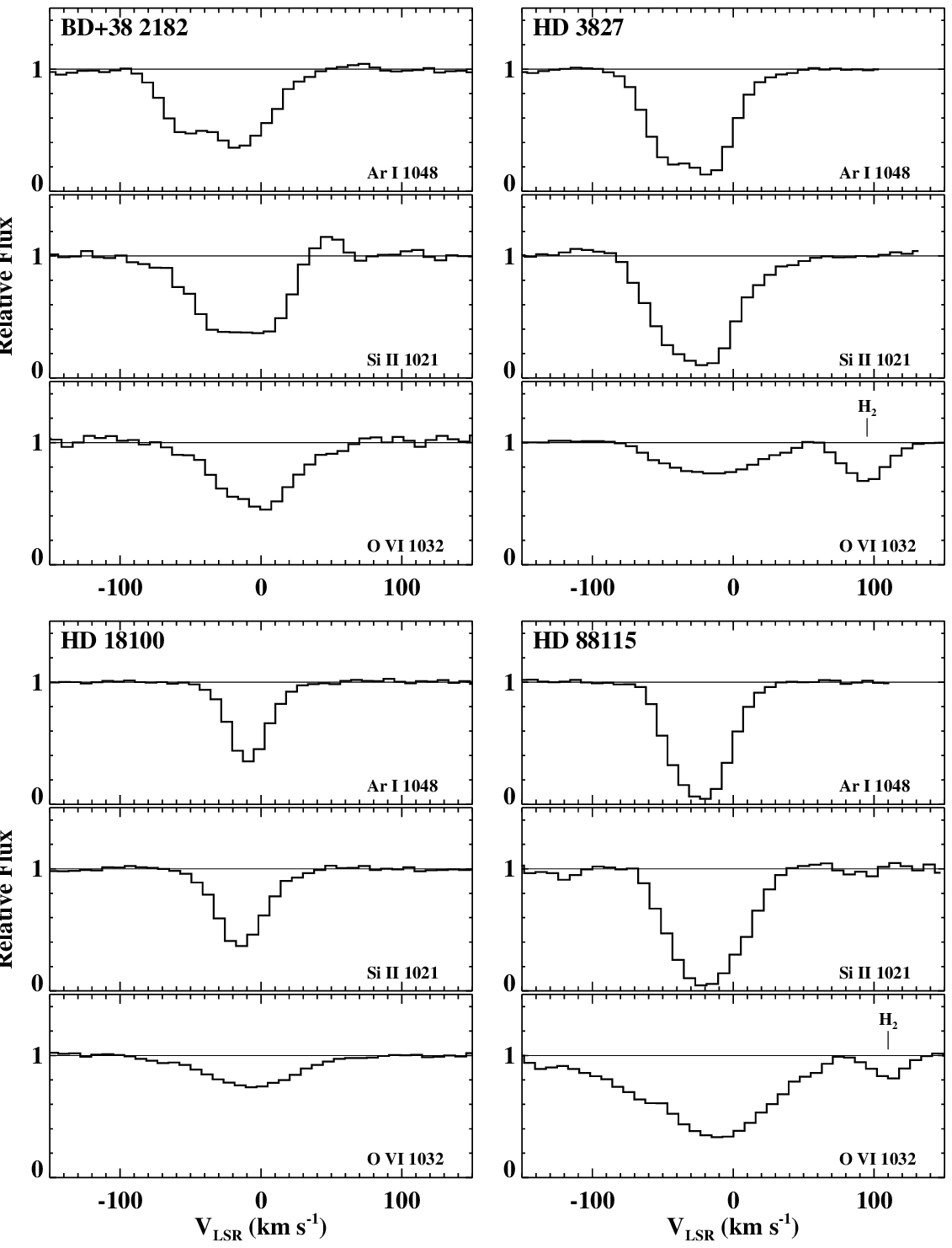}}
\figcaption{  }
\end{center}
\end{figure}
\addtocounter{figure}{-1}

\clearpage
\begin{figure}[htbp] 
\begin{center}
\rotatebox{0}{
\epsscale{0.8}
\plotone{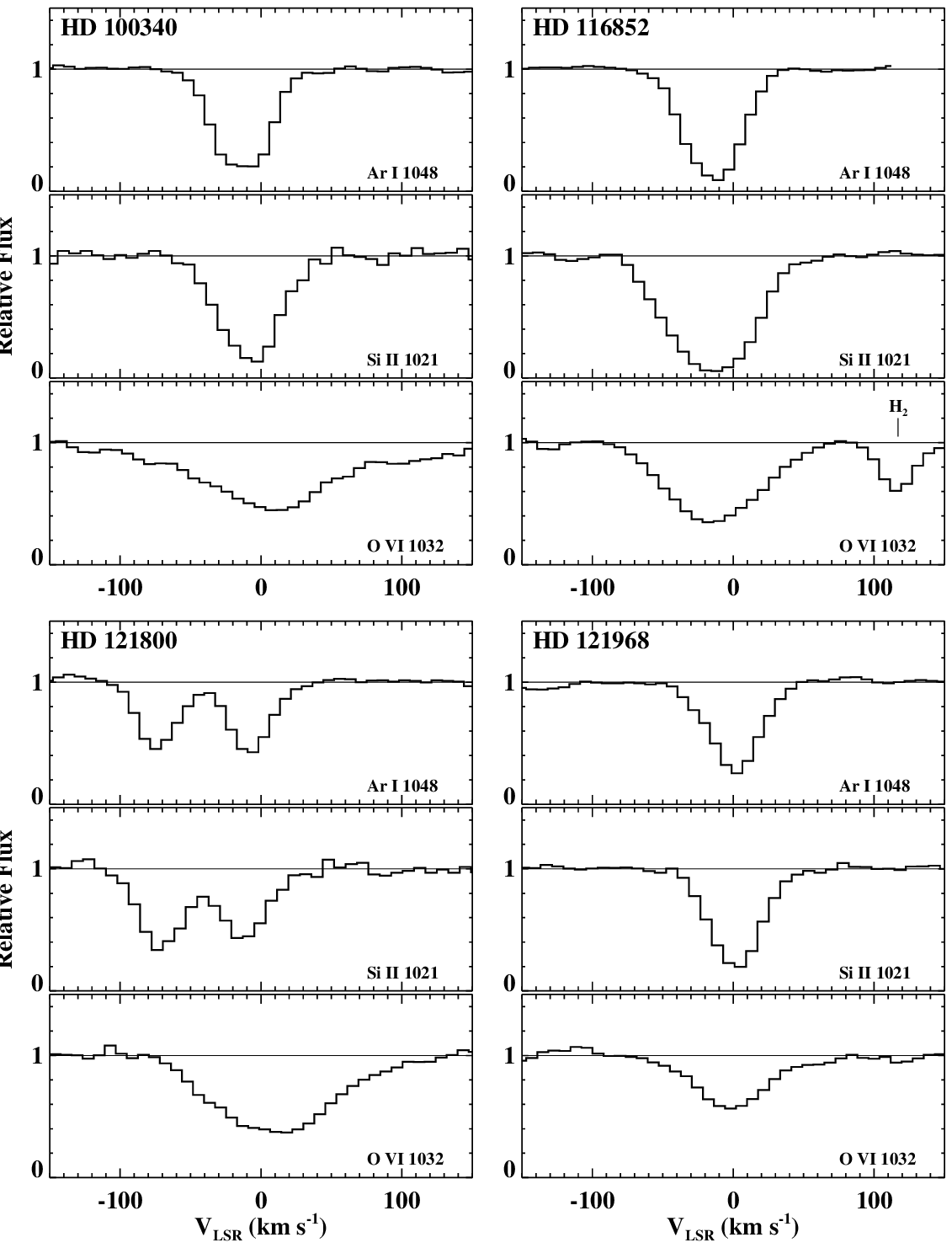}}
\figcaption{  }
\end{center}
\end{figure}
\addtocounter{figure}{-1}

\clearpage
\begin{figure}[htbp] 
\begin{center}
\rotatebox{0}{
\epsscale{0.8}
\plotone{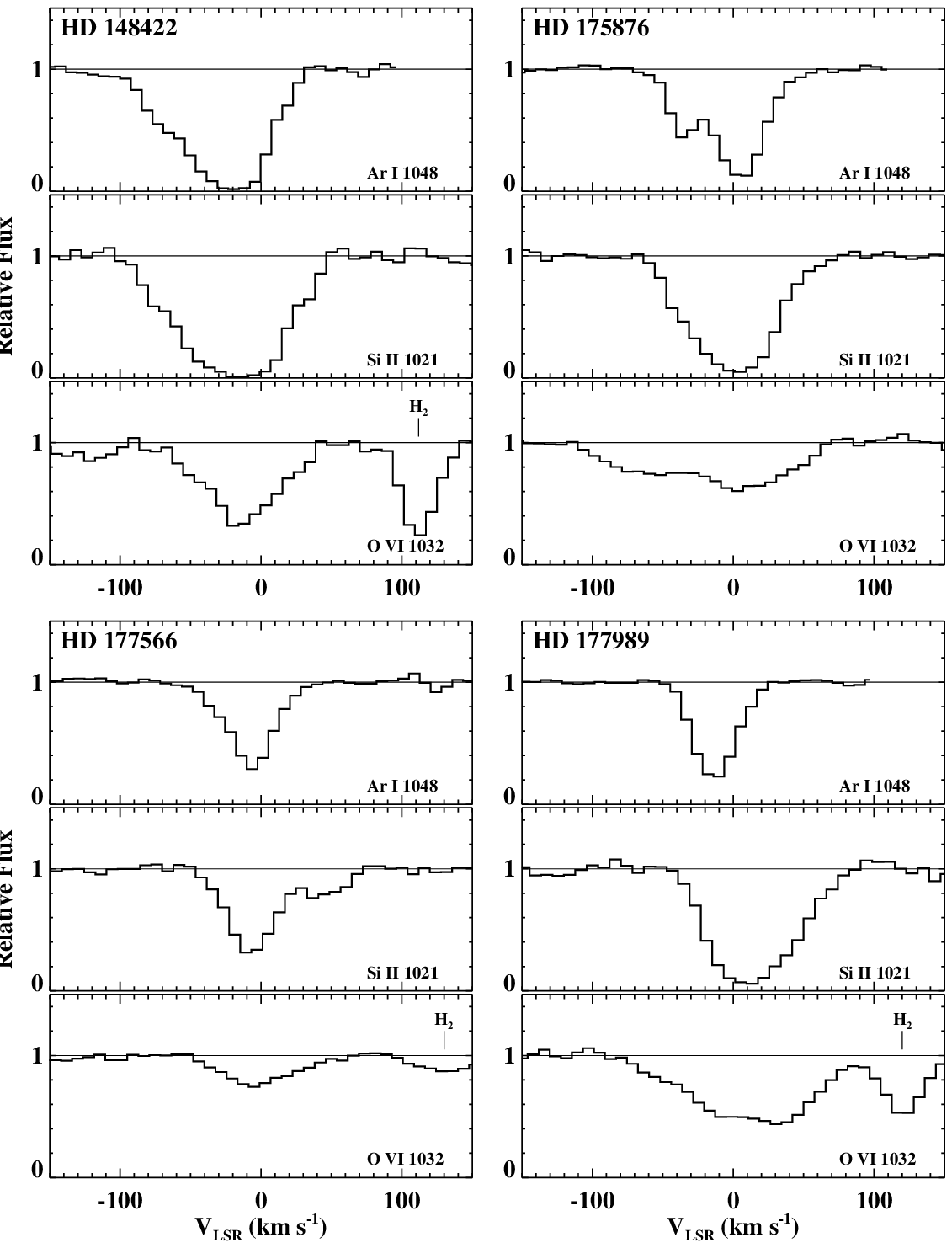}}
\figcaption{  }
\end{center}
\end{figure}
\addtocounter{figure}{-1}

\clearpage
\begin{figure}[htbp] 
\begin{center}
\rotatebox{0}{
\epsscale{0.8}
\plotone{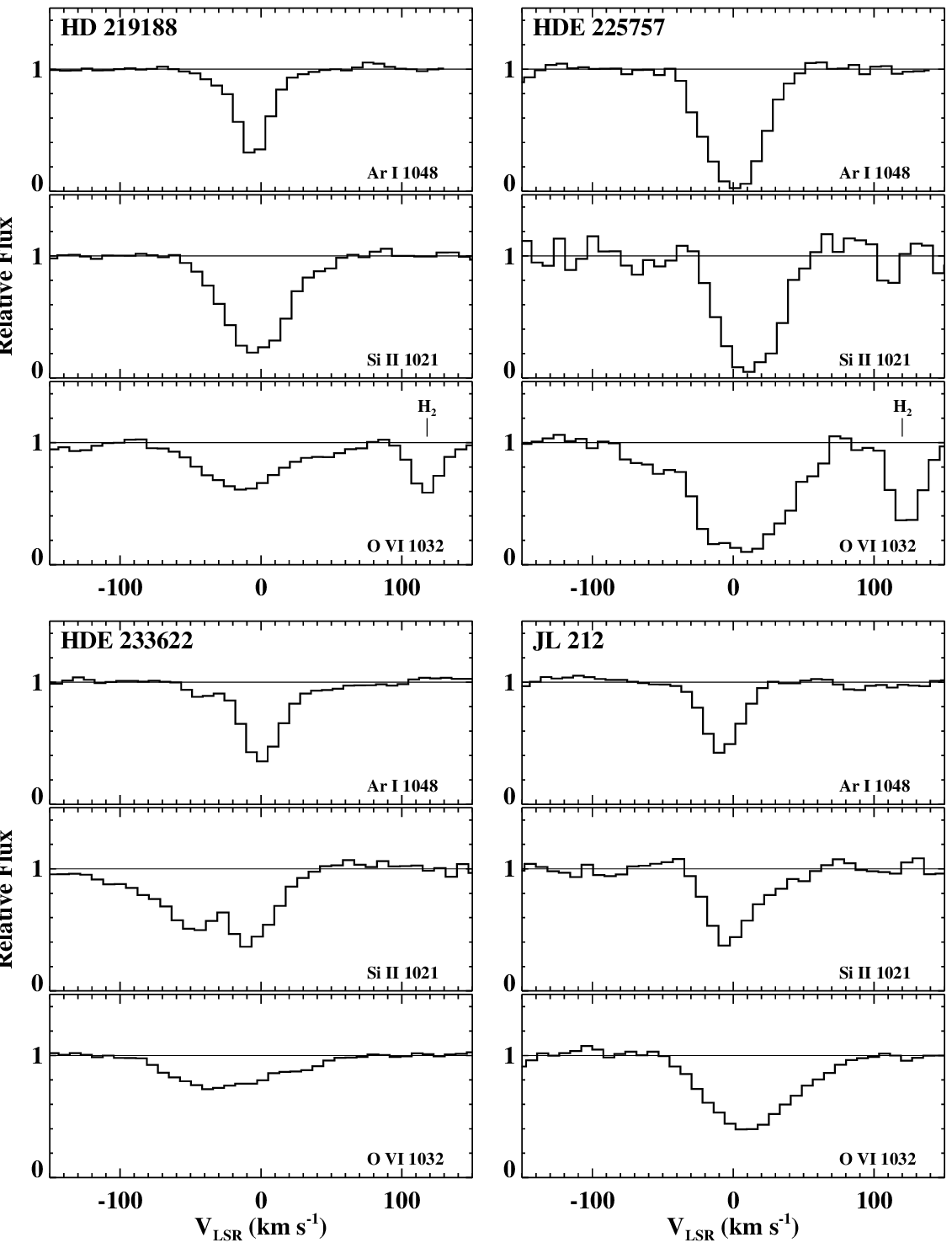}}
\figcaption{  }
\end{center}
\end{figure}
\addtocounter{figure}{-1}

\clearpage
\begin{figure}[htbp] 
\begin{center}
\rotatebox{0}{
\epsscale{0.8}
\plotone{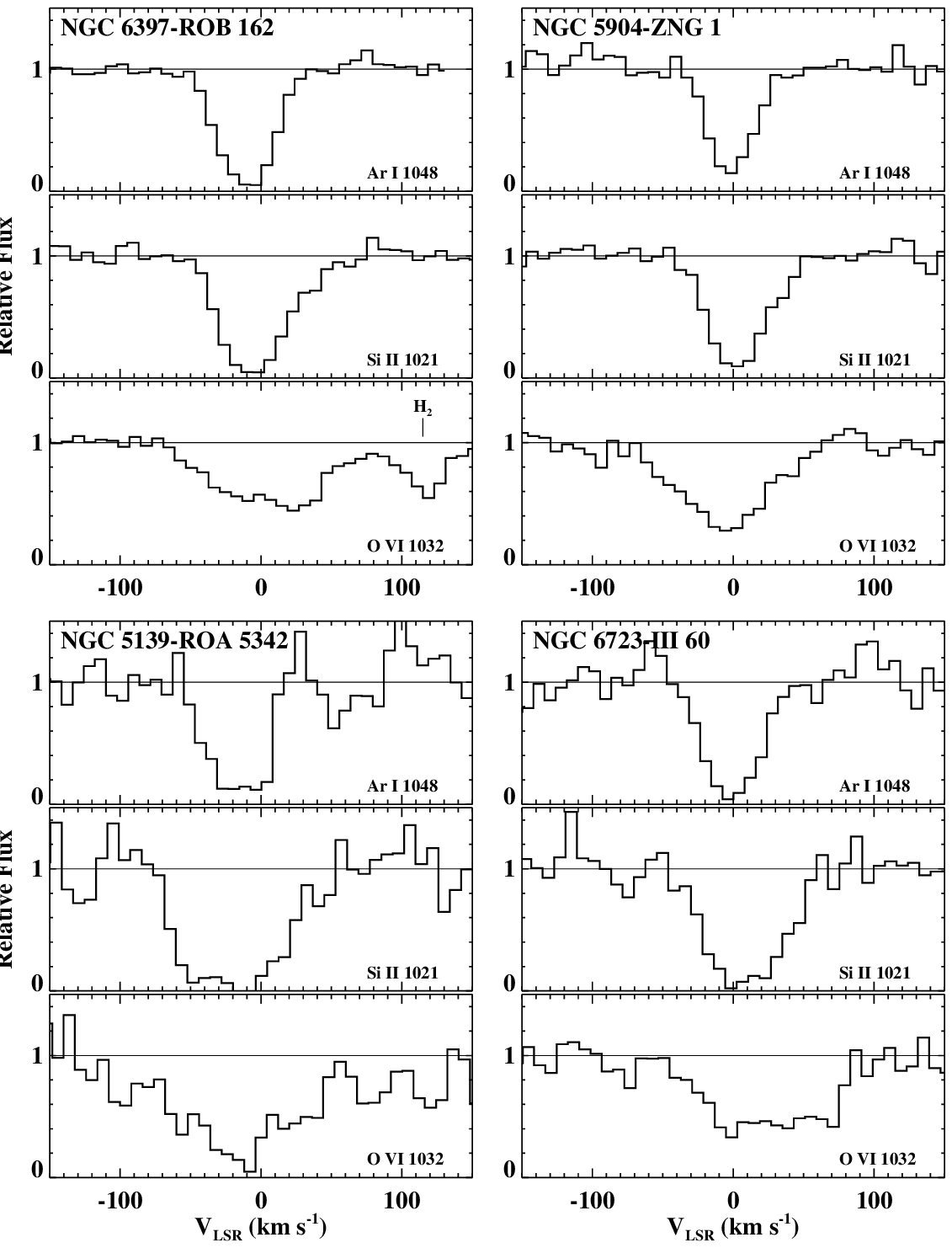}}
\figcaption{The comparison of low-ionization states (\ion{Ar}{1} $\lambda$1048.22 
and \ion{Si}{2} $\lambda$1020.70) with the \ion{O}{6} absorption at 1031.93 \AA\  toward the 
halo star sight lines. We label the position of the H$_2$ R(4)$_{6-0}$ line
if it is clearly present. A velocity correction of +10 \kms\ was applied for the profiles of \ion{Ar}{1} $\lambda$1048.22 
in the cases of LWRS observations (see Table~\ref{tab2}) to account for the systematic calibration errors present
in the $FUSE$ spectra reduced by CALFUSE v1.8.7 (see \S\ref{section:Obs}). 
\label{lines}}
\end{center}
\end{figure}

\clearpage
\begin{figure}[htbp] 
\begin{center}
\rotatebox{0}{
\epsscale{1.0}
\plotone{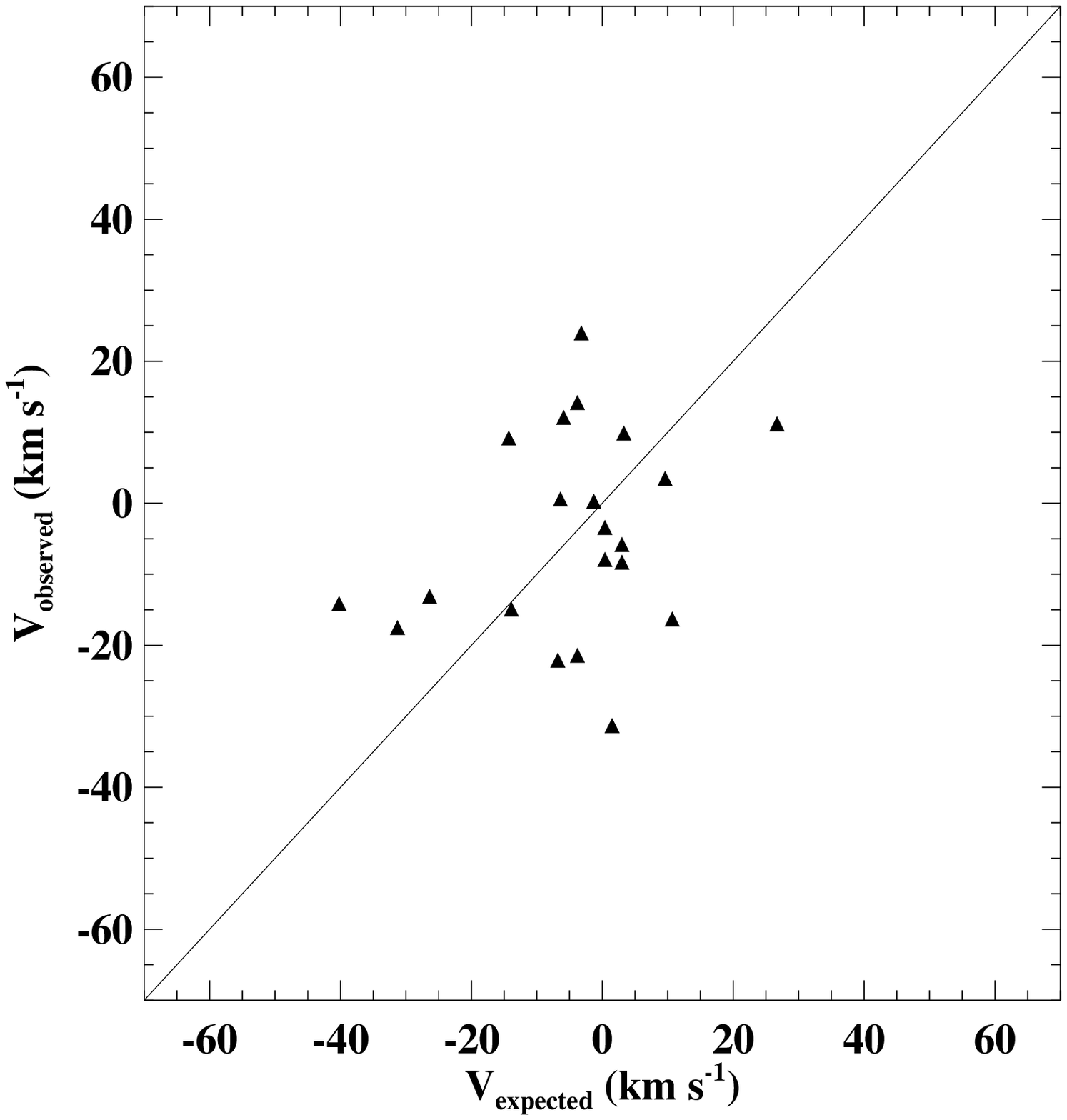}}
\figcaption{The measured \ion{O}{6} line centroid velocities plotted against the velocities expected from the
Galactic rotation assuming co-rotation of disk and halo gas and the Clemens (1985) Galactic rotation
curve. The solid line represents a 1:1 correlation between the two quantities. HD~97991 was omitted, 
because only an upper limit for the \ion{O}{6} column was measured. 
\label{clemens}}
\end{center}
\end{figure}

\clearpage
\begin{figure}[htbp] 
\begin{center}
\rotatebox{0}{
\epsscale{1.0}
\plotone{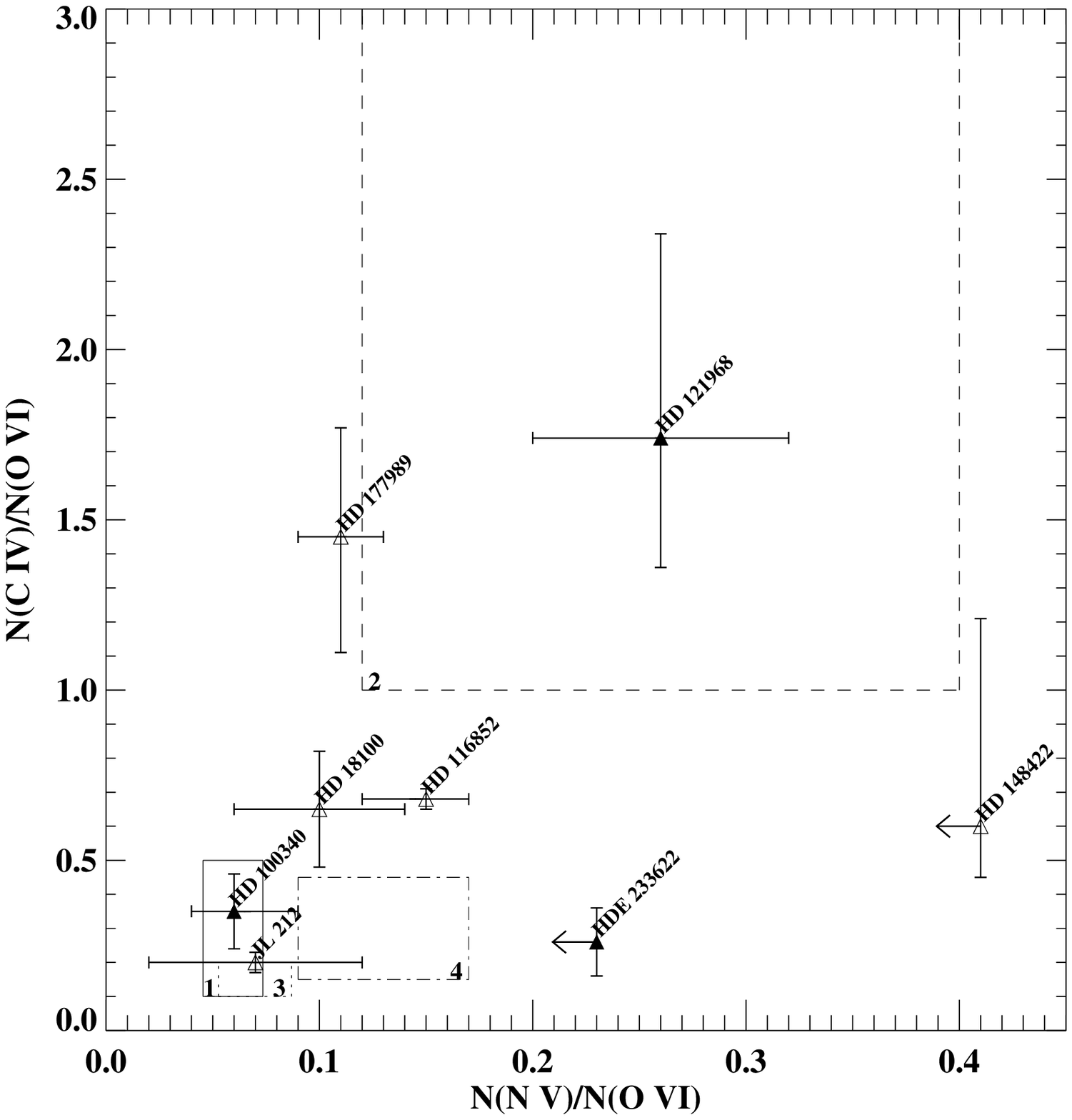}}
\figcaption{The N(\ion{C}{4})/N(\ion{O}{6}) ratios plotted against the N(\ion{N}{5})/N(\ion{O}{6}) 
ratios. Lower and upper limits are shown by arrows pointing in the appropriate directions. Sight lines
with $b \geq$ 0$^{\circ}$ and $b <$ 0$^{\circ}$ are displayed by solid and open symbols, respectively.
The values that are predicted by a model for high ion production are within the corresponding rectangle 
in the figure. We displayed the predictions of the following models: 1: (CGF) cooling Galactic fountain model 
R. Benjamin (2002, private communication), solid lines; 2: (TML) turbulent mixing layer model of 
Slavin {\it et al}. (1993), dashed lines; 3: (SNR) cooling supernova remnant model of Slavin \& Cox (1992),
dotted lines; 4: (CI) thermal conduction model of Borkowski {\it et al}. (1990), dash-dotted lines. 
\label{CIVvsNV}}
\end{center}
\end{figure}

\clearpage
\begin{figure}[htbp] 
\begin{center}
\rotatebox{0}{
\epsscale{1.0}
\plotone{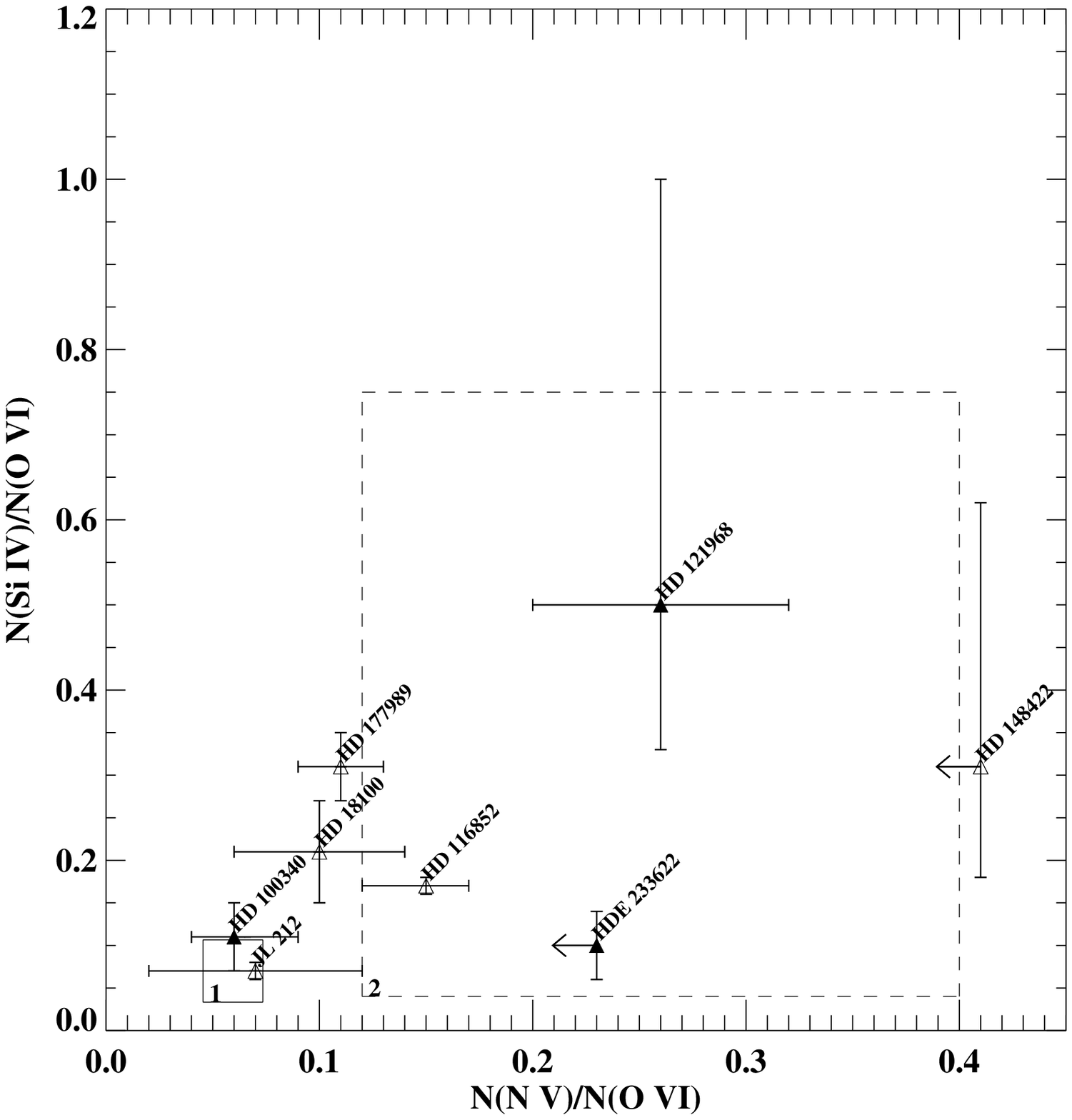}}
\figcaption{Same as Figure~\ref{CIVvsNV}, but for N(\ion{Si}{4})/N(\ion{O}{6}) as a 
function of N(\ion{N}{5})/N(\ion{O}{6}). The predictions of the SNR and CI models for the 
N(\ion{Si}{4})/N(\ion{O}{6}) ratios are too low to display. 
\label{SiIVvsNV}}
\end{center}
\end{figure}

\clearpage
\begin{figure}[htbp] 
\begin{center}
\rotatebox{0}{
\epsscale{1.0}
\plotone{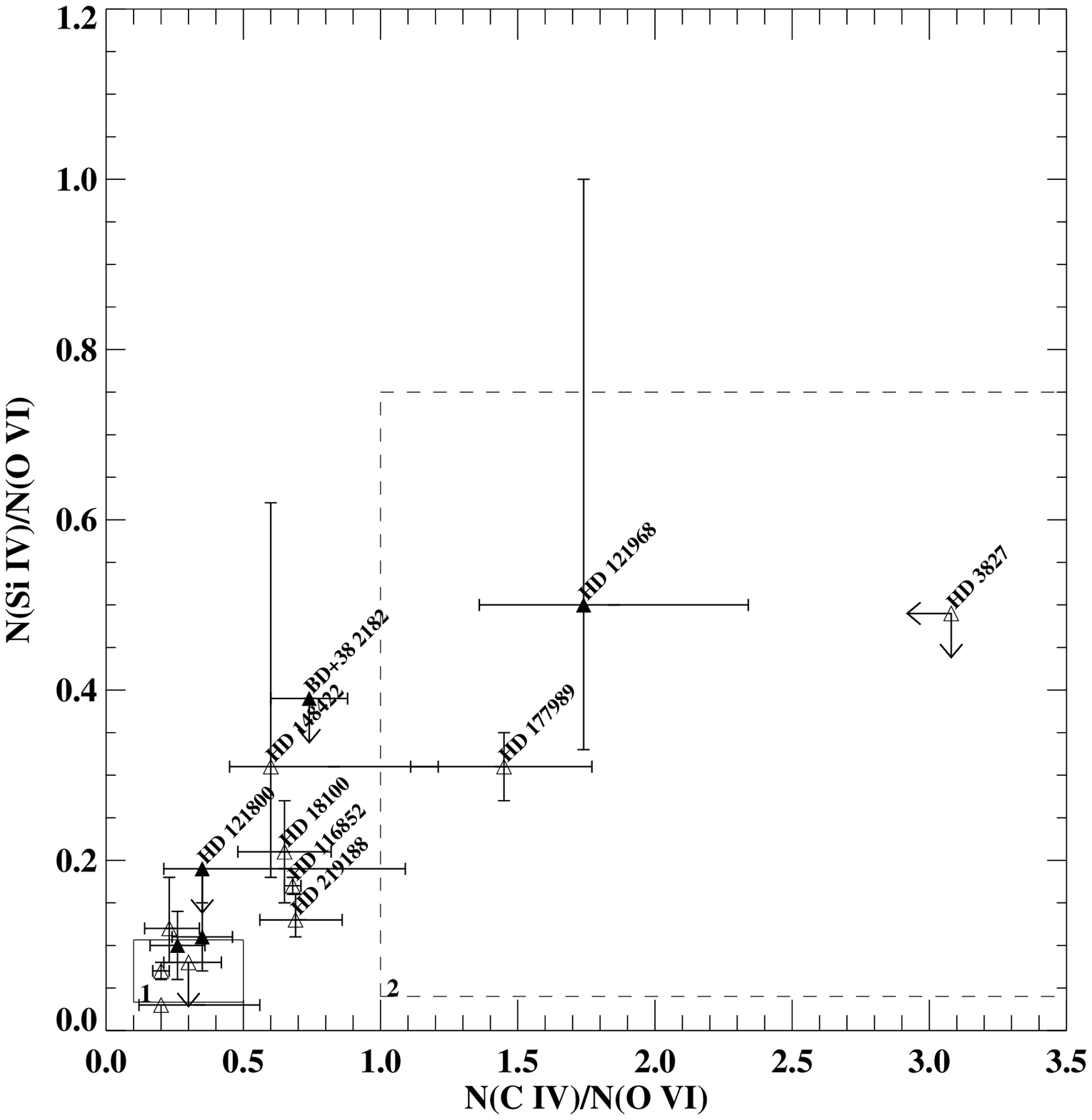}}
\figcaption{Same as Figure~\ref{CIVvsNV}, but for N(\ion{Si}{4})/N(\ion{O}{6}) as a 
function of N(\ion{C}{4})/N(\ion{O}{6}). The predictions of the SNR and CI models for the 
N(\ion{Si}{4})/N(\ion{O}{6}) ratios are too low to display. We did not label sight lines with 
N(\ion{Si}{4})/N(\ion{O}{6}) $\leq$ 0.15 and N(\ion{C}{4})/N(\ion{O}{6}) $\leq$ 0.5 for clarity. 
\label{SiIVvsCIV}}
\end{center}
\end{figure}

\clearpage
\begin{figure}[htbp]
\begin{center} 
\rotatebox{0}{
\epsscale{0.9}
\plotone{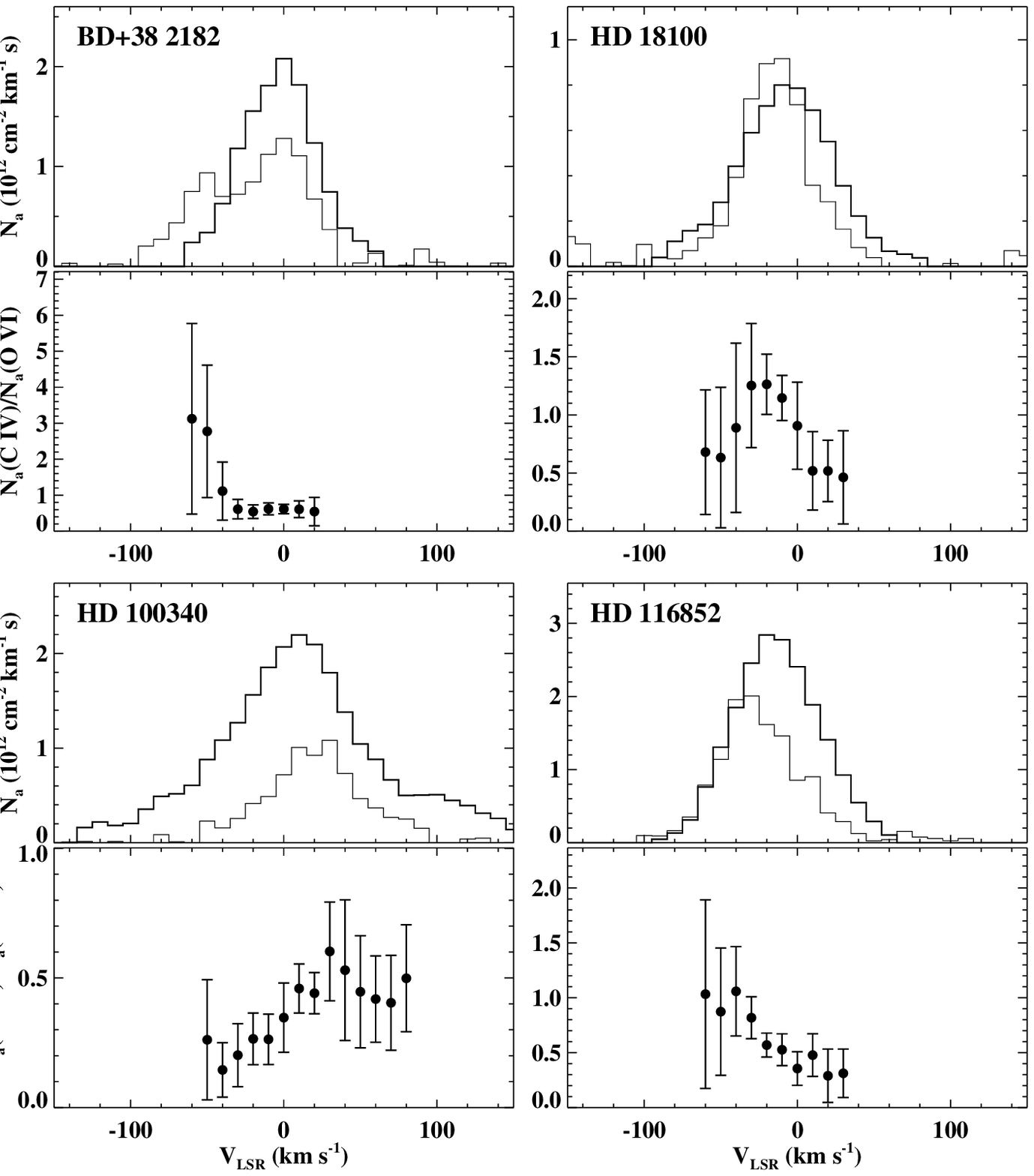}}
\figcaption{  }
\end{center}
\end{figure}
\addtocounter{figure}{-1}

\clearpage
\begin{figure}[htbp] 
\begin{center} 
\rotatebox{0}{
\epsscale{0.9}
\plotone{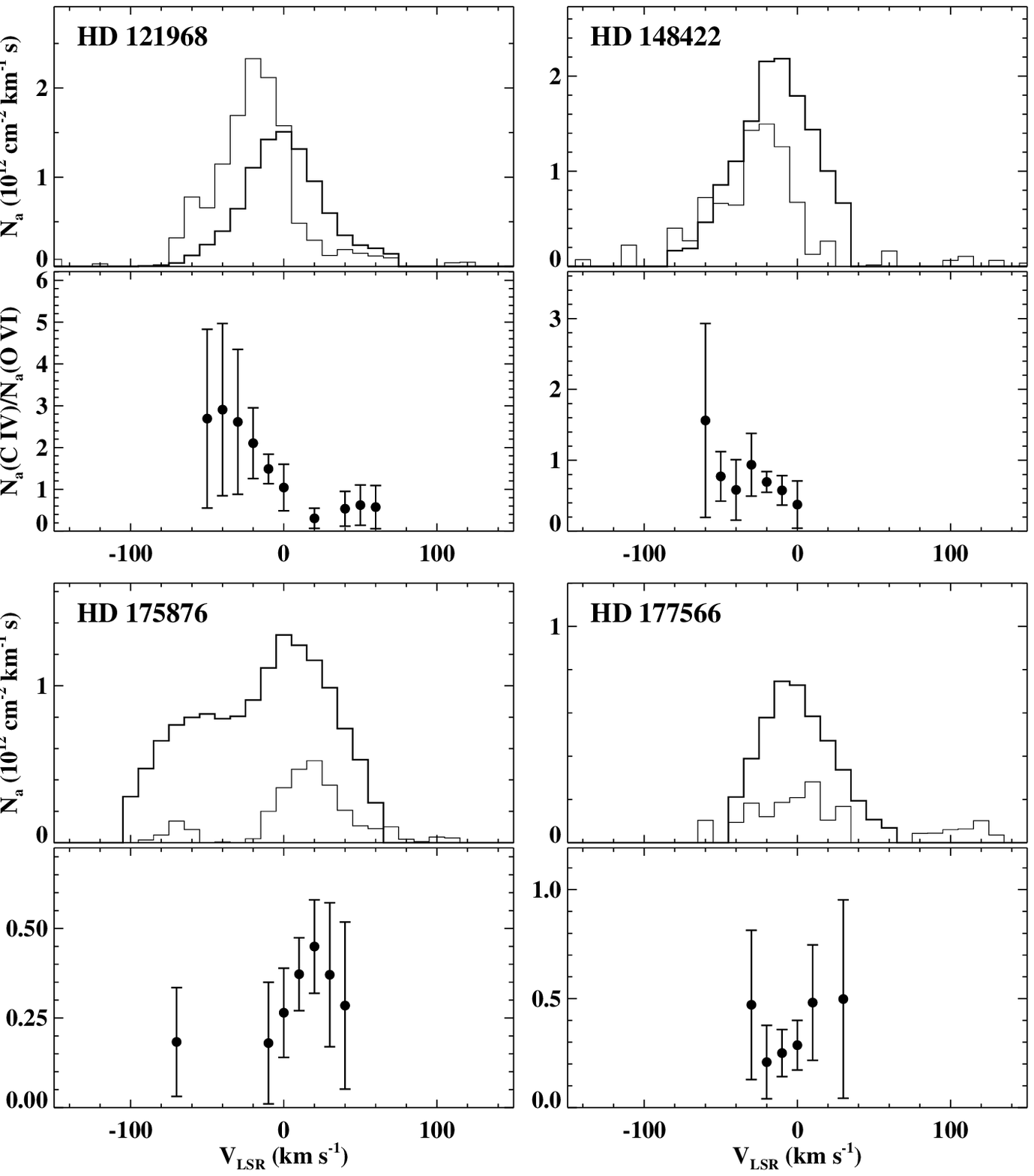}}
\figcaption{  }
\end{center}
\end{figure}
\addtocounter{figure}{-1}

\clearpage
\begin{figure}[htbp] 
\begin{center} 
\rotatebox{0}{
\epsscale{0.9}
\plotone{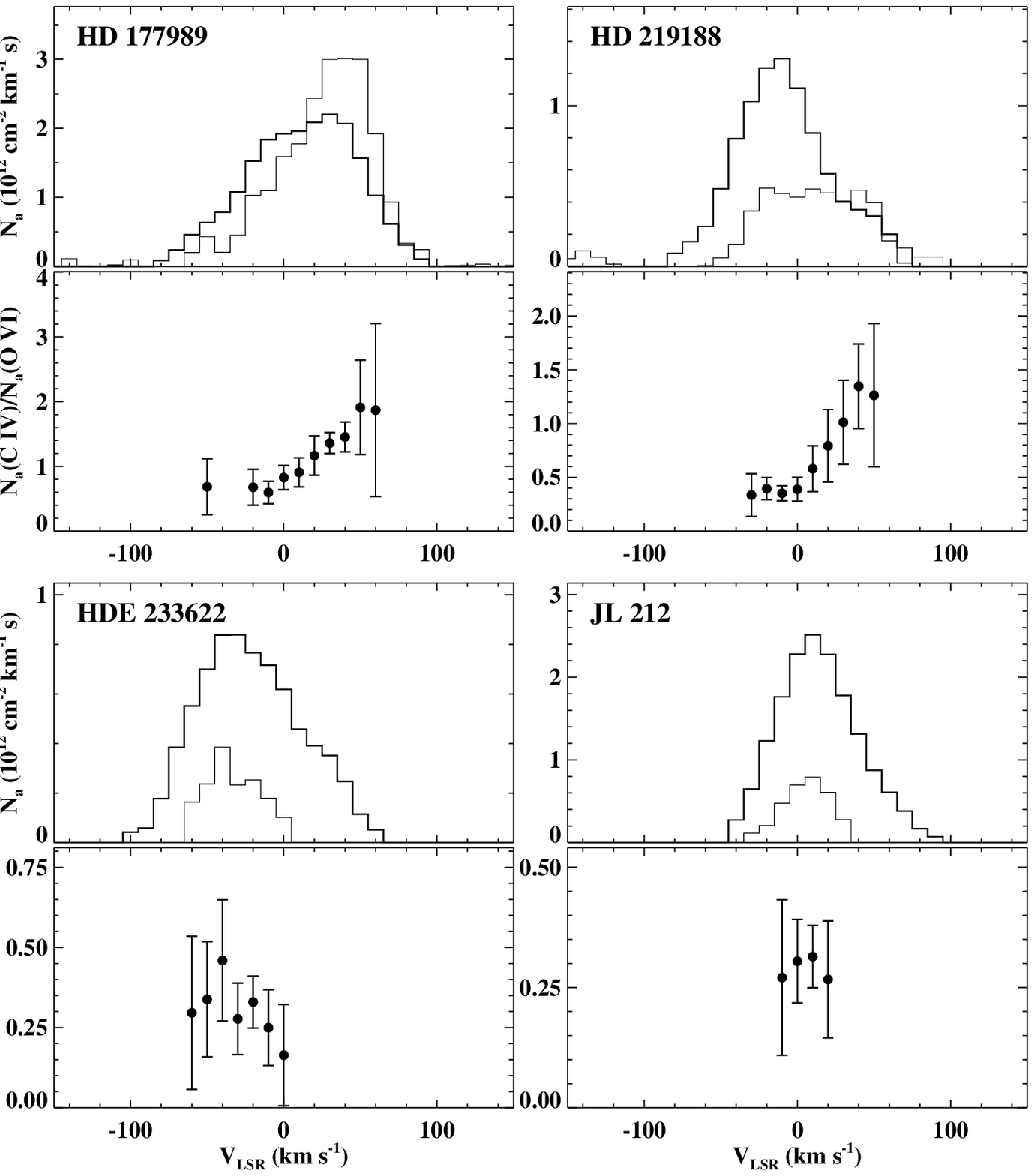}}
\figcaption{  
Upper panels: Apparent column density profiles of \ion{O}{6} $\lambda$1031.93 
(thick line) and \ion{C}{4} $\lambda$1548.20 (thin lines) toward sight lines with observed 
\ion{C}{4} absorption. The \ion{C}{4} $\lambda$1548.20 profiles are from Savage {\it et al}. 
(2001a), except for HDE~233622 and JL~212, which were extracted from STIS archival spectra.
We rebinned both profiles to 10 \kms\ velocity bins. 
Lower panels: The \ion{C}{4} to \ion{O}{6} ratio as a function of LSR velocity. The ratios are
based on the N$_a$($v$) profiles displayed in the upper panel. The error bars reflect the statistical errors
and the uncertainties in the LSR velocity zero-point. The ratios are plotted only if their uncertainties
do not exceed the ratio.
\label{CDratio}}
\end{center}
\end{figure}

\clearpage
\begin{figure}[htbp] 
\begin{center}
\rotatebox{0}{
\epsscale{0.9}
\plotone{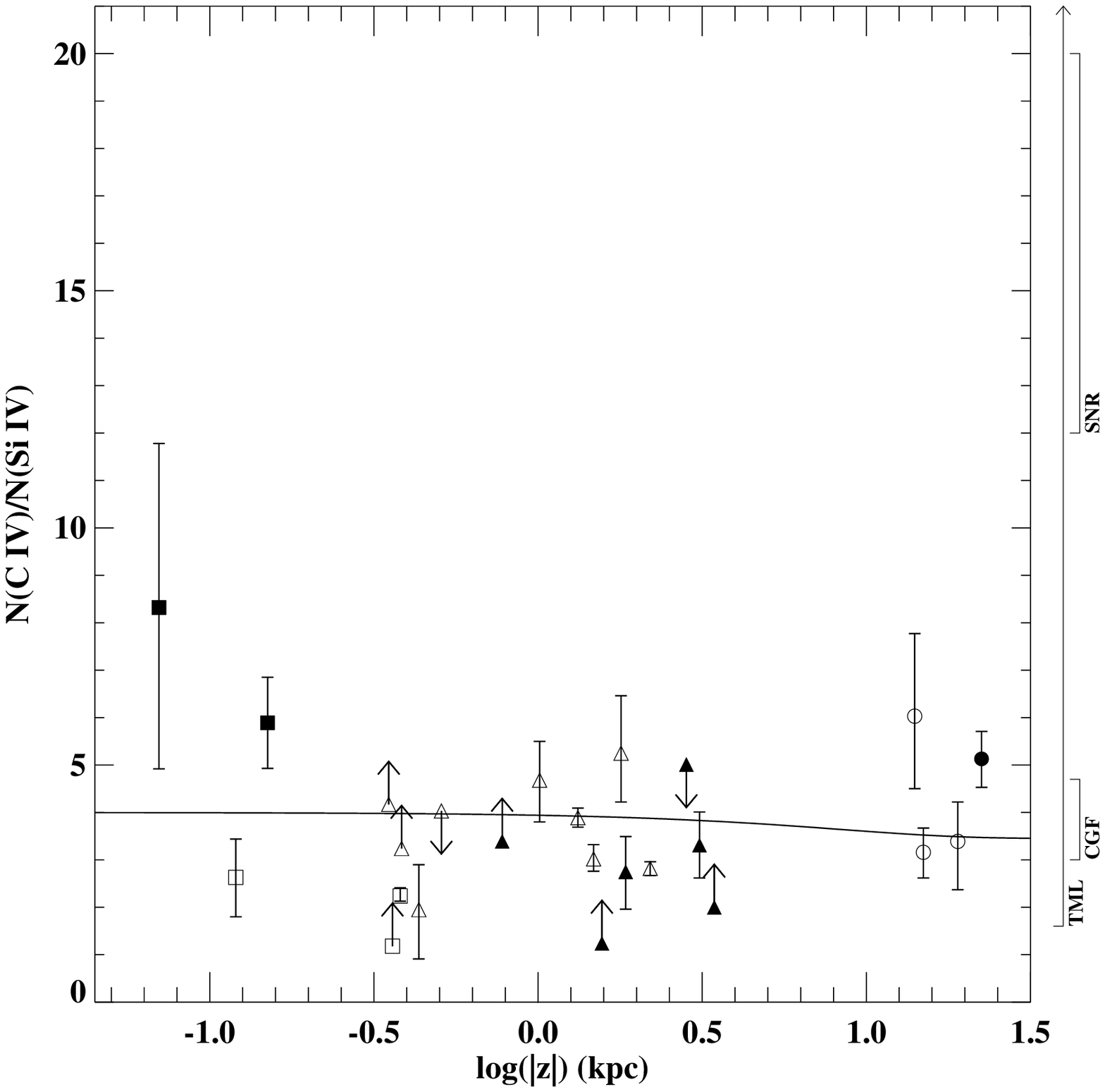}}
\figcaption{The N(\ion{C}{4})/N(\ion{Si}{4}) ratios as a function of distance from the Galactic 
plane. Triangles indicate our measurements, while the ratios toward extragalactic targets (Savage {\it et al}. 2002; 
Hoopes {\it et al}. 2002) are shown by circles. We also display the ratios toward the six stars from 
the disk sample of Spitzer (1996) as squares. Lower and upper limits are indicated 
by the appropriate arrows.  Sight lines with $b <$~0$^{\circ}$ and $b \geq$~0$^{\circ}$ are shown with open 
and solid symbols, respectively. The predicted N(\ion{C}{4})/N(\ion{Si}{4}) ratios, assuming simple exponential
distributions, are plotted by a solid line (see \S6.2). 
The ion ratios allowed by the various theories of high ion production are indicated on the right-hand side 
of the figure.  The values and the references for the theoretical predictions are the same as those for 
Figure~\ref{CIVvsNV}. References for the \ion{Si}{4} and \ion{C}{4} column densities for disk and
extragalactic sight lines can be found in Spitzer (1996), Savage {\it et al}. (2002), and 
Savage {\it et al}. (1997). 
\label{CIVpSiIVvsZ}}
\end{center}
\end{figure}

\clearpage
\begin{figure}[htbp] 
\begin{center}
\rotatebox{0}{
\epsscale{0.9}
\plotone{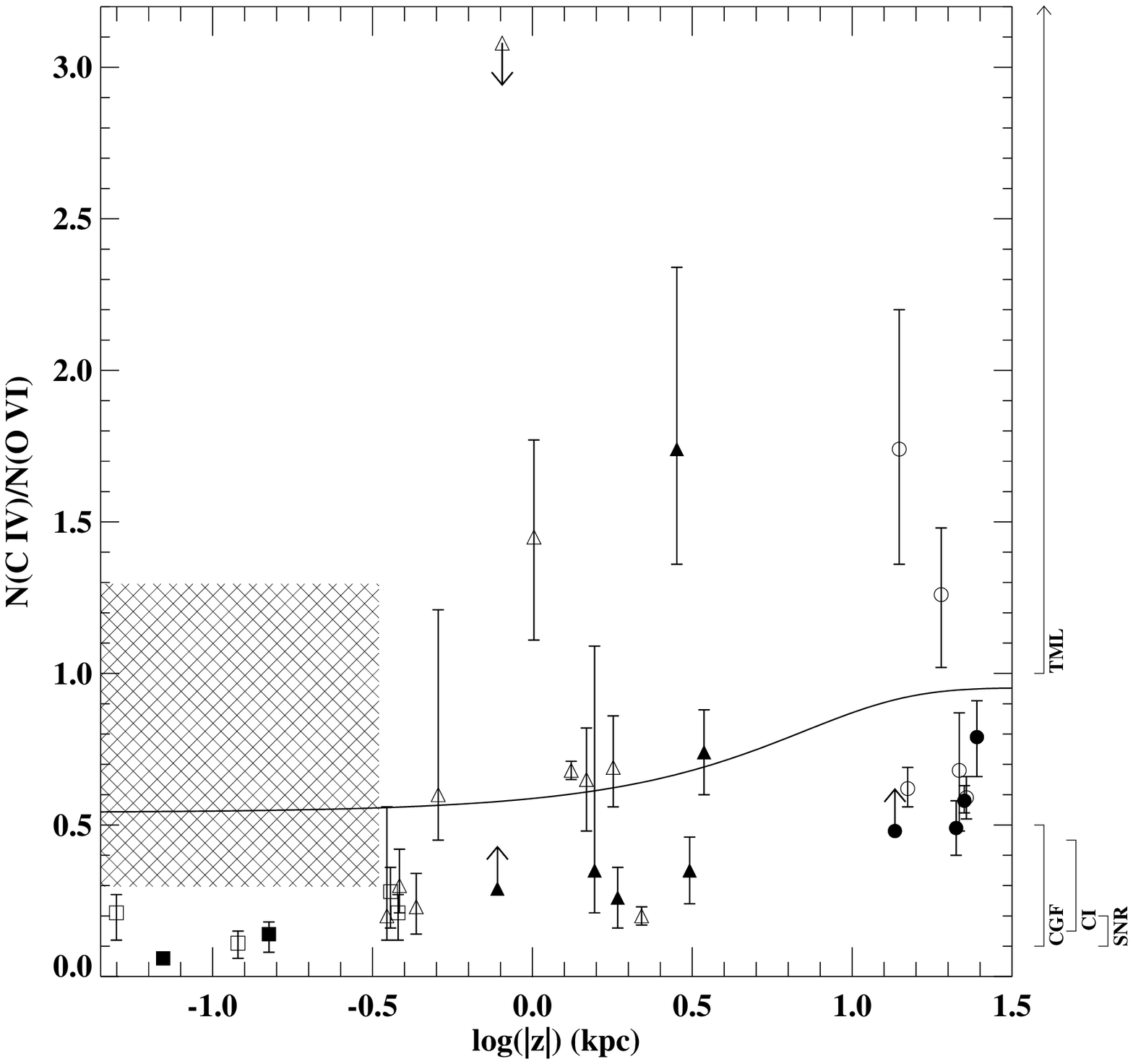}}
\figcaption{Same as Figure~\ref{CIVpSiIVvsZ} for N(\ion{C}{4})/N(\ion{O}{6}). The cross-hatched
region at $log(|z|) \leq$ -0.5 indicates the $\pm$1$\sigma$ range of N(\ion{C}{4})/N(\ion{O}{6})
in the disk survey of E. Jenkins (2002, private communication).
\label{CIVvsZ}}
\end{center}
\end{figure}

\clearpage
\begin{figure}[htbp] 
\begin{center}
\rotatebox{0}{
\epsscale{0.9}
\plotone{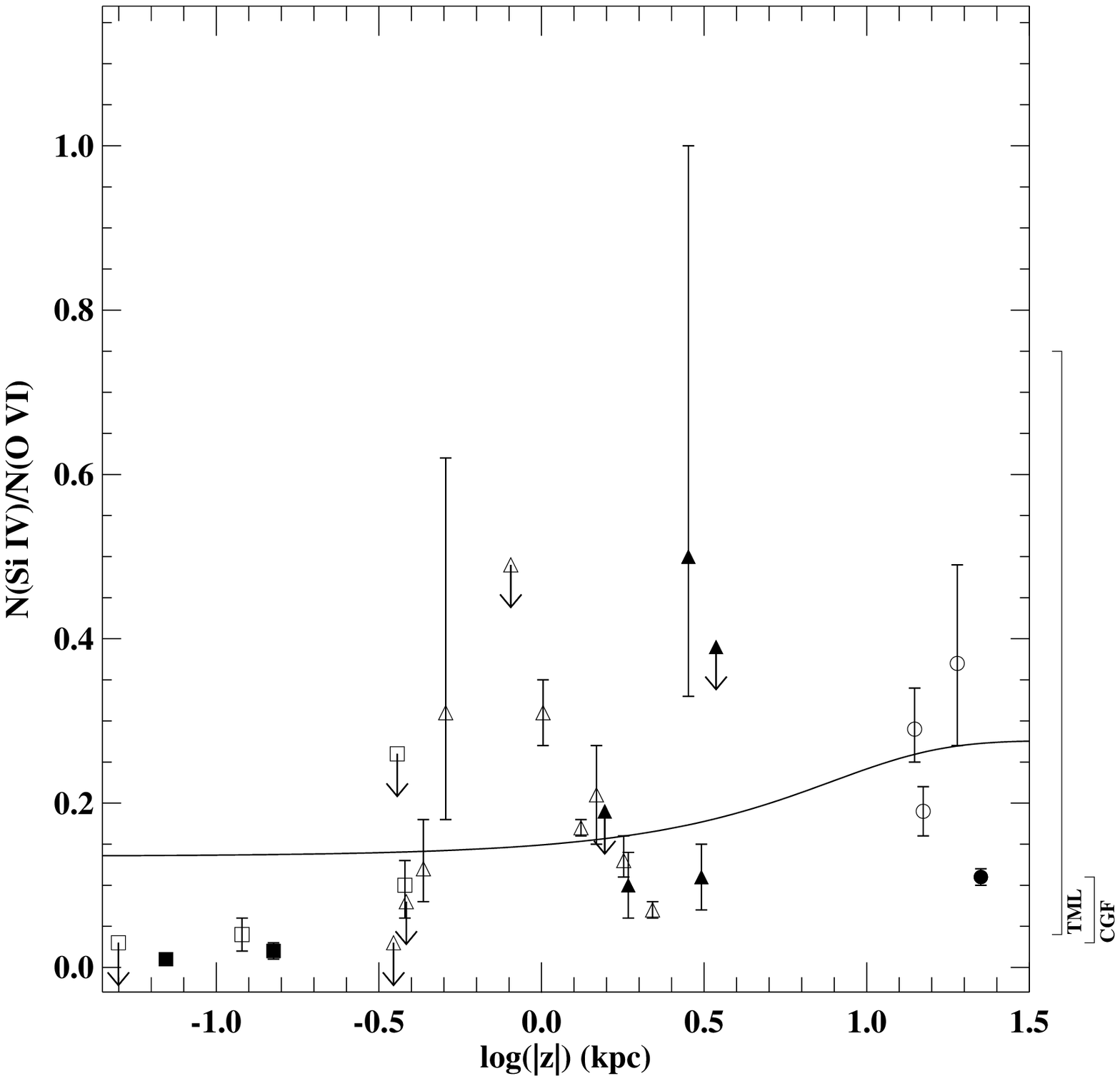}}
\figcaption{Same as Figure~\ref{CIVpSiIVvsZ} for N(\ion{Si}{4})/N(\ion{O}{6}).
\label{SiIVvsZ}}
\end{center}
\end{figure}

\clearpage
\begin{figure}[htbp] 
\begin{center}
\rotatebox{0}{
\epsscale{0.9}
\plotone{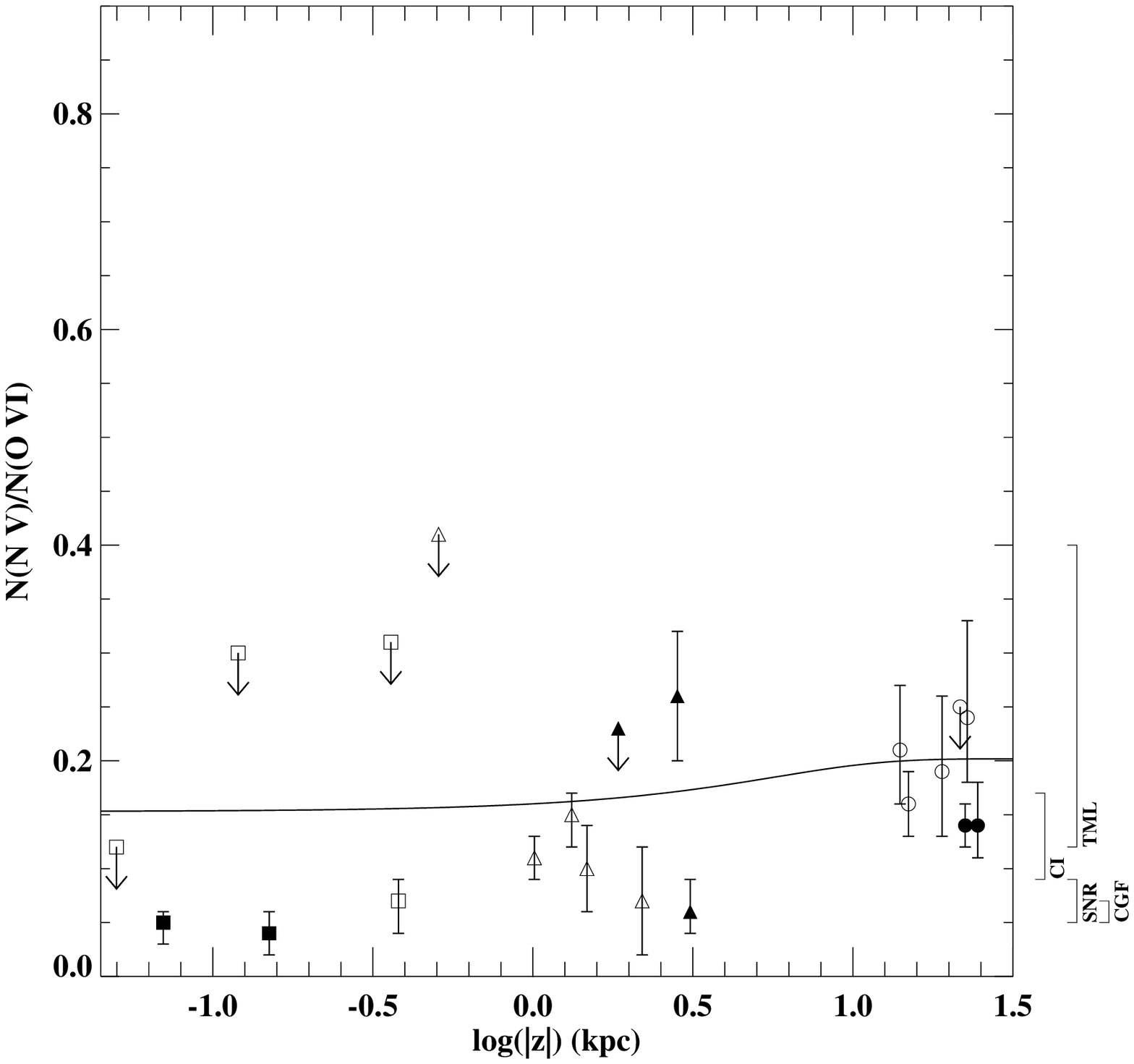}}
\figcaption{Same as Figure~\ref{CIVpSiIVvsZ} for N(\ion{N}{5})/N(\ion{O}{6}).
References for the \ion{N}{5} column densities for the disk and extragalactic sight lines can be found 
in Spitzer (1996), Savage {\it et al}. (2002), and Savage {\it et al}. (1997). 
\label{NVvsZ}}
\end{center}
\end{figure}

\clearpage
\begin{figure}[htbp]
\begin{center}
\vspace{2cm} 
\rotatebox{0}{
\epsscale{1.0}
\plotone{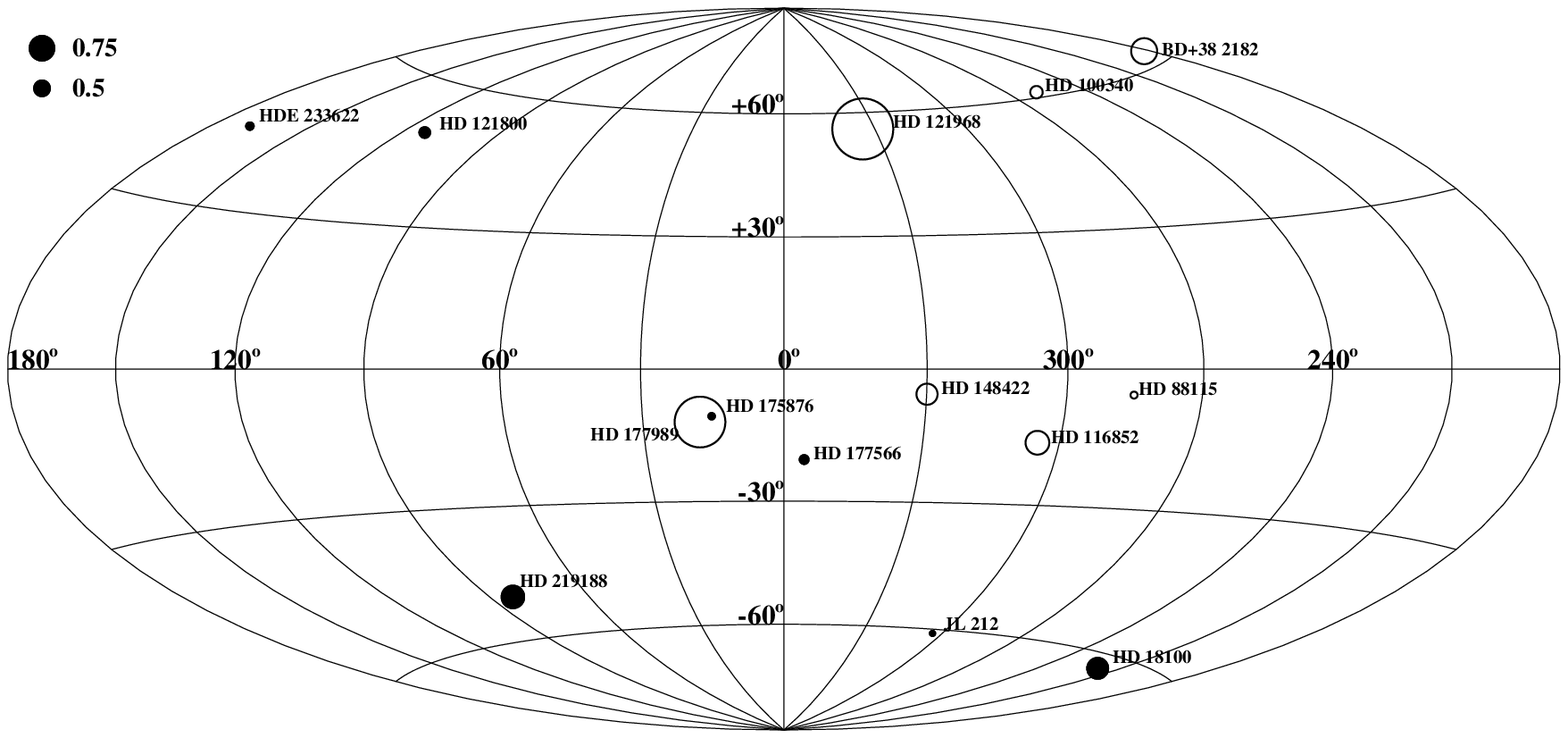}}
\vspace{2cm}
\figcaption{The N(\ion{C}{4})/N(\ion{O}{6}) ratios toward our sight lines, displayed in a Hammer-Aitoff 
projection. The Galactic Center is in the middle of the Figure and Galactic longitude increases to the left. 
The diameter of each circle is linearly proportional to the value of the ratio, and the 
scaling is displayed in the upper left corner.  
Open and closed symbols represent measurements for stars with $d >$ 3~kpc and $d \leq$ 3~kpc, respectively.
The name of each object is printed beside its location. NGC~6397-ROB~162, NGC~5904-ZNG~1, NGC~5139-ROA~5342, 
and NGC~6723-III 60 were shortened to ROB~162, ZNG~1, ROA~5342, and III 60, respectively.
\label{CIV_per_OVIgalaxy}}
\end{center}
\end{figure}

\clearpage
\begin{deluxetable}{lcrrcrrrr}
\tabletypesize{\footnotesize}
\tablewidth{0pc} 
\tablecaption{Target Stars. \label{tab1}}
\tablehead{
Star & Spectral & $v \, sin(i)$ &   V~~ & E(B-V) &      $l$     &      $b$     &   d  &   z  \\
ID   & type     &  km s$^{-1}$  & (mag) &        & ($^{\circ}$) & ($^{\circ}$) & (pc) & (pc) \\
}
\startdata
BD+38~2182        & B3V      & $\ldots$ & 11.25 & 0.00 & 182.16 & +62.21 &  3890 &  3441 \\
HD~3827           & B0.7V    &    125   &  8.01 & 0.02 & 120.79 & -23.23 &  2040 &  -804 \\
HD~18100          & B1V      &    265   &  8.46 & 0.02 & 217.93 & -62.73 &  1660 & -1476 \\
HD~88115          & B1.5IIn  &    245   &  8.30 & 0.16 & 285.32 &  -5.53 &  3640 &  -351 \\
HD~97991          & B1V      & $\ldots$ &  7.41 & 0.04 & 262.34 & +51.73 &   990 &   777 \\
HD~100340         & B1V      &    275   & 10.11 & 0.02 & 258.85 & +61.23 &  3540 &  3103 \\
HD~116852         & O9III    &    136   &  8.47 & 0.22 & 304.88 & -16.13 &  4760 & -1322 \\
HD~121800         & B1.5V    &    121   &  9.11 & 0.08 & 113.01 & +49.76 &  2050 &  1565 \\
HD~121968         & B1V      &    160   & 10.25 & 0.09 & 333.97 & +55.84 &  3420 &  2830 \\
HD~148422         & B1Ia     &     81   &  8.60 & 0.28 & 329.92 &  -5.60 &  5200 &  -507 \\
HD~175876         & O6.5IIIn &    266   &  6.95 & 0.22 &  15.28 & -10.58 &  2360 &  -433 \\
HD~177566         & PAGB     & $\ldots$ & 10.20 & 0.07 & 355.55 & -20.42 &  1100 &  -384 \\
HD~177989         & B0III    &    150   &  9.33 & 0.25 &  17.81 & -11.88 &  4910 & -1011 \\
HD~219188         & B0.5II-I &    197   &  7.05 & 0.14 &  83.03 & -50.17 &  2330 & -1789 \\
HDE~225757        & B1IIIn   & $\ldots$ & 10.59 & 0.22 &  69.64 &  +4.85 &  6630 &   560 \\
HDE~233622        & B2V      &    240   & 10.01 & 0.03 & 168.17 & +44.23 &  2650 &  1849 \\
JL~212            & B2       &    219   & 10.20 & 0.13 & 303.63 & -61.03 &  2510 & -2196 \\
NGC~6397-ROB~162  & PAGB     & $\ldots$ & 13.23 & 0.20 & 337.90 & -11.40 &  2300 &  -500 \\
NGC~5904-ZNG~1    & PAGB     & $\ldots$ & 14.20 & 0.03 &   3.49 & +47.40 &  7500 &  5500 \\
NGC~5139-ROA~5342 & PAGB     & $\ldots$ & 15.89 & 0.12 & 308.45 & +15.40 &  5300 &  1400 \\
NGC~6723-III~60   & PAGB     & $\ldots$ & 15.61 & 0.05 & 359.69 & -16.70 &  8800 & -2600 \\
vZ~1128           & PAGB     & $\ldots$ & 15.03 & 0.01 &  42.50 & +78.68 & 10200 & 10000 \\
\enddata
\tablenotetext{~}{ Original sources for the listed quantities can be found in 
Savage \& Sembach (1994) and Sembach \& Savage (1992). The visual magnitudes of the 
PAGB-s (except vZ1128) are
from Alcaino \& Liller (1980), Bohlin {\it et al}. (1983), Dickens, Brodie, \& Bingham (1988), 
and Menzies (1974).
Their values of E(B-V) and distances are from Heber \& Kudritzki (1986) and Harris (1996).}
\end{deluxetable}

\clearpage
\begin{deluxetable}{lcrrcc}
\tabletypesize{\footnotesize}
\tablewidth{0pc} 
\tablecaption{Rejected Stars. \label{tab6}}
\tablehead{
Star       & Spectral\tablenotemark{a}   &      $l$     &      $b$     &  FUSE Archival  & Category\tablenotemark{b}  \\
ID         & type       & ($^{\circ}$) & ($^{\circ}$) &       ID                &           \\
}
\startdata
BARNARD29  & PAGB       &    58.97     &    +40.93    &  P1015201                   &  3b   \\
BD+35~4258 & B0.5Vn     &    77.19     &     -4.74    &  P1017901/P1017902          &  3b   \\
BD+48~3437 & B1Iab      &    93.56     &     -2.06    &  P1018401/P1018402/P1018403 &  3a/c \\
CPD-691743 & B0.5IIIn   &   303.71     &     -7.35    &  P1013701                   &  2a/c \\
CPD-721184 & B0III      &   299.15     &    -10.94    &  S5140102                   &  3a   \\
CPD-741569 & O9.5V      &   317.02     &    -18.68    &  P1015301                   &  3a   \\
HD~73      & B1.5IV     &   114.17     &    -18.69    &  P1010101/P1010102          &  3a   \\
HD~12740   & B1.5II     &   135.30     &    -11.91    &  P1010701                   &  3a   \\ 
HD~22586   & B2III      &   264.19     &    -50.36    &  P1011101                   &  3a   \\
HD~34656   & O7IIf      &   170.04     &     +0.27    &  P1011301                   &  2b/a \\
HD~66788   & O8V        &   245.43     &     +2.05    &  P1011801                   &  2a/b \\
HD~92702   & B1Iab      &   286.15     &     +0.96    &  S5130301                   &  3a/c \\
HD~93840   & BN1Ib      &   282.14     &    +11.10    &  P1012701                   &  3a   \\
HD~119069  & B1III      &   312.05     &    +16.12    &  P1014001                   &  3b/a \\
HD~119608  & B1Ib       &   320.35     &    +43.13    &  P1014201                   &  3a   \\
HD~125924  & B2IV       &   338.16     &    +48.28    &  P1014701                   &  3a   \\
HD~146813  & B1.5       &    85.71     &    +43.81    &  P1014901                   &  3a   \\
HD~149881  & B0.5III    &    31.37     &    +36.23    &  P1015102                   &  3a   \\
HD~152218  & O9.5IVn    &   343.53     &     +1.28    &  P1015402                   &  2b/a \\
HD~156359  & O9.7Ib-II  &   328.68     &    -14.52    &  P1015502                   &  2d   \\
HD~158243  & B1Iab      &   337.59     &    -10.64    &  P1015601                   &  3d   \\
HD~160993  & B1Iab      &   345.61     &     -8.56    &  P1015701                   &  3d   \\
HD~163522  & B1Ia       &   349.57     &     -9.09    &  P1015801                   &  3d   \\
HD~163758  & O6.5Iaf    &   355.36     &     -6.10    &  P1015901                   &  2d/b \\
HD~164816  & O9.5III-IV &     6.06     &     -1.20    &  P1016001                   &  3a   \\
HD~167402  & B0Ib       &     2.26     &     -6.39    &  P1016201                   &  3a/d \\
HD~168941  & O9.5II-III &     5.82     &     -6.31    &  P1016501/P1016502          &  3d   \\
HD~172140  & B0.5III    &     5.28     &    -10.61    &  P1016601/P1016602/P1016603 &  3a   \\
HD~173502  & B0.5III    &     5.36     &    -12.27    &  P1016701                   &  3a/b \\
HD~175754  & O8IIf      &    16.39     &     -9.92    &  P1016802                   &  2d/b \\
HD~178487  & B0Ib       &    25.78     &     -8.56    &  P1017201                   &  3a/b \\
HD~179407  & B0.5Ib     &    24.02     &    -10.40    &  P1017301                   &  3a/b \\
HD~183899  & B2III      &    13.07     &    -20.14    &  P1017601                   &  3a   \\
HD~195455  & B0.5III    &    20.27     &    -32.14    &  P1017801                   &  2a   \\
HD~201638  & B0.5Ib     &    80.29     &     -8.44    &  P1018001                   &  3a   \\
HD~214080  & B1Ib       &    44.80     &    -56.92    &  P1018502                   &  3a/d \\
HD~215733  & B1II       &    85.16     &    -36.35    &  P1018602                   &  3a   \\
HD~218915  & O9.5Iab    &   108.06     &     -6.89    &  P1018801                   &  3d/b \\
LS~4825    & B1Ib-II    &     1.67     &     -6.63    &  P1016101                   &  3c   \\
\enddata
\tablenotetext{a}{The original sources for spectral types are in Savage \& Sembach (1994) and 
Sembach \& Savage (1992).}
\tablenotetext{b}{The category assigned to the star. See \S\ref{section:Sample} for details. The letters
indicate the reason for rejection; a: complex stellar continuum; b: blending with unidentified 
narrow stellar lines or strong HD $\lambda$1031.91 contamination; 
c: low S/N ratio; d: presence of strong stellar wind features near or at the \ion{O}{6} 1031.93 line.}
\end{deluxetable}

\clearpage
\begin{deluxetable}{lccrr}
\tabletypesize{\footnotesize}
\tablewidth{0pc} 
\tablecaption{Description of the Observations. \label{tab2}}
\tablehead{
Star & FUSE Archival & Aperture/Mode\tablenotemark{a} & Total Exposure Time 
& S/N\tablenotemark{b} \\
ID   &      ID       &                                &        per Observation (s)           
& ratio                \\
}
\startdata
BD+38~2182        & P1012801 & LWRS/TTAG & 12226 & 55 \\
HD~3827           & P1010302 & MDRS/HIST & 2885  & 92 \\
HD~18100          & P1011002 & MDRS/HIST & 1932  & 69 \\
HD~88115          & P1012301 & LWRS/HIST & 4512  & 69 \\
HD~97991          & P1013001 & LWRS/HIST &   59  & 14 \\
HD~100340         & P1013201 &  HRS/HIST & 5281  & 66 \\
HD~116852         & P1013801 & LWRS/HIST & 7212  & 77 \\
HD~121800         & P1014401 & LWRS/HIST & 3987  & 65 \\
HD~121968         & P1014501 & LWRS/HIST & 9301  & 54 \\
HD~148422         & P1015001/P1015002/P1015003 
& LWRS/TTAG & 2063/3026/3243 & 28 \\
HD~175876         & P1016902 & MDRS/HIST & 1932  & 85 \\
HD~177566         & P1017003 & LWRS/HIST & 7764  & 74 \\
HD~177989         & P1017101 & LWRS/HIST & 10289 & 65 \\
HD~219188         & P1018902 & MDRS/HIST & 1660  & 72 \\
HDE~225757        & P1017701/P1017703 & LWRS/TTAG 
& 5805/6351      & 26 \\
HDE~233622        & P1012102 & LWRS/HIST & 4662  & 89 \\
JL~212            & P1010401 & LWRS/HIST & 5437  & 62 \\
NGC~6397-ROB~162  & A0260201 & LWRS/TTAG & 11500 & 31 \\
NGC~5904-ZNG~1    & A1080303 & LWRS/TTAG & 3900  & 20 \\
NGC~5139-ROA~5342 & A1080101 & LWRS/TTAG & 7400  & 11 \\
NGC~6723-III~60   & A1080201 & LWRS/TTAG & 5400  & 11 \\
vZ~1128           & P1014101/P1014102/P1014103
& LWRS/TTAG & 9151/8248/15610 & 25 \\
\enddata
\tablenotetext{a}{TTAG: time-tagged mode; HIST: histogram mode \\
LWRS: 30~$\times$~30 arcsec; MDRS: 4~$\times$~20 arcsec; HRS: 1.25~$\times$~20 arcsec }
\tablenotetext{b}{The S/N ratio per resolution element ($\sim$20~km~s$^{-1}$)
measured around the \ion{O}{6} $\lambda$1031.93 line. } 
\end{deluxetable}

\clearpage
\begin{deluxetable}{lccccrc}
\tabletypesize{\footnotesize}
\tablewidth{0pc} 
\tablecaption{Description of the \ion{O}{6} Absorption.\label{tab4}}
\tablehead{
Star & $\left[ l , b \right]$ & $\left[ v_- , v_+ \right]$\tablenotemark{a} & W$_{\lambda}$\tablenotemark{b} & $\log N$ & $\log  N_{\perp} $\tablenotemark{c} & $\overline{n}$\tablenotemark{d} \\
ID   &                        &       km s$^{-1}$                           &          m\AA                  &          &                                     & (10$^{-8}$ cm$^{-3}$)           \\
}
\startdata
BD+38~2182        & $\left[ 182.2 , +62.2 \right]$ & $\left[  -75 ,  74 \right]$ & 123$\pm$14 & 14.10$\pm$0.07          & 14.05       & 1.05$\pm$0.17 \\
HD~3827           & $\left[ 120.8 , -23.2 \right]$ & $\left[  -88 ,  53 \right]$ &  71$\pm$ 8 & 13.80$\pm$0.04          & 13.40       & 1.00$\pm$0.09 \\
HD~18100          & $\left[ 217.9 , -62.7 \right]$ & $\left[ -101 ,  87 \right]$ &  70$\pm$11 & 13.77$^{+0.10}_{-0.13}$ & 13.72       & 1.15$\pm$0.30 \\
HD~88115          & $\left[ 285.3 ,  -5.5 \right]$ & $\left[ -152 ,  75 \right]$ & 250$\pm$12 & 14.44$\pm$0.02          & 13.42       & 2.45$\pm$0.11 \\
HD~97991          & $\left[ 262.3 , +51.7 \right]$ &           NA                &  $\leq$61  & $\leq$13.69             & $\leq$13.58 & $\leq$1.60    \\
HD~100340         & $\left[ 258.9 , +61.2 \right]$ & $\left[ -142 , 156 \right]$ & 246$\pm$23 & 14.28$^{+0.11}_{-0.16}$ & 14.22       & 1.74$\pm$0.52 \\
HD~116852\tablenotemark{e} & $\left[ 304.9 , -16.1 \right]$ & $\left[  -96 ,  68 \right]$ & 179$\pm$11 & 14.30$\pm$0.02 & 13.74       & 1.36$\pm$0.06 \\
HD~121800         & $\left[ 113.0 , +49.8 \right]$ & $\left[  -84 , 136 \right]$ & 227$\pm$19 & 14.40$\pm$0.03          & 14.28       & 3.97$\pm$0.27 \\
HD~121968         & $\left[ 334.0 , +55.8 \right]$ & $\left[  -96 ,  84 \right]$ &  98$\pm$18 & 13.97$\pm$0.06          & 13.89       & 0.88$\pm$0.12 \\
HD~148422\tablenotemark{e} & $\left[ 329.9 ,  -5.6 \right]$ & $\left[  -90 ,  43 \right]$ & 133$\pm$25 & 14.14$\pm$0.09 & 13.13       & 0.86$\pm$0.18 \\
HD~175876\tablenotemark{e} & $\left[  15.3 , -10.6 \right]$ & $\left[ -114 ,  66 \right]$ & 146$\pm$11 & 14.14$^{+0.10}_{-0.13}$ & 13.40 & 1.90$\pm$0.49 \\
HD~177566         & $\left[ 355.6 , -20.4 \right]$ & $\left[  -52 ,  66 \right]$ &  51$\pm$ 8 & 13.65$^{+0.06}_{-0.08}$ & 13.19       & 1.32$\pm$0.21 \\
HD~177989\tablenotemark{e} & $\left[  17.8 , -11.9 \right]$ & $\left[  -95 , 101 \right]$ & 196$\pm$10 & 14.31$\pm$0.06 & 13.62       & 1.35$\pm$0.19 \\
HD~219188\tablenotemark{e} & $\left[  83.0 , -50.2 \right]$ & $\left[  -85 ,  79 \right]$ &  99$\pm$ 8 & 13.97$\pm$0.06 & 13.86 & 1.30$\pm$0.18 \\
HDE~225757        & $\left[  69.6 ,  +4.9 \right]$ & $\left[  -84 ,  72 \right]$ & 253$\pm$27 & 14.57$\pm$0.04          & 13.50       & 1.82$\pm$0.17 \\
HDE~233622        & $\left[ 168.2 , +44.2 \right]$ & $\left[ -108 ,  72 \right]$ &  82$\pm$10 & 13.87$^{+0.13}_{-0.19}$ & 13.71       & 0.91$\pm$0.32 \\
JL~212            & $\left[ 303.6 , -61.0 \right]$ & $\left[  -49 , 100 \right]$ & 152$\pm$13 & 14.21$\pm$0.07          & 14.15       & 2.09$\pm$0.34 \\
NGC~6397-ROB~162  & $\left[ 337.9 , -11.4 \right]$ & $\left[  -73 , 111 \right]$ & 174$\pm$22 & 14.25$\pm$0.06          & 13.55       & 2.51$\pm$0.35 \\
NGC~5904-ZNG~1    & $\left[   3.5 , +47.4 \right]$ & $\left[  -70 ,  67 \right]$ & 181$\pm$34 & 14.41$\pm$0.08          & 14.28       & 1.11$\pm$0.21 \\
NGC~5139-ROA~5342 & $\left[ 308.5 , +15.4 \right]$ & $\left[ -112 ,  56 \right]$ & 252$\pm$83 & 14.54$^{+0.10}_{-0.12}$ & 13.96       & 2.12$\pm$0.53 \\
NGC~6723-III~60   & $\left[ 359.7 , -16.7 \right]$ & $\left[  -49 ,  87 \right]$ & 207$\pm$67 & 14.37$^{+0.10}_{-0.12}$ & 13.83       & 0.86$\pm$0.22 \\
vZ~1128\tablenotemark{f} & $\left[  42.5 , +78.7 \right]$ & $\left[ -160 , 100 \right]$ & 260$\pm$ 7 & 14.49$\pm$0.03          & 14.48       & 0.98$\pm$0.07 \\
                  &                                &                             &            &                         &             &               \\
Average$\pm \sigma$ & $\ldots$ & $\ldots$ &  $\ldots$  & 14.17$\pm$0.28 & 13.77$\pm$0.37 & 1.56$\pm$0.75 \\
Median            &         $\ldots$               &          $\ldots$           &  $\ldots$  & 14.25                   & 13.74       & 1.35          \\
\enddata
\tablenotetext{a}{Velocity limits for integration of $N_a(v)$.}
\tablenotetext{b}{Equivalent widths are based on the \ion{O}{6} 1031.93 \AA~line.}
\tablenotetext{c}{$N_{\perp} = N $sin$ |b|$}
\tablenotetext{d}{$\overline{n}$= N/d, the average \ion{O}{6} density along the sight line.}
\tablenotetext{e}{Substantial HD~R(0) $\lambda$1031.91 contamination in the \ion{O}{6} absorption at 1031.93~\AA\ . See 
\S\ref{section:AnMeth} for a discussion of the HD decontamination process.}
\tablenotetext{f}{See Howk, Sembach, \& Savage (2002) for details.}
\end{deluxetable}

\clearpage
\begin{deluxetable}{lcrcrrcrrcrrcrr}
\tabletypesize{\footnotesize}
\tablewidth{0pc} 
\tablecaption{Kinematical Description of the Low- and High-Ionization States.\label{tab3}}
\tablehead{
Star & &  & & \multicolumn{2}{c}{\ion{Ar}{1}} & & \multicolumn{2}{c}{\ion{O}{6}} & & \multicolumn{2}{c}{\ion{C}{4}} & 
& \multicolumn{2}{c}{\ion{Si}{4}} \\
\cline{5-6} \cline{8-9} \cline{11-12} \cline{14-15}
ID   & & $v_{exp}$\tablenotemark{a} & & $<v>$\tablenotemark{b} & $b$\tablenotemark{b} & & $<v>$\tablenotemark{b} & 
$b$\tablenotemark{b} & & $<v>$\tablenotemark{b} & $b$\tablenotemark{b} & & $<v>$\tablenotemark{b} & $b$\tablenotemark{b} \\
}
\startdata
BD+38~2182        & &   0.4 & & -27.1 & 36.6 & &  -3.4    & 35.2     & & -21.6    & 42.3     & & -11.8    & 47.8     \\
HD~3827           & & -13.9 & & -28.1 & 29.6 & & -14.9    & 39.0     & & $\ldots$ & $\ldots$ & & $\ldots$ & $\ldots$ \\
HD~18100          & &   3.0 & &  -8.0 & 19.8 & &  -8.3    & 45.0     & & -14.2    & 27.6     & & -14.1    & 20.8     \\
HD~88115          & &  -6.8 & & -23.2 & 21.2 & & -22.1    & 59.2     & & -18.5    & 32.1     & & $\ldots$ & $\ldots$ \\
HD~97991          & &   1.6 & &  -2.7 & 24.9 & & $\ldots$ & $\ldots$ & &  3.5     & 41.9     & & $\ldots$ & $\ldots$ \\
HD~100340         & &   3.3 & & -13.2 & 23.3 & &   9.9    & 78.3     & &  18.6    & 44.5     & &  13.2    & 36.6     \\
HD~116852         & & -26.4 & & -14.3 & 22.1 & & -13.1    & 38.7     & & -25.9    & 36.8     & & -28.7    & 34.1     \\
HD~121800         & &  -5.9 & &  -6.9\tablenotemark{c}  & 18.6\tablenotemark{c} & &  12.1   & 52.0     & & -38.2    & 56.3     & & -47.9    & 50.2     \\
HD~121968         & &  -6.4 & &   0.7 & 22.3 & &   0.6    & 38.2     & & -21.5    & 29.8     & & -19.1    & 34.9     \\
HD~148422         & & -40.2 & & -26.8 & 33.5 & & -14.1    & 35.5     & & -30.0    & 34.5     & & -35.9    & 31.7     \\
HD~175876         & &  10.7 & &  -3.2 & 30.2 & & -16.3    & 59.2     & &  21.2    & 28.4     & &  18.9    & 25.7     \\
HD~177566         & &  -1.3 & &  -8.1 & 22.9 & &   0.3    & 32.6     & &  -3.0    & 29.6     & & $\ldots$ & $\ldots$ \\
HD~177989         & &  26.7 & & -12.7 & 17.3 & &  11.2    & 49.5     & &  27.3    & 41.9     & &  25.5    & 36.9     \\
HD~219188         & &   0.4 & &  -6.3 & 19.5 & &  -7.9    & 43.8     & &  10.7    & 40.6     & &   8.7    & 36.3     \\
HDE~225757\tablenotemark{d}        & &   9.6 & &   1.0 & 20.1 & &   3.5    & 38.2     & & $\ldots$ & $\ldots$ & & $\ldots$ & $\ldots$ \\
HDE~233622        & &  -3.8 & &  -0.1 & 27.8 & & -21.4    & 47.2     & & $\ldots$ & $\ldots$ & & $\ldots$ & $\ldots$ \\
JL~212            & &  -3.8 & &  -8.4 & 18.4 & &  14.2    & 37.0     & &  28.2    & 22.8     & &  20.4    & 22.1     \\
NGC~6397-ROB~162\tablenotemark{d}  & & -14.3 & &  -10.0 & 19.2 & &   9.2    & 48.2     & & $\ldots$ & $\ldots$ & & $\ldots$ & $\ldots$ \\
NGC~5904-ZNG~1    & &   3.0 & &  -0.9 & 18.7 & &  -5.8    & 36.5     & & $\ldots$ & $\ldots$ & & $\ldots$ & $\ldots$ \\
NGC~5139-ROA~5342\tablenotemark{d} & & -31.3 & & -15.8 & 20.3 & & -17.5    & 43.2     & & $\ldots$ & $\ldots$ & & $\ldots$ & $\ldots$ \\
NGC~6723-III~60\tablenotemark{d}   & &  -3.2 & &  -0.7 & 19.6 & &  24.0    & 44.3     & & $\ldots$ & $\ldots$ & & $\ldots$ & $\ldots$ \\
vZ~1128\tablenotemark{e}           & &   1.5 & & -13.1 & 20.6 & & -31.3    & 52.8     & & $\ldots$ & $\ldots$ & & $\ldots$ & $\ldots$ \\
\enddata
\tablenotetext{~}{Data for \ion{Ar}{1} and \ion{O}{6} are from this investigation, while those for \ion{C}{4} and \ion{Si}{4} 
are from Savage, {\it et al}. (2001a).}
\tablenotetext{a}{The expected Galactic rotational velocity based on the Galactic rotation curve of
Clemens (1985).}
\tablenotetext{b}{Based on the \ion{Ar}{1} $\lambda$1048.22, \ion{O}{6} $\lambda$1031.93, 
\ion{C}{4} $\lambda$1548.20, and \ion{Si}{4} $\lambda$1393.76 lines.  Quantities  $<v>$ and $b$ are 
in km~s$^{-1}$ and were calculated by Equations~\ref{eq:<v>} and \ref{eq:bvalue}.}
\tablenotetext{c}{ 
Only the component at lower $|v|_{LSR}$ was used to calculate $<v>$ and $b$. See Figure~\ref{lines}.}
\tablenotetext{d}{The LSR velocities are not well defined.}  
\tablenotetext{e}{See Howk {\it et al}. (2002a) for details.}
\end{deluxetable}

\clearpage
\begin{deluxetable}{lccccc}
\tabletypesize{\scriptsize}
\tablewidth{0pc} 
\tablecaption{Ion Ratios of \ion{Si}{4}, \ion{C}{4}, \ion{N}{5}, and \ion{O}{6}. \label{tab5}}
\tablehead{
Star & & $\frac{ N(C~IV) }{ N(Si~IV) }$ & $\frac{ N(C~IV) }{ N(O~VI) }$ & 
$\frac{ N(Si~IV) }{ N(O~VI) }$ & $\frac{ N(N~V) }{ N(O~VI) }$ \\
ID                          & &                      &                         &                      &                      \\
}
\startdata
BD+38~2182                  & & $\geq$2.00           & 0.74$\pm$0.14           & $\leq$0.39           &  $\ldots$           \\
HD~3827                     & &  $\ldots$            & $\leq$3.08              & $\leq$0.49           &  $\ldots$            \\
HD~18100\tablenotemark{b}   & & 3.02$\pm$(0.30,0.26)\tablenotemark{a} & 0.65$\pm$0.17 & 0.21$\pm$0.06 & 0.10$\pm$0.04        \\
HD~88115                    & & $\geq$4.17           & 0.20$\pm$(0.36, 0.08)   & $\leq$0.03           &  $\ldots$            \\
HD~97991                    & & $\geq$3.39           & $\geq$0.28              & $\ldots$             &  $\ldots$            \\
HD~100340\tablenotemark{b}  & & 3.31$\pm$(0.70,0.69) & 0.35$\pm$0.11           & 0.11$\pm$0.04        & 0.06$\pm$(0.03,0.02) \\
HD~116852\tablenotemark{c}  & & 3.89$\pm$0.20        & 0.68$\pm$0.03           & 0.17$\pm$0.01        & 0.15$\pm$(0.02,0.03) \\
HD~121800                   & & $\geq$1.23           & 0.35$\pm$(0.74,0.14)    & $\leq$0.19           &  $\ldots$            \\
HD~121968\tablenotemark{d}  & & $\leq$5.00           & 1.74$\pm$(0.60,0.38)    & 0.50$\pm$(0.50,0.17) & 0.26$\pm$0.06        \\
HD~148422\tablenotemark{e}  & & $\leq$4.03           & 0.60$\pm$(0.61,0.15)    & 0.31$\pm$(0.31,0.13) & $\leq$0.41           \\
HD~175876                   & & 1.95$\pm$(0.95,1.04) & 0.23$\pm$(0.11,0.09)    & 0.12$\pm$(0.06,0.04) &  $\ldots$            \\
HD~177566                   & & $\geq$3.24           & 0.30$\pm$(0.12,0.09)    & $\leq$0.08           &  $\ldots$            \\
HD~177989\tablenotemark{f}  & & 4.68$\pm$(0.82,0.88) & 1.45$\pm$(0.32,0.34)    & 0.31$\pm$0.04        & 0.11$\pm$0.02        \\
HD~219188                   & & 5.25$\pm$(1.21,1.03) & 0.69$\pm$(0.17,0.13)    & 0.13$\pm$(0.03,0.02) &  $\ldots$            \\
HDE~233622\tablenotemark{g} & & 2.75$\pm$(0.74,0.79) & 0.26$\pm$0.10           & 0.10$\pm$0.04        & $\leq$0.23           \\
JL~212\tablenotemark{g}     & & 2.82$\pm$(0.14,0.15) & 0.20$\pm$0.03           & 0.07$\pm$0.01        & 0.07$\pm$0.05        \\
                            & &                      &                         &                      &                      \\
Average\tablenotemark{h}  $\pm$1$\sigma$ & &    3.46$\pm$1.09     &    0.60$\pm$0.47        &     0.20$\pm$0.13    & 0.12$\pm$0.07        \\
                            & &                      &                         &                      &                      \\
Models~:                    & &                      &                         &                      &                      \\
CGF\tablenotemark{i}        & &      3.0-4.7         &       0.1-0.5           &       0.03-0.11      &  0.05-0.07           \\
TML\tablenotemark{j}        & &       1.6-90         &       1.0-8.4           &       0.04-0.75      &   0.12-0.4           \\
SNR\tablenotemark{i}        & &       12-20          &       0.1-0.2           &      $\leq$0.01      &  0.05-0.09           \\
CI\tablenotemark{k}         & &       27-50          &      0.15-0.45          &      $\leq$0.02      &  0.09-0.17           \\
\enddata
\tablenotetext{~}{Data for \ion{C}{4} and \ion{Si}{4} are always from Savage {\it et al}. (2001a) unless
noted otherwise.}
\tablenotetext{a}{ 3.02$\pm$(0.30,0.26)= 3.02$^{+0.30}_{-0.26}$.}
\tablenotetext{b}{\ion{Si}{4}, \ion{C}{4}, and \ion{N}{5} column densities are from Savage~\&~Sembach~(1994).}
\tablenotetext{c}{\ion{Si}{4}, \ion{C}{4}, and \ion{N}{5} column densities are from Sembach~\&~Savage~(1994).}
\tablenotetext{d}{\ion{N}{5} column density is from Sembach,~\&~Savage~(1992).}
\tablenotetext{e}{\ion{N}{5} column density is from Sembach {\it et al}. (1997).}
\tablenotetext{f}{\ion{Si}{4}, \ion{C}{4}, and \ion{N}{5} column densities are from 
Savage {\it et al}. (2001b).}
\tablenotetext{g}{\ion{Si}{4}, \ion{C}{4}, and \ion{N}{5} column densities were measured by 
the authors of the present article using STIS observations.}
\tablenotetext{h}{The lower and upper limits were not included in the calculation.}
\tablenotetext{i}{The Cooling Galactic Fountain (CGF) model of R. Benjamin (2002, private communication) 
and the cooling SNR model of Slavin \& Cox (1992). The quoted numbers are from Sembach {\it et al}. (1997).}
\tablenotetext{j}{The Turbulent Mixing Layer (TML) model of Slavin {\it et al}. (1993). The quoted ranges cover all
values that are predicted by the TML models for the entrainment velocities, abundances, and postmixed gas temperatures 
used by Slavin {\it et al}. (1993).}
\tablenotetext{k}{The magnetized Conductive Interface (CI) model of Borkowski {\it et al}. (1990). The values are 
appropriate for magnetic field inclinations between 0$^{\circ}$ and 85$^{\circ}$ (relative to the front normal)
and for a conduction front age of 2.5$\times$10$^5$ years.}
\end{deluxetable}

\clearpage
\begin{deluxetable}{lcrrccl}
\tabletypesize{\footnotesize}
\tablewidth{0pc} 
\tablecaption{Closely Aligned Galactic and Extragalactic Sight Lines. \label{tab7}}
\tablehead{
Star/AGN        &      $\left[ l, b \right]$     &   $\Delta$\tablenotemark{a}   &    d    & 
$\left[ v_- , v_+ \right]$\tablenotemark{b} & $\log N$ & Note\tablenotemark{c} \\
ID              &  & ($^{\circ}$) &  (pc)   &  (km s$^{-1}$)  &  & \\
}  
\startdata
HD~175876       & $\left[  15.3 , -10.6 \right]$ & \nodata &  2360   & $\left[ -114 ,  66 \right]$ & 14.14$^{+0.10}_{-0.13}$ & MW  \\
HD~177989       & $\left[  17.8 , -11.9 \right]$ &    2.8  &  4910   & $\left[  -95 , 101 \right]$ & 14.31$\pm$0.06          & MW  \\
                &                                &         &         &                             &                         &     \\
                &                                &         &         &                             &                         &     \\
vZ~1128         & $\left[  42.5 , +78.7 \right]$ & \nodata & 10200   & $\left[ -160 , 100 \right]$ & 14.49$\pm$0.03          & MW  \\
PG~1402+261     & $\left[  33.0 , +73.5 \right]$ &    5.7  & \nodata & $\left[ -120 , 115 \right]$ & 14.53$\pm$0.05          & MW  \\
                &                                &         &         &                             &                         &     \\
                &                                &         &         &                             &                         &     \\
HD~219188       & $\left[  83.0 , -50.2 \right]$ & \nodata &  2330   & $\left[  -85 ,  79 \right]$ & 13.97$\pm$0.06          & MW  \\
NGC~7469        & $\left[  83.1 , -45.5 \right]$ &    4.7  & \nodata & $\left[  -65 ,  45 \right]$ & 13.96$\pm$0.09          & MW  \\
                &                                &         &         & $\left[ -370 ,-235 \right]$ & 14.18$\pm$0.12          & MS  \\
                &                                &         &         & $\left[ -235 ,-120 \right]$ & 14.22$\pm$0.16          & LG  \\
NGC~7714        & $\left[  88.2 , -55.6 \right]$ &    6.2  & \nodata & $\left[  -60 ,  45 \right]$ & 13.85$\pm$0.15          & MW  \\
                &                                &         &         & $\left[ -310 ,-230 \right]$ & 14.13$\pm$0.13          & MS  \\
                &                                &         &         &                             &                         &     \\
                &                                &         &         &                             &                         &     \\
HD~121800       & $\left[ 113.0 , +49.8 \right]$ & \nodata &  2050   & $\left[  -84 , 136 \right]$ & 14.40$\pm$0.03          & MW  \\ 
PG~1351+640     & $\left[ 111.9 , +52.0 \right]$ &    2.4  & \nodata & $\left[ -100 , 100 \right]$ & 14.38$\pm$0.07          & MW  \\
                &                                &         &         & $\left[ -160 ,-100 \right]$ & 13.67$\pm$0.24          & C   \\
                &                                &         &         & $\left[  100 , 160 \right]$ & 13.44$\pm$0.26          & Oth \\
Mrk~279         & $\left[ 115.0 , +46.9 \right]$ &    3.2  & \nodata & $\left[ -115 , 100 \right]$ & 14.41$\pm$0.03          & MW  \\
                &                                &         &         & $\left[ -210 ,-115 \right]$ & 13.67$\pm$0.14          & C   \\
                &                                &         &         &                             &                         &     \\
                &                                &         &         &                             &                         &     \\
HDE~233622      & $\left[ 168.2 , +44.2 \right]$ & \nodata &  2650   & $\left[ -108 ,  72 \right]$ & 13.87$^{+0.13}_{-0.19}$ & MW  \\
Mrk~106         & $\left[ 161.1 , +42.9 \right]$ &    5.3  & \nodata & $\left[ -100 , 100 \right]$ & 14.45$\pm$0.09          & MW  \\
                &                                &         &         & $\left[ -150 ,-100 \right]$ & 13.81$\pm$0.22          & A   \\
Mrk~116         & $\left[ 160.5 , +44.8 \right]$ &    5.5  & \nodata & $\left[ -125 , 110 \right]$ & 14.21$\pm$0.06          & MW  \\
                &                                &         &         &                             &                         &     \\
                &                                &         &         &                             &                         &     \\
BD+38~2182      & $\left[ 182.2 , +62.2 \right]$ & \nodata &  3890   & $\left[  -75 ,  74 \right]$ & 14.10$\pm$0.07          & MW  \\
Mrk~421         & $\left[ 179.8 , +65.0 \right]$ &    3.0  & \nodata & $\left[ -130 , 100 \right]$ & 14.39$\pm$0.04          & MW  \\
                &                                &         &         & $\left[  100 , 185 \right]$ & 13.51$\pm$0.22          & EPn \\
HS~1102+3441    & $\left[ 188.6 , +66.2 \right]$ &    4.9  & \nodata & $\left[ -140 ,  95 \right]$ & 14.71$\pm$0.08          & MW  \\
                &                                &         &         & $\left[   95 , 210 \right]$ & 14.30$\pm$0.12          & EPn \\
Ton~1187        & $\left[ 188.3 , +55.4 \right]$ &    7.5  & \nodata & $\left[  -90 ,  65 \right]$ & 14.35$\pm$0.08          & MW  \\
                &                                &         &         &                             &                         &     \\
                &                                &         &         &                             &                         &     \\
HD~18100        & $\left[ 217.9 , -62.7 \right]$ & \nodata &  1660   & $\left[ -101 ,  87 \right]$ & 13.77$^{+0.10}_{-0.13}$ & MW  \\
HE~0238-1904    & $\left[ 200.5 , -63.6 \right]$ &    7.9  & \nodata & $\left[ -120 , 110 \right]$ & 14.33$\pm$0.09          & MW  \\ 
                &                                &         &         &                             &                         &     \\
                &                                &         &         &                             &                         &     \\
HD~100340       & $\left[ 258.9 , +61.2 \right]$ & \nodata &  3540   & $\left[ -142 , 156 \right]$ & 14.28$^{+0.11}_{-0.16}$ & MW  \\
Mrk~734         & $\left[ 244.8 , +63.9 \right]$ &    7.0  & \nodata & $\left[  -35 , 140 \right]$ & 14.55$\pm$0.08          & MW  \\
                &                                &         &         & $\left[  140 , 275 \right]$ & 14.10$\pm$0.19          & EPn \\
                &                                &         &         &                             &                         &     \\
                &                                &         &         &                             &                         &     \\
JL~212          & $\left[ 303.6 , -61.0 \right]$ & \nodata &  2510   & $\left[  -49 , 100 \right]$ & 14.21$\pm$0.07          & MW  \\
Fairall~9       & $\left[ 295.1 , -57.8 \right]$ &    5.4  & \nodata & $\left[ -110 , 100 \right]$ & 14.38$\pm$0.09          & MW  \\
                &                                &         &         & $\left[  100 , 275 \right]$ & 14.33$\pm$0.10          & MS  \\
                &                                &         &         &                             &                         &     \\
                &                                &         &         &                             &                         &     \\
HD~177566       & $\left[ 355.6 , -20.4 \right]$ & \nodata &  1100   & $\left[  -52 ,  66 \right]$ & 13.65$^{+0.06}_{-0.08}$ & MW  \\
NGC~6723-III~60 & $\left[ 359.7 , -16.7 \right]$ &    5.4  &  8800   & $\left[  -49 ,  87 \right]$ & 14.37$^{+0.10}_{-0.12}$ & MW  \\
Tol~1924-416    & $\left[ 356.9 , -24.1 \right]$ &    3.9  & \nodata & $\left[  -70 ,  95 \right]$ & 14.62$\pm$0.05          & MW  \\
\enddata
\tablenotetext{~}{\footnotesize Data for the thick disk and the high velocity absorption toward extragalactic sight 
lines are from Savage {\it et al}. (2002) and Sembach {\it et al}. (2002), respectively. }
\tablenotetext{a}{Angular separation is measured from the first sight line in each group.}
\tablenotetext{b}{Velocity limits for integration of $N_a(v)$.}
\tablenotetext{c}{Identification of the \ion{O}{6} absorption: MW= Milky Way thick disk/low halo, A= Complex A, C= Complex C,
EPn= Extreme Positive (North), LG= Local Group, MS= Magellanic Stream, Oth= Other. See Sembach {\it et al}. 
(2002) for details on HVC identification. }
\end{deluxetable}

\end{document}